\documentclass[journal]{IEEEtran}
\ifCLASSINFOpdf
   \usepackage[pdftex]{graphicx}
   \graphicspath{{../pdf/}{../jpeg/}}
   \DeclareGraphicsExtensions{.pdf,.jpeg,.png}
\else
   \usepackage[dvips]{graphicx}
   \graphicspath{{../eps/}}
   \DeclareGraphicsExtensions{.eps}
\fi

\usepackage{booktabs}
\usepackage{multirow}
\usepackage{amsmath}
\usepackage{amsthm}
\usepackage{amsfonts}
\usepackage[linesnumbered,boxed,ruled,commentsnumbered]{algorithm2e}

\usepackage{algpseudocode}
\usepackage{array}
\ifCLASSOPTIONcompsoc
  \usepackage[caption=false,font=footnotesize,labelfont=sf,textfont=sf]{subfig}
\else
  \usepackage[caption=false,font=footnotesize]{subfig}
\fi
\usepackage{cite}
\usepackage{url}
\usepackage{color}
\usepackage{epstopdf}
\usepackage{overpic}

\newtheorem{lemma}{Lemma}
\newtheorem{Def}{Define}

\begin{document}
%
\title{Minimum-volume Multichannel Nonnegative Matrix Factorization For Blind Source Separation}
%
%
%

\author{\IEEEauthorblockN{Jianyu Wang,
Shanzheng Guan,
Shupei Liu,
Xiao-Lei Zhang, \textit{Senior Member, IEEE} }
\thanks{M. Shell was with the Department
of Electrical and Computer Engineering, Georgia Institute of Technology, Atlanta,
GA, 30332 USA e-mail: (see http://www.michaelshell.org/contact.html).}
\thanks{J. Doe and J. Doe are with Anonymous University.}
\thanks{Manuscript received April 19, 2005; revised August 26, 2015.}}

%
%

\markboth{Journal of \LaTeX\ Class Files,~Vol.~14, No.~8, August~2015}%
{Shell \MakeLowercase{\textit{et al.}}: Bare Demo of IEEEtran.cls for IEEE Journals}
%



\maketitle

\begin{abstract}
 Multichannel blind audio source separation aims to recover the latent sources from their multichannel mixtures without supervised information.
  One state-of-the-art blind audio source separation method, named independent low-rank matrix analysis (ILRMA), unifies independent vector analysis (IVA) and nonnegative matrix factorization (NMF).
 However, the spectra matrix produced from NMF may not find a compact spectral basis. It may not guarantee the identifiability of each source as well.
 To address this problem, here we propose to enhance the identifiability of the source model by a minimum-volume prior distribution. We further regularize a multichannel NMF (MNMF) and ILRMA respectively with the minimum-volume regularizer.
 The proposed methods maximize the posterior distribution of the separated sources, which ensures the stability of the convergence.
 Experimental results demonstrate the effectiveness of the proposed methods compared with auxiliary independent vector analysis, MNMF, ILRMA and its extensions.
\end{abstract}


%
\IEEEpeerreviewmaketitle

\section{Introduction}









\IEEEPARstart{B}{lind} source separation (BSS) is a technique of separating source components from a given multichannel mixture without any knowledge about the mixing system or microphone positions. Most BSS methods aim to cluster the time-frequency units of the spectrogram of the mixture into different sources.
A promising approach of multichannel BSS to achieve the above goal is to represent the hierarchical generative process of the time-frequency spectrogram of the mixture by a source model and a spatial model, where the source model represents the generative process of source spectrograms, and the spatial model represents the mixing process of the sources.

This paper focuses on nonnegative matrix factorization (NMF) based multichannel BSS \cite{ozerov2009multichannel,sawada2013multichannel,kitamura2016determined,mogami2019independent}.
It usually decomposes the spectrogram of a mixture into several spectral bases and temporal activations.
Existing NMF-based BSS methods usually have the following major problems.
First, because the NMF decomposition is an NP-hard problem, it is difficult to obtain a meaningful representation of the spectral bases.
Second, they may not guarantee that the spectral structure of each source is identifiable.
For example, different random initializations of NMF may produce dramatically different spectral structures of separated sources after the separation.
Finally, the non-sparse solution of standard NMF may lose some local information of the sources.

Simplex volume minimization {\cite{fu2016robust}}, which learns an identifiable spectral basis, provides a reliable estimation to the source model of BSS.
To our knowledge, it has not been explored in multichannel BSS yet, due to maybe the difficult mathematical derivation and convergence analysis of the simplex volume minimization.

\subsection{Contributions}
In this paper, we aim to explore the \textit{minimum-volume} (MinVol) prior for multichannel BSS. Specifically, we apply the MinVol prior as a regularizer for multichannel nonnegative matrix factorization (MNMF) \cite{sawada2013multichannel} and independent low-rank matrix analysis (ILRMA) \cite{kitamura2016determined}, which are named m-MNMF and m-ILRMA, respectively.
Because the object function of MinVol is to minimize $\left|\mathbf{W}^T\mathbf{W}\right|$ where $\mathbf{W}$ is the basis matrix of NMF and $\left|\cdot\right|$ denotes the determinant operator for a nonsingular matrix, it is formulated as a complicated optimization problem.
To overcome this difficulty, we design two auxiliary function for the object functions of m-MNMF and m-ILRMA respectively, and combine them with the maximum a posterior (MAP) estimation. Each auxiliary function is solved by iteratively updating the demixing matrix, spectrogram basis, and temporal activations in the function. The proposed methods improve the performance of MNMF and ILRMA with theoretically guaranteed advantages to the source model as follows:
\begin{itemize}
  \item MinVol improves the identifiability of the separated spectrograms that are produced from the source model.

  \item MinVol improves the sparseness levels of the factorized spectral basis matrices of the source model, which enhances the learning ability of the source model to capture the local information of sources.

  \item MinVol enhances the orthogonality of the factorized spectral basis matrices of the source model, which leads the spectral basis matrices to a rigorous clustering interpretation.
\end{itemize}

In this paper, we first introduce some related work and preliminaries in the following two subsections, then present the proposed MinVol prior distribution, as well as m-MNMF and m-ILRMA in Section \ref{Algorithm}.
Section \ref{Experiment} presents the experimental results. Finally, Section \ref{Conclusion} concludes our findings.

\subsection{Related work}

{
A multichannel BSS method is composed of a spatial model and a source model.
A mixture sound is usually represented as a sum of multiple source signals convolved with the room impulse responses of the corresponding source directions. It is equivalent to an instantaneous mix up in the frequency domain.
The mixture is usually separated by the spatial model, where the phase difference between microphones is important for the demixing system.
Common algorithms for multichannel BSS are independent component analysis (ICA) \cite{comon1994independent} and its extensions such as independent vector analysis (IVA) \cite{kim2006blind}.
They make a statistical independence assumption between the sources. However, they do not utilize the spectral structures of the source signals.
}


Recently, the importance of source models has been fully aware. According to the difference of the source models, modern multichannel BSS methods can be categorized mainly into the NMF-based methods, probability-based methods, and deep neural network (DNN) based methods, which will be introduced in the next three subsections respectively.

\subsubsection{NMF-based models}

The original ICA and IVA employ a spherical multivariate Laplace distribution as the source model to ensure higher-order correlations between the frequency bins in each source. However, these source models do not fully utilize the spectral structure of sources. As we know, the spectral structure may significantly help improve the BSS performance if properly incorporated into source models.

To overcome this weakness,
NMF \cite{lee1999learning,lee2001algorithms}, which is a nonnegative-parts-based low-rank decomposition of an observed nonnegative data matrix, can be used as a source model. It generates a "{clustering-friendly}" latent spectrogram basis for each source by introducing a low-rank structure into the source model.
Generally, NMF-based BSS models adopt Itakura-Saito divergence to evaluate the reconstruction error between the mixture and the estimated sources.
MNMF \cite{ozerov2009multichannel,sawada2013multichannel}, which is an extension of the NMF methods, estimates the mixing system of convolutive mixtures in a similar way to ICA and IVA, which is used for the clustering of spectrogram bases.
It consists of a low-rank source model and a full-rank spatial model. The full-rank spatial model is capable of representing a wide variety of source directivity under an echoic condition.

However, MNMF tends to get stuck in bad local optima, since that a large number of unconstrained spatial covariance matrices are needed to be estimated iteratively.
To address this problem, Kitamura \textit{et al.} \cite{kitamura2016determined,kitamura2018determined} proposed ILRMA. It makes a rank-1 assumption to the spatial model. It performs well for directional sources in practice. Essentially, the spatial model and source model of ILRMA are independent vector analysis (IVA) \cite{kim2006blind} and NMF respectively, which are optimized iteratively.

In the original MNMF and ILRMA, the observed signal is assumed to follow a time-variant multivariant complex Gaussian distribution.
Recently, the methods \textit{t}-MNMF \cite{kitamura2016student} and \textit{t}-ILRMA \cite{mogami2017independent} use the isotropic complex Cauchy distribution \cite{7336900} and its generalization---complex student's \textit{t}-distribution \cite{yoshii2016student} respectively to replace the original complex Gaussian distribution.
Because the Student's \textit{t}-distribution belongs to the family of the $\alpha$-stable distribution which is more suitable for modeling complex-valued signals than the complex Gaussian distribution, it is suitable for audio source modeling {\cite{mogami2017independent}}.
Moreover, Kitamura \textit{et al.} \cite{kitamura2018generalized,ikeshita2018independent,mogami2019independent,kamo2020joint} developed a complex generalized Gaussian distribution for ILRMA, which takes \textit{t}-ILRMA and ILRMA as its special cases.
To reduce the huge computational cost of the spatial covariance matrices,
Sekiguchi \textit{et al.} \cite{sekiguchi2019fast,sekiguchi2020fast} proposed a fast MNMF, which restricts the covariance matrices to jointly-diagonalizable full-rank matrices in a frequency-wise manner.
However, its source separation performance was not improved and the physical meaning of the joint-diagonalization process was unclear \cite{kamo2020regularized}. To address this issue,
Kamo \textit{et al.} \cite{kamo2020regularized} proposed FastMNMF with a new regularization, where the authors declared that the regularization can be applied to ILRMA as well.




\subsubsection{Probability-based models}

If the frequency bins of each source are sparsely distributed, the source spectrograms can be assumed to be disjoint with each other in most time-frequency units.
Under this assumption, Otsuka \textit{et al.} \cite{otsuka2013bayesian} proposed a Bayesian mixture model, called hierarchical latent Dirichlet allocation (LDA) \cite{blei2003latent}, to classify each time-frequency unit into one source only, and classify each source into a single direction. However, it does not build a source model, which is insufficient in utilizing the spectral structure of sources.

To overcome this weakness, probabilistic models were employed to build priors for the distributions of the parameters of the source model.
Itakura \textit{et al.} \cite{itakura2016unified} improve the LDA-based method \cite{otsuka2013bayesian} by combining the low-rank structure of the NMF-based source model. The method iteratively updates the spectrogram basis and temporal activations of the source model, and the variables of the LDA-based spatial model.
Itakura \textit{et al.} \cite{itakura2017bayesian} further introduced an anechoic spatial correlation matrix as a prior distribution of a real spatial correlation matrix for each direction, which avoids the impulse response assumption in previous studies.
Recently, Itakura \textit{et al.} \cite{itakura2018bayesian} proposed a unified Bayesian framework for multichannel BSS and incorporated prior knowledge of the microphone array into BSS. Based on the fundamental categorization of probabilistic models which can be categorized to mixture models and factor models, they proposed four methods for joint modeling the source and spatial models: factor-factor model, mixture-factor model, factor-mixture model, and mixture-mixture model.
The above models jointly estimate low-rank sources and spatial covariances on the fly. However, the low-rank assumption does not always hold for speech spectra. To remedy this problem,
Sekiguchi \textit{et al.} \cite{sekiguchi2018bayesian} proposed a semi-supervised method based on an extension of MNMF which consists of a deep generative model called variational auto-encoder (VAE) for speech spectra and a standard low-rank model for noise spectra.
Narisetty \textit{et al.} \cite{narisetty2019bayesian} used Bayesian non-parametric modeling of sources to avoid parameter tuning.

\subsubsection{DNN-based models}

To provide a highly accurate estimation to the parameters of the source model, supervised DNN has been introduced into multichannel BSS for the estimation to the source model \cite{sekiguchi2018bayesian, kameoka2018semi, sekiguchi2019semi, makishima2019independent, seki2019generalized, kameoka2019supervised, li2020determined}.
Sekiguchi \textit{et al.} \cite{sekiguchi2018bayesian,sekiguchi2019semi} proposed a deep pre-trained generative model of speech spectra and an NMF-based generative model of noise spectra for multichannel speech enhancement.
Makishima \textit{et al.} \cite{makishima2019independent} proposed independent deeply learned matrix analysis (IDLMA), which utilizes mutually independent DNN source models for the separation.
Kameoka \textit{et al.} \cite{kameoka2018semi, inoue2019joint, seki2019generalized, seki2019underdetermined, kameoka2019supervised} proposed a multichannel variational autoencoder (MVAE), which uses a conditional VAE to estimate the power spectrograms of sources.
Although the convergence of the optimization of MVAE is guaranteed, its computational complexity is high. Moreover, the accuracy of the source classification of MVAE is unsatisfied. To solve the problems,
Li \textit{et al.} \cite{li2019fast} employed an auxiliary classifier VAE, which is an information-theoretic extension of the conditional VAE, to learn the generative model of source spectrograms.
Togami \cite{togami2019multi} trained a source model by bidirectional long short-term memory networks with the multichannel Itakura-Saito distance as the training objective.
Li \textit{et al.} \cite{li2020determined} modeled power spectrograms of sources by a star generative adversarial network (StarGAN).
Although more and more DNN models were used in BSS, these models require clean sources for pre-training, which is out of the focus of this paper.
Therefore, we will not discuss and compare with the DNN-based models anymore.


\section{Problem formulation}

In this section, we formulate the BSS problem.
Suppose the short-time Fourier transform (STFT) of a multichannel mixture is $\mathbf{x}_{ij} = [x_{ij1},\dots,x_{ijm},\dots,x_{tfM}]^T\in\mathbb{C}^{M}$, where $i = 1,\dots,I$, $j = 1,\dots,J$, and $m = 1, \dots, M$ are the indices of the frequency bins, time frames, and microphones, respectively.
The complex spectrograms of source signals are defined as $\mathbf{s}_{ij} = [s_{ij1},\dots,s_{ijn},\dots,s_{ijN}]^T \in \mathbb{C}^N$, where $N$ is the number of sources and $n = 1,\dots,N$ is the index of the $n$th source, and $^T$ denotes the transpose operator.



\begin{figure*}[htb]
\vspace{-0.3cm}
\begin{minipage}[b]{1.0\linewidth}
  \centering
  \centerline{\includegraphics[width=16.5cm]{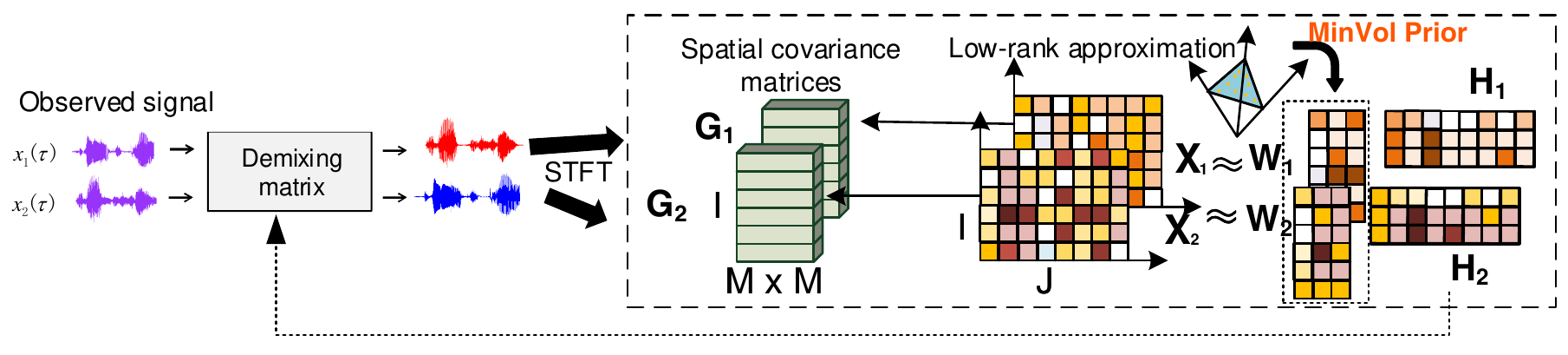}}
\end{minipage}
\vspace{-0.8cm}
\caption{Principle of the proposed m-MNMF algorithm.}
\label{fig:modelMNMF}
\vspace{-0.3cm}
\end{figure*}

We assume the mixing process in the frequency domain is instantaneous, and each source of the mixture is a point source. Then, the mixture and its sources have the following connection:
\begin{equation}\label{MixSys}
\begin{split}
 \mathbf{x}_{ij} = \mathbf{A}_i \mathbf{s}_{ij}
\end{split}
\end{equation}
where $\mathbf{A}_i = [\mathbf{a}_{i1},\dots,\mathbf{a}_{in},\dots,\mathbf{a}_{iN}] \in \mathbb{C}^{M \times N}$ is the mixing matrix at the $i$th frequency bin.
Similar to \cite{ozerov2009multichannel,sawada2013multichannel,itakura2018bayesian}, we assume that ${s}_{ijn}$ follows a zero-mean complex Gaussian distribution as follows:
\begin{equation}\label{sourceCGausDis}
\begin{split}
 s_{ijn} \sim \mathcal{N}_{\mathbb{C}}\left( 0, \lambda_{ijn} \right)
\end{split}
\end{equation}
where $\lambda_{ijn}$ is a power spectrum density of the source $n$ at time $j$ and frequency $f$.
Substituting \eqref{sourceCGausDis} into \eqref{MixSys}, the observation ${x}_{ijm}$ is found to follow the complex Gaussian distribution as follows:
\begin{equation}\label{MixCGausDis}
\begin{split}
 x_{ijn} \sim \mathcal{N}_{\mathbb{C}}\left( 0, \lambda_{ijn}\mathbf{G}_{in} \right)
\end{split}
\end{equation}
where $\mathbf{G}_{in} = \mathbf{a}_{in}\mathbf{a}_{in}^H$

{
The problem of source separation is to find an estimation of $(\mathbf{A}_i)^{-1}$, denoted as $\mathbf{D}_i = [\mathbf{d}_{i,1},\dots,\mathbf{d}_{i,N}]^H$, where $\mathbf{D}_i \in \mathbb{C}^{N \times M}$, such that when we apply $\mathbf{D}_i$ to $\mathbf{x}_{ij}$, we obtain the separated signal:
\begin{equation}\label{Demixing}
\begin{split}
 \mathbf{y}_{ij} = \mathbf{D}_i \mathbf{x}_{ij}
\end{split}
\end{equation}
where $^H$ denotes the Hermitian transpose, and $\mathbf{y}_{ij}$ is an estimation of $\mathbf{s}_{ij}$.
Here, we emphasize that MNMF is suitable for both the underdetermined situation ($M<N$) and the determined situation ($M=N$), while ILRMA is only suitable for the determined situation.
}



\section{proposed methods}
\label{Algorithm}

In this section, we first propose the MinVol based source model in Section \ref{subsec:minvol}, and then present the MinVol regularized MNMF and ILRMA respectively in Sections \ref{subsec:MNMF} and \ref{subsec:ILRMA}.


\subsection{Minimum-volume prior distribution for source models}\label{subsec:minvol}

We propose a minimum-volume prior distribution for the source model of the NMF-based BSS models.
Specifically, we formulate the generative process of source power spectrograms $\boldsymbol{\lambda} = [\boldsymbol{\lambda}_1,\dots,\boldsymbol{\lambda}_n,\dots,\boldsymbol{\lambda}_N] = \{\lambda_{ijn}\}_{i,j,n=1}^{I,J,N}$ as follows: $\boldsymbol{\lambda}$ is generated by a basis {spectra} $\mathbf{W} = [\mathbf{W}_1,\dots,\mathbf{W}_n,\dots,\mathbf{W}_N] = \{ w_{nik} \}_{n,i,k=1}^{N,I,K}$ and activations $\mathbf{H} = [\mathbf{H}_1,\dots,\mathbf{H}_n,\dots,\mathbf{H}_N] = \{ h_{nkj} \}_{n,k,j=1}^{N,K,J}$, where $K$ is the number of the bases of the basis matrix $\mathbf{W}$.
The power spectrogram of each source is decomposed into basis spectra and temporal activations by low-rank factorization:
\begin{equation}\label{SourceFac}
\begin{split}
\lambda_{ijn} = \sum_{k=1}^K w_{nik}h_{nkj}
\end{split}
\end{equation}

The above connection can construct a likelihood:
\begin{equation}\label{LikelihoodSourceFac}
\begin{split}
p(\boldsymbol{\lambda}_n|\mathbf{W}_n,\mathbf{H}_n) = \prod_{i=1}^I \prod_{j=1}^J \delta(\lambda_{ijn} - \sum_{k=1}^K w_{nik}h_{nkj})
\end{split}
\end{equation}
In many existing decomposition methods, the prior over $w_{nik}$ is constructed as a unified distribution over the non-negative real numbers,
\begin{equation}\label{priorW}
\begin{split}
p(w_{nik}) & = \lim_{u_w\rightarrow \infty} \frac{1}{u_w} \mathbb{I} [0\leq w_{nik} \leq u_w]  \\
& \propto \mathbb{I}[w_{nik} \geq 0]
\end{split}
\end{equation}
where $\mathbb{I} [\cdot]$ denotes an indicator function, which has the value one when its argument is true and zero otherwise.
The prior for $h_{nkj}$ is chosen as uniform between zero and one:
\begin{equation}\label{priorH}
\begin{split}
p(h_{nkj}) = \mathbb{I}[0\leq h_{nkj} \leq 1]
\end{split}
\end{equation}

Under the Bayes' rule, the posterior density of $w_{nik}$ and $h_{nkj}$ is given by:
\begin{equation}\label{GenerativeLambda}
\begin{split}
p(\mathbf{W}_n,\mathbf{H}_n|\boldsymbol{\lambda}_{n}) = & \frac{1}{Z} \prod_{i=1}^I \prod_{j=1}^J \delta(\lambda_{ijn} - \sum_{k=1}^K w_{nik}h_{nkj}) \\
& \times \mathbb{I}[w_{nik} \geq 0] \mathbb{I}[0\leq h_{nkj} \leq 1]
\end{split}
\end{equation}
where $Z$ is a normalization constant.

Many algorithms have been developed to find a unique and identifiable factorization for NMF, e.g. \cite{huang2013non, fu2016robust, fu2018identifiability, fu2019nonnegative, leplat2020blind}.
We are interested in the MinVol criterion among these algorithms.
MinVol is motivated by the nice geometrical interpretation of the constraints in \cite{fu2018identifiability}.
Under the MinVol constraints, all the data points lie in a \textit{convex hull} spanned by the spectrogram basis.
For convenience, we here present a probabilistic Bayesian formulation of a prior about \textit{the volume of the data simplex}:
\begin{equation}\label{VolumePrior}
\begin{split}
p(\mathbf{W}|\gamma) \propto \exp(-\gamma \log|\mathbf{W}^T\mathbf{W} + \eta \mathbf{I}|)
\end{split}
\end{equation}
where $|\cdot|$ is the determinant operator on a matrix, and $\gamma$ is a parameter that reflects the influence of the prior to the likelihood function.
We choose MinVol as a prior distribution of the spectrogram basis. It encourages the simplex spanned by the estimated spectrograms to be small, and constrains each element of the spectrogram basis to be non-negative.

The posterior density of the source model based on the volume of the data simplex can be represented as follows:
\begin{equation}\label{GenerativeLambdaMinVol}
\begin{split}
p(\mathbf{W}_n,\mathbf{H}_n|\boldsymbol{\lambda}_{n}) = & \frac{1}{Z} \prod_{i=1}^I \prod_{j=1}^J \delta(\lambda_{ijn} - \sum_{k=1}^K w_{nik}h_{nkj}) \\
& \times \exp(-\gamma \log|\mathbf{W}_n^T\mathbf{W}_n + \eta \mathbf{I}|) \\
& \times \mathbb{I}[w_{nik} \geq 0] \mathbb{I}[0\leq h_{nkj} \leq 1]
\end{split}
\end{equation}
Maximizing the likelihood of \eqref{GenerativeLambdaMinVol} is equivalent to the volume-minimization problem of the data simplex.

\subsection{MNMF with minimum-volume regularizer}\label{subsec:MNMF}

\subsubsection{Preliminary}

{
MNMF decomposes the spatial covariance matrix of each source into a weighted sum of direction-dependent matrices for joint source separation and location, which can be formulated as the following maximization problem:}
\begin{equation}\label{Likelihood_MNMF1}
\begin{split}
\log & [p(\mathbf{X}|\mathbf{W},\mathbf{H},\mathbf{G})p(\mathbf{W})p(\mathbf{H})]\\
 = & \sum_{i=1}^{I} \sum_{j=1}^{J} \log \mathcal{N}_{\mathbb{C}}(\mathbf{x}_{ij}|\mathbf{0},\hat{\mathbf{X}}_{ij})\\
 = & \sum_{i=1}^{I} \sum_{j=1}^{J} \left( -\mathrm{tr}\Big( \hat{\mathbf{X}}_{ij}^{-1} \mathbf{X}_{ij} \Big) - \log|\hat{\mathbf{X}}_{ij}| \right)  + \mathrm{const}
\end{split}
\end{equation}
where $\mathbf{X}_{ij}=\mathbf{x}_{ij}\mathbf{x}_{ij}^H$, $\hat{\mathbf{X}}_{ij}=\sum_n\lambda_{ij}\mathbf{G}_{in}$, and `$\mathrm{const}$' represents a constant.

{
It is usually solved by a multiplicative update rule which iteratively updates one of the parameters according to the conditional posterior distribution with the other parameters fixed:}
\begin{equation}\label{Likelihood_MNMFw}
\begin{split}
w_{ik} \leftarrow w_{ik} \sqrt{\frac{\sum_j h_{kj} \mathrm{tr}\left( \hat{\mathbf{X}}_{ij}^{-1} \mathbf{X}_{ij} \hat{\mathbf{X}}_{ij}^{-1} \mathbf{G}_{in} \right)}{\sum_j h_{kj} \mathrm{tr}\left( \hat{X}_{ij}^{-1} \mathbf{G}_{in} \right)} }
\end{split}
\end{equation}
\begin{equation}\label{Likelihood_MNMFh}
\begin{split}
h_{kj} \leftarrow h_{kj} \sqrt{\frac{\sum_i w_{ik} \mathrm{tr}\left( \hat{\mathbf{X}}_{ij}^{-1} \mathbf{X}_{ij} \hat{\mathbf{X}}_{ij}^{-1} \mathbf{G}_{in} \right)}{\sum_i w_{ik} \mathrm{tr}\left( \hat{X}_{ij}^{-1} \mathbf{G}_{in} \right)} }
\end{split}
\end{equation}
To update $\mathbf{G}_{ik}$, Sawada \textit{et al.} solve am algebraic Riccati equation:
\begin{equation}\label{Likelihood_MNMFG}
\begin{split}
\mathbf{G}_{in} \mathbf{A} \mathbf{G}_{in} = \mathbf{B}
\end{split}
\end{equation}
with $\mathbf{A}$ and $\mathbf{B}$ defined as:
\begin{equation}\label{Likelihood_MNMFAB}
\begin{split}
\mathbf{A} = \sum_j h_{kj} \hat{\mathbf{X}}_{ij}^{-1}, \quad \mathbf{B} = \mathbf{G}_{in}^\star \left( \sum_j h_{kj} \hat{\mathbf{X}}_{ij}^{-1} \right) \mathbf{G}_{in}^\star
\end{split}
\end{equation}
where $\mathbf{G}^\star$ is the old value of the variable $\mathbf{G}$ calculated in the previous step.

\subsubsection{Objective function of m-MNMF}

To remedy the non-unique identifiable problem of the source model of MNMF, here we propose m-MNMF. Fig. \ref{fig:modelMNMF} shows a conceptual model of m-MNMF, which is described as follows. 

The likelihood function of the unknown variables $\mathbf{W}$,$\mathbf{H}$,$\mathbf{G}$ of m-MNMF is formulated as:
\begin{equation}\label{Likelihood_MNMF}
\begin{split}
\log & [p(\mathbf{X}|\mathbf{W},\mathbf{H},\mathbf{G})p(\mathbf{W}|\gamma)p(\mathbf{H})]\\
 = & \sum_{i=1}^{I} \sum_{j=1}^{J} \log \mathcal{N}_{\mathbb{C}}(\mathbf{x}_{ij}|\mathbf{0},\hat{\mathbf{X}}_{ij}) - \sum_{n=1}^N \gamma\log|\mathbf{W}_n^T\mathbf{W}_n| \\
 = & \sum_{i=1}^{I} \sum_{j=1}^{J} \left( -\mathrm{tr}\Big( \hat{\mathbf{X}}_{ij}^{-1} \mathbf{X}_{ij} \Big) - \log|\hat{\mathbf{X}}_{ij}| \right) \\
& - \sum_{n=1}^N \gamma\log|\mathbf{W}_n^T\mathbf{W}_n + \eta \mathbf{I}| + \mathrm{const}
\end{split}
\end{equation}
where $\mathbf{X}_{ij}=\mathbf{x}_{ij}\mathbf{x}_{ij}^H$, $\hat{\mathbf{X}}_{ij}=\sum_{n=1}^N \lambda_{ijn}\mathbf{G}_{ni}$, and `$\mathrm{const}$' is a constant.

{Problem \eqref{Likelihood_MNMF} is intractable. Therefore, we propose to maximize its lower bound instead.}
To derive a lower bound for \eqref{Likelihood_MNMF}, we use the following two inequalities \cite{yoshii2016student} to relax { the logarithmic determinant term in \eqref{Likelihood_MNMF}}:

First, for a convex function $f(\mathbf{Z}) = -\log|\mathbf{Z}|$ with $\mathbf{Z}\geq \mathbf{0}$ being a positive semi-definite matrix, we have the following lower bound at an arbitrary positive semi-definite matrix $\mathbf{U} \geq \mathbf{0}$:
\begin{equation}\label{TaylorExp}
\begin{split}
f(\mathbf{Z}) = -\log|\mathbf{Z}| \geq -\log|\mathbf{U}| - \mathrm{tr}(\mathbf{U}^{-1}\mathbf{Z}) + M
\end{split}
\end{equation}
where the equality holds when $\mathbf{U}=\mathbf{Z}$.

Second, for a concave function $g(\mathbf{Z}) = -\mathrm{tr}(\mathbf{Z}^{-1}\mathbf{A})$ with any matrix $\mathbf{A} \geq \mathbf{0}$, we have the following lower bound:
\begin{equation}\label{ConcaveIneq}
\begin{split}
g(\{\mathbf{Z}_l\}_{l=1}^{L}) = - \mathrm{tr}\Bigg( \bigg( \sum_{l=1}^L \mathbf{Z}_l \bigg)^{-1} \mathbf{A} \Bigg) \geq -\sum_{l=1}^L \mathrm{tr}(\mathbf{Z}_l^{-1}\boldsymbol{\Phi}_l \mathbf{A} \boldsymbol{\Phi}_l^H)
\end{split}
\end{equation}
where $\{\mathbf{Z}_l\}_{l=1}^{L}$ is a set of arbitrary matrices, $\{ \boldsymbol{\Phi}_l \}_{l=1}^L$ is a set of auxiliary matrices that satisfies $\sum_l \boldsymbol{\Phi}_l = \mathbf{I}$, and the equality holds when $\boldsymbol{\Phi}_k = \mathbf{Z}_k(\sum_{l^\prime} \mathbf{Z}_{l^\prime})^{-1}$.

Substituting the two inequalities \eqref{TaylorExp} and \eqref{ConcaveIneq} into \eqref{Likelihood_MNMF} derives the following lower bound of \eqref{Likelihood_MNMF}, denoted as $\mathcal{L}$:
\begin{equation}\label{LikelihoodBound}
\begin{split}
\log & [p(\mathbf{X}|\mathbf{W},\mathbf{H},\mathbf{G})p(\mathbf{W}|\gamma)p(\mathbf{H})] \\
\geq & \sum_{i=1}^{I} \sum_{j=1}^{J} \Big( -\mathrm{tr}\big( \hat{\mathbf{X}}_{ij} \mathbf{U}_{ij}^{-1} \big) - \log|\mathbf{U}_{ij}| + M \Big) \\
& - \sum_{i=1}^I \sum_{j=1}^J \sum_{n=1}^N \mathrm{tr}\bigg( \hat{\mathbf{X}}_{ijn}^{-1} \boldsymbol{\Phi}_{ijn} \mathbf{X}_{ij} \boldsymbol{\Phi}_{ijn}^H \bigg) \\
& + \sum_{n=1}^N \Big( - \log|\mathbf{V}^{-1}| - \mathrm{tr}(\mathbf{V}\mathbf{W}_n^T \mathbf{W}_n) + K \Big) \\
= & - \sum_{i=1}^{I} \sum_{j=1}^{J} \sum_{n=1}^N \lambda_{ijn}\mathrm{tr}\big( {\mathbf{G}}_{ni} \mathbf{U}_{ij}^{-1} \big) - \sum_{i=1}^I \sum_{j=1}^J \log|\mathbf{U}_{ij}| \\
& - \sum_{i=1}^I \sum_{j=1}^J \sum_{n=1}^N \lambda_{ijn}^{-1} \mathrm{tr}\bigg( {\mathbf{G}}_{ni}^{-1} \boldsymbol{\Phi}_{ijn} \mathbf{X}_{ij} \boldsymbol{\Phi}_{ijn}^H \bigg) \\
& + \gamma \sum_{n=1}^N \Big( - \log|\mathbf{V}^{-1}| - \mathrm{tr}(\mathbf{V}\mathbf{W}_n^T \mathbf{W}_n) \Big) + \mathrm{const} = \mathcal{L}  \\
\end{split}
\end{equation}
where $\boldsymbol{\lambda}$ is a function of $\mathbf{H}$ and $\mathbf{W}$ defined in \eqref{SourceFac}, and $\mathbf{U}_{ij}$, $\boldsymbol{\Phi}_{ijn}$, and $\mathbf{V}$ are auxiliary variables.
The above lower bound is a tight one when $\mathbf{U}_{ij}$, $\boldsymbol{\Phi}_{ijn}$ and $\mathbf{V}$ satisfy:
\begin{eqnarray}
  &&\mathbf{U}_{ij} = \hat{\mathbf{X}}_{ij}\label{eq:xx1}\\
  &&\boldsymbol{\Phi}_{ijn} = \hat{\mathbf{X}}_{ijn} \hat{\mathbf{X}}_{ij}^{-1}\label{eq:xx2}\\
  &&\mathbf{V} = (\mathbf{W}_n^T\mathbf{W}_n + \eta\mathbf{I})^{-1}\label{eq:xx3}
\end{eqnarray}
The objective function of m-MNMF is to maximize $\mathcal{L} $.

\subsubsection{Optimization of m-MNMF}

m-MNMF is optimized by the multiplicative updating (MU) rule, which optimizes $\mathbf{H}_n$, $\mathbf{W}_n$, and $\mathbf{G}_{ni}$, alternatively.

Given $\mathbf{W}_n$ and $\mathbf{G}_{ni}$ fixed, $\mathbf{H}_n$ is calculated as follows. Letting the partial derivative of \eqref{LikelihoodBound} with respect to $h_{nkj}$ equal to zero derives:
\begin{equation}\label{DerivativeW}
\begin{split}
& \sum_{i=1}^I   h_{nkj}^{-2} w_{nik}^{-1} \mathrm{tr}\bigg( {\mathbf{G}}_{ni}^{-1} \boldsymbol{\Phi}_{ijn} \mathbf{X}_{ij} \boldsymbol{\Phi}_{ijn}^H \bigg) \\
& - \sum_{i=1}^I   w_{nik} \mathrm{tr}\big( {\mathbf{G}}_{ni} \mathbf{U}_{ij}^{-1} \big) = 0
\end{split}
\end{equation}
Substituting \eqref{eq:xx1} and \eqref{eq:xx2} into \eqref{DerivativeW} derives:
\begin{equation}\label{AuxiliaryRelationship_h}
\begin{split}
h_{nkj}^\star a^h h_{nkj}^\star = h_{nkj} b^h h_{nkj}
\end{split}
\end{equation}
with $h_{nkj}^\star$ as the old value of the variable $h_{nkj}$ calculated in the previous step, and $a^h$ and $b^h$ short for:
\begin{equation}\label{numeratorH}
\begin{split}
a^h = \sum_{i=1}^I w_{nik} \mathrm{tr}\bigg( {\mathbf{G}}_{ni} \hat{\mathbf{X}}_{ij}^{-1} \mathbf{X}_{ij} \hat{\mathbf{X}}_{ij}^{-1} \bigg)
\end{split}
\end{equation}
\begin{equation}\label{denominatorH}
\begin{split}
b^h = \sum_{i=1}^I w_{nik} \mathrm{tr}\big( {\mathbf{G}}_{ni} \hat{\mathbf{X}}_{ij}^{-1} \big)
\end{split}
\end{equation}
Solving \eqref{AuxiliaryRelationship_h} obtains $h_{nkj}$ as:
\begin{equation}\label{Update_H}
\begin{split}
h_{nkj} \leftarrow h_{nkj}^\star \sqrt{\frac{a_h}{b_h}}
\end{split}
\end{equation}

Given $\mathbf{H}_n$ and $\mathbf{G}_{ni}$ fixed, $\mathbf{W}_n$ is calculated as follows.
Because $\mathrm{tr}(\mathbf{V}\mathbf{W}_n^T\mathbf{W})$ of \eqref{LikelihoodBound} is quadratic and not separable,
we optimize a compact lower-bound of \eqref{LikelihoodBound} with an approximate separable auxiliary function. Specifically, we assume that $\mathbf{V}$ can be decomposed as $\mathbf{V} = \mathbf{V}^+ -\mathbf{V}^-$ with $\mathbf{V}^+ = \max{(\mathbf{V},\mathbf{0})}$ and $\mathbf{V}^- = \max{(-\mathbf{V},\mathbf{0})}$. Suppose that the diagonal matrix $\boldsymbol{\Omega}(\mathbf{w}_{ni}^T) = \mathrm{Diag} \Big( 2\frac{[\mathbf{V}^+\mathbf{w}_{ni}^T + \mathbf{V}^-\mathbf{w}_{ni}^T]}{[\mathbf{w}_{ni}^T]} \Big)$, and $\frac{[\mathbf{A}]}{[\mathbf{B}]}$ is the component-wise division between $\mathbf{A}$ and $\mathbf{B}$, then we obtain the lower-bound of $\mathcal{L}$ with respect to $w_{nik}$ as:
\begin{equation}\label{Likelihood_W}
\begin{split}
\mathcal{L}_{w_{nik}} & =  -\sum_{i=1}^I \sum_{j=1}^J \sum_{n=1}^N \Big( \sum_{k=1}^K w_{nik}h_{nkj} \Big) \mathrm{tr}(\mathbf{G}_{ni}\mathbf{U}_{ij}^{-1}) \\
& -\sum_{i=1}^I \sum_{j=1}^J \sum_{n=1}^N \Big( \sum_{k=1}^K w_{nik}h_{nkj} \Big)^{-1} \mathrm{tr}(\mathbf{G}_{ni}^{-1} \boldsymbol{\Phi}_{jin} \mathbf{X}_{ij} \boldsymbol{\Phi}_{ijn}^H) \\
& - \gamma \sum_{n=1}^N \sum_{i=1}^I \Big[ \hat{\mathbf{w}}_{ni} \mathbf{V} \hat{\mathbf{w}}_{ni}^T + 2\Delta{\hat{\mathbf{w}}_{ni}} \mathbf{V} \hat{\mathbf{w}}_{ni}^T \\
& + \Delta{\hat{\mathbf{w}}_{ni}} \boldsymbol{\Omega}{(\hat{\mathbf{w}}_{ni}^T)} \Delta{\hat{\mathbf{w}}_{ni}}^T \Big]
\end{split}
\end{equation}
where {\textbf{$\mathbf{w}_{ni}\in \mathbb{R}^{1\times K}$ is the $i$th row of $\mathbf{W}_n$}}, $\Delta{\hat{\mathbf{w}}_{ni}} = \hat{\mathbf{w}}_{ni} - \mathbf{w}_{ni}$.
We let the partial derivative of $\mathcal{L}_{w_{nik}}$ to zero:
\begin{equation}\label{DerivativeLikelihood_W}
\begin{split}
\frac{\partial \mathcal{L}_{w_{nik}}}{\partial w_{nik}} = &  w_{nik}^{-2} h_{nkj}^{-1} \mathrm{tr}\bigg( {\mathbf{G}}_{ni}^{-1} \boldsymbol{\Phi}_{ijn} \mathbf{X}_{ij} \boldsymbol{\Phi}_{ijn}^H \bigg) \\
& - h_{nkj} \mathrm{tr}\big( {\mathbf{G}}_{ni} \mathbf{U}_{ij}^{-1} \big) + 2\gamma [\mathbf{V}\hat{\mathbf{w}}_{ni}^T]_k \\
& + 2 \gamma \bigg[ \mathrm{Diag}\Big( \frac{\mathbf{V}^+ \hat{\mathbf{w}}_{ni}^T + \mathbf{V}^- \hat{\mathbf{w}}_{ni}^T}{\hat{\mathbf{w}}_{ni}} \Big) \bigg]_k w_{nik} \\
& - 2 \gamma \bigg[ \mathrm{Diag} \Big( \frac{\mathbf{V}^+ \hat{\mathbf{w}}_{ni}^T + \mathbf{V}^- \hat{\mathbf{w}}_{ni}^T}{\hat{\mathbf{w}}_{ni}^T} \Big) \bigg]_k \hat{w}_{nik} = 0
\end{split}
\end{equation}
and further make the following abbreviations for clarity:
\begin{equation}\label{CoeparameterW_a}
\begin{split}
a = 2 \gamma \bigg[ \mathrm{Diag}\Big( \frac{\mathbf{V}^+ \hat{\mathbf{w}}_{ni}^T + \mathbf{V}^- \hat{\mathbf{w}}_{ni}^T}{\hat{\mathbf{w}}_{ni}^T} \Big) \bigg]_k
\end{split}
\end{equation}
\begin{equation}\label{CoeparameterW_b}
\begin{split}
b = & 2\gamma [\mathbf{V}\hat{\mathbf{w}}_{ni}^T]_k - h_{nkj} \mathrm{tr}\big( {\mathbf{G}}_{ni} \mathbf{U}_{ij}^{-1} \big) \\
& - 2 \gamma \bigg[ \mathrm{Diag} \Big( \frac{\mathbf{V}^+ \hat{\mathbf{w}}_{ni}^T + \mathbf{V}^- \hat{\mathbf{w}}_{ni}^T}{\hat{\mathbf{w}}_{ni}^T} \Big) \bigg]_k \hat{w}_{nik}
\end{split}
\end{equation}
\begin{equation}\label{CoeparameterW_d}
\begin{split}
d = h_{nkj}^{-1} \mathrm{tr}\bigg( {\mathbf{G}}_{ni}^{-1} \boldsymbol{\Phi}_{ijn} \mathbf{X}_{ij} \boldsymbol{\Phi}_{ijn}^H \bigg)
\end{split}
\end{equation}
Then, \eqref{DerivativeLikelihood_W} can be rewritten as:
\begin{equation}\label{eq:xx4}
\begin{split}
aw_{nik}^3 + bw_{nik}^2 + d = 0
\end{split}
\end{equation}
We employ the {\textit{cubic roots}} procedure \cite{rechtschaffen200892} to solve problem \eqref{eq:xx4}.

Given $\mathbf{H}_n$ and $\mathbf{W}_n$ fixed, the spatial model $\mathbf{G}_{ni}$ is calculated as follows.
We let the partial derivative of $\mathcal{L}$ with respect to $\mathbf{G}_{ni}$ equal to zero:
\begin{equation}\label{DerivativeLikelihood_G}
\begin{split}
\sum_{j=1}^J \lambda_{ijn}^{-1} \mathbf{G}_{ni}^{-1} \boldsymbol{\Phi}_{ijn} \mathbf{X}_{ij} \boldsymbol{\Phi}_{ijn}^H \mathbf{G}_{ni}^{-1} - \sum_{j=1}^J \lambda_{ijn} \mathbf{U}_{ij}^{-1} = \mathbf{0}
\end{split}
\end{equation}
where $\mathbf{0}$ is an all-zero matrix of size $M\times M$. Substituting $\mathbf{U}_{ij}$ and $\boldsymbol{\Phi}_{ijn}$ into \eqref{DerivativeLikelihood_G} derives:
\begin{equation}\label{AuxiliaryRelationship_G}
\begin{split}
\mathbf{G}_{ni}^\star \mathbf{A}_{\mathbf{G}} \mathbf{G}_{ni}^\star = \mathbf{G}_{ni} \mathbf{B}_{\mathbf{G}} \mathbf{G}_{ni}
\end{split}
\end{equation}
where $\mathbf{G}_{ni}^\star$ is value of $\mathbf{G}_{ni}$ at the previous step, $\mathbf{A}_{\mathbf{G}}$ and $\mathbf{B}_{\mathbf{G}}$ are short for
\begin{equation}\label{numeratorG}
\begin{split}
\mathbf{A}_{\mathbf{G}} = \sum_{j=1}^J \lambda_{ijn} \hat{\mathbf{X}}_{ij}^{-1} \mathbf{X}_{ij} \hat{\mathbf{X}}_{ij}^{-1}
\end{split}
\end{equation}
\begin{equation}\label{denominatorG}
\begin{split}
\mathbf{B}_{\mathbf{G}} = \sum_{j=1}^J \lambda_{ijn} \hat{\mathbf{X}}_{ij}^{-1}
\end{split}
\end{equation}
Equation \eqref{AuxiliaryRelationship_G} has a closed-form updating rule for $\mathbf{G}_{ni}$:
\begin{equation}\label{Update_G}
\begin{split}
\mathbf{G}_{ni} \leftarrow \mathbf{G}_{ni}^\star (\mathbf{G}_{ni}^\star\mathbf{A}_{\mathbf{G}}\mathbf{G}_{ni}^\star) \sharp (\mathbf{B}_{\mathbf{G}})^{-1}
\end{split}
\end{equation}
where $\mathbf{A}\sharp\mathbf{B}$ is the geometric mean of two positive semi-definite matrices $\mathbf{A}$ and $\mathbf{B}$:
\begin{equation}\label{GeometricMean}
\begin{split}
\mathbf{A}\sharp\mathbf{B} = \mathbf{A}^{\frac{1}{2}}\big( \mathbf{A}^{-\frac{1}{2}} \mathbf{B} \mathbf{A}^{-\frac{1}{2}}\big)^{\frac{1}{2}} \mathbf{A}^{\frac{1}{2}} = \mathbf{A}(\mathbf{A}^{-1}\mathbf{B})^{\frac{1}{2}}
\end{split}
\end{equation}

The above algorithm is summarized in Algorithm \ref{alg:tDNMF}.

\begin{algorithm}[t]
\label{alg:tDNMF}
\small
\setlength{\tabcolsep}{0.1mm}{
  \SetKwInOut{Input}{\textbf{Input}}   
  \SetKwInOut{Output}{\textbf{Output}}
  \caption{m-MNMF. }
  \Input{
         Mixture $\mathbf{x}_{ij}$,
         number of sources $N$,
         \textit{MaxIteration},
         hyperparameter $\eta \geq  0$.\\
        }
    \Output{Separated signal $\mathbf{s}_{ij}$.}
    \BlankLine
    Initialize: $\mathbf{W}_n$,
          $\mathbf{H}_n$,
          $\mathbf{G}_{ni}$;\\
    \For{\text{iteration} = 1 \text{to MaxIteration}}
        {
        \For{n=1 \text{to} N}
        {
            \For{i=1 \text{to} I}
            {
                \For{k=1 \text{to} K}
                {
                 Update $h_{nkj}$ by \eqref{Update_H};\\
                 Update $w_{nik}$ by solving \eqref{eq:xx4}; \\
                }
                 \For{j=1 \text{to} J}
                 {
                 Compute $\boldsymbol{\Phi}_{ijn} = \hat{\mathbf{X}}_{ijn}\hat{\mathbf{X}}_{ij}^{-1}$, $\hat{\mathbf{X}}_{ij} = \sum_{n=1}^N \hat{\mathbf{X}}_{ijn}$;\\
                 }
            }
            Compute $\mathbf{V}_n = (\mathbf{W}_n^T \mathbf{W}_n + \delta \mathbf{I})^{-1}$; \\
        }
        \For{n=1 \text{to} N}
        {
        \For{i = 1 \text{to} I}
        {
        Update spatial covariance matrix $\mathbf{G}_{ni}$ by \eqref{numeratorG}, \eqref{denominatorG}, \eqref{Update_G}; \\
                 \For{j=1 \text{to} J}
                 {
                 Compute $\boldsymbol{\Phi}_{ijn} = \hat{\mathbf{X}}_{ijn}\hat{\mathbf{X}}_{ij}^{-1}$, $\hat{\mathbf{X}}_{ij} = \sum_{n=1}^N \hat{\mathbf{X}}_{ijn}$;\\
                 }
                 }
        }

          }
    Compute $\mathbf{s}_{ij}$ by multichannel Wiener filter.
    }

\end{algorithm}

\subsection{ILRMA with minimum-volume regularizer}\label{subsec:ILRMA}

\subsubsection{Preliminary}

{
ILRMA utilizes the assumption of the invertibility of mixing matrix $\mathbf{A}_i$ to transform the spatial optimization of MNMF into the estimation problem of the demixing matrix $\mathbf{D}_i$.
We note that ILRMA cannot be applied to the underdetermined BSS problem because the mixing matrix $\mathbf{A}_i$ must be invertible.
ILRMA employs a flexible source model to estimate the demixing matrix $\mathbf{D}_i$ in a stable manner as in AuxIVA \cite{ono2011stable}.
When $\mathbf{G}_{ni}$ is a rank-1 matrix given by $\mathbf{G}_{ni} = \mathbf{a}_{in}^H\mathbf{a}_{in}$, $\hat{\mathbf{X}}_{ij}$ can be calculated by:
\begin{equation}\label{Xhat_ILRMA}
\begin{split}
\hat{\mathbf{X}}_{ij} & = \sum_{n=1}^N \lambda_{ijn} \mathbf{a}_{in}^H \mathbf{a}_{in} \\
& = \mathbf{A}_i \boldsymbol{\Lambda}_{ij} \mathbf{A}_i^H \\
& = \mathbf{D}_i^{-1} \boldsymbol{\Lambda}_{ij} \mathbf{D}_i^{-H}
\end{split}
\end{equation}
where $\boldsymbol{\Lambda}_{ij} = \mathrm{Diag}(\lambda_{ij1},\dots,\lambda_{ijN})$ is a diagonal matrix.
}

By substituting \eqref{Xhat_ILRMA} into the cost function of MNMF \eqref{Likelihood_MNMF1}, we obtain:
\begin{equation}\label{ObjectF_ILRMA}
\begin{split}
 \log & [p(\mathbf{X}|\mathbf{W},\mathbf{H},\mathbf{G})p(\mathbf{W})p(\mathbf{H})]\\
= & - \sum_{i=1}^I \sum_{j=1}^J \mathrm{tr}\Big( \mathbf{s}_{ij}^H \mathbf{D}_i^{-H} \big( \mathbf{D}_{i}^H \boldsymbol{\Lambda}_{ij}^{-1} \mathbf{D}_{i} \big) \mathbf{D}_{i}^{-1} \mathbf{s}_{ij} \Big) \\
& + J\sum_{i=1}^I \log|\mathbf{D}_{i}\mathbf{D}_i^H| - \sum_{i=1}^I \sum_{j=1}^J \log|\boldsymbol{\Lambda}_{ij}| + \mathrm{const}
\end{split}
\end{equation}

The demixing matrix $\mathbf{D}_i$ of the spatial model in ILRMA is updated based on the rules of AuxIVA which can be represented as follows:
\begin{equation}\label{Update_ILRMAD}
\begin{split}
 \mathbf{G}_{ni} & = \frac{1}{J}\sum_j\frac{1}{\lambda_{ijn}}\mathbf{x}_{ij}\mathbf{x}_{ij}^h \\
 \mathbf{d}_{in} & \leftarrow (\mathbf{D}_i\mathbf{G}_{ni})^{-1}\mathbf{e}_m \\
 \mathbf{d}_{in} & \leftarrow \mathbf{d}_{in}(\mathbf{d}_{in}^h\mathbf{G}_{ni}\mathbf{d}_{in})^{-\frac{1}{2}}
\end{split}
\end{equation}

The parameters of the source model $\mathbf{W_n}$ and $\mathbf{T}_n$ are updated by MU:
\begin{equation}\label{Update_ILRMAD1}
\begin{split}
 w_{nik} \leftarrow w_{nik} \sqrt{\frac{\sum_j | y_{ijn} |^2 h_{nkj} \left( \sum_k w_{nik} h_{nkj} \right)^{-2}}{\sum_j h_{nkj} \left( \sum_k w_{nik} h_{nkj} \right)^{-1}} }
\end{split}
\end{equation}
\begin{equation}\label{Update_ILRMAD2}
\begin{split}
 h_{nkj} \leftarrow h_{nkj} \sqrt{\frac{\sum_i | y_{ijn} |^2 w_{nik} \left( \sum_k w_{nik} h_{nkj} \right)^{-2}}{\sum_i w_{nik} \left( \sum_k w_{nik} h_{nkj} \right)^{-1}} }
\end{split}
\end{equation}

\begin{figure*}[htb]
\vspace{-0.3cm}
\begin{minipage}[b]{1.0\linewidth}
  \centering
  \centerline{\includegraphics[width=16.5cm]{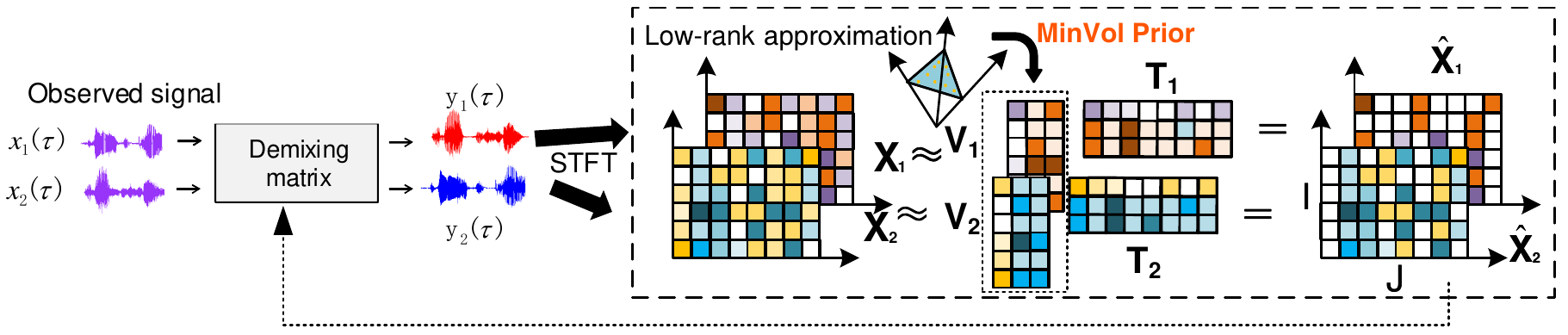}}
\end{minipage}
\vspace{-0.8cm}
\caption{Principle of the proposed m-ILRMA algorithm.}
\label{fig:model}
\vspace{-0.3cm}
\end{figure*}

\subsubsection{Objective function of m-ILRMA}

{To remedy the non-unique identifiable problem of the source model of ILRMA, here we propose m-ILRMA.} Fig. \ref{fig:model} shows a conceptual model of m-ILRMA. Specifically, substituting {\eqref{Xhat_ILRMA}} into \eqref{Likelihood_MNMF} derives the objective of m-ILRMA:
\begin{equation}\label{Likelihood_ILRMA}
\begin{split}
\log & [p(\mathbf{X}|\mathbf{W},\mathbf{H},\mathbf{G})p(\mathbf{W}|\gamma)p(\mathbf{H})]\\
= & \sum_{i=1}^{I} \sum_{j=1}^{J} \left( -\mathrm{tr}\Big( \hat{\mathbf{X}}_{ij}^{-1} \mathbf{X}_{ij} \Big) - \log|\hat{\mathbf{X}}_{ij}| \right) \\
& - \sum_{n=1}^N \gamma\log|\mathbf{W}_n^T\mathbf{W}_n + \eta \mathbf{I}| + \mathrm{const} \\
= & - \sum_{i=1}^I \sum_{j=1}^J \mathrm{tr}\Big( \mathbf{s}_{ij}^H \mathbf{D}_i^{-H} \big( \mathbf{D}_{i}^H \boldsymbol{\Lambda}_{ij}^{-1} \mathbf{D}_{i} \big) \mathbf{D}_{i}^{-1} \mathbf{s}_{ij} \Big) \\
& + J\sum_{i=1}^I \log|\mathbf{D}_{i}\mathbf{D}_i^H| - \sum_{n=1}^N \gamma\log|\mathbf{W}_n^T\mathbf{W}_n + \delta \mathbf{I}| \\
& - \sum_{i=1}^I \sum_{j=1}^J \log|\boldsymbol{\Lambda}_{ij}| + \mathrm{const} \\
= & - \sum_{i=1}^I \sum_{j=1}^J \sum_{n=1}^N \bigg( \frac{|s_{ijn}|^2}{\lambda_{ijn}} + \log \lambda_{ijn} \bigg) + \mathrm{const} \\
& + J \sum_{i=1}^I \log|\mathbf{D}_i\mathbf{D}_i^H|  - \sum_{n=1}^N \gamma \log{|\mathbf{W}_n^T \mathbf{W}_n + \delta \mathbf{I}|}
\end{split}
\end{equation}
where $s_{ijn} = \mathbf{d}_{in}^h\mathbf{x}_{ij}$.

{Because the full rank spatial model in MNMF has more parameters than the rank-1 spatial model in ILRMA, ILRMA is less sensitive to its parameter initialization, so as to the advantage of m-LIRMA over m-MNMF. As will be shown in the experiments, m-ILRMA achieves better separation performance than m-MNMF.}

\subsubsection{Optimization of m-ILRMA}

m-ILRMA is optimized by the multiplicative updating (MU) rule, which optimizes $\mathbf{H}_n$, $\mathbf{W}_n$, and $\mathbf{G}_{ni}$, alternatively.

Given $\mathbf{H}_n$ and $\mathbf{G}_{ni}$ fixed, the objective \eqref{Likelihood_ILRMA} with respect to $\mathbf{W}_n$ is a difficult maximization problem. In order to solve this problem, we propose to maximize a lower-bound of \eqref{Likelihood_ILRMA} by an auxiliary function proposed in \cite{fevotte2011algorithms}. Specifically, we first define a $Q$ function by Lemma \ref{lemma:1}.

\begin{lemma}\label{lemma:1}
Let $\hat{s}_{ijn} = \sum_k\hat{w}_{nik}h_{nkj}$ and $\hat{s}_{ijn} \geq 0$, $\hat{w}_{nik} \geq 0$. Then, the following $Q$ function:
\begin{equation}\label{auxi_NMF}
\begin{split}
 & Q(\mathbf{w}_{ni}^T|\hat{\mathbf{w}}_{ni}^T) = \left[ \sum_k \frac{\hat{{w}}_{nik}h_{nkj}}{\hat{s}_{ijn}} \check{\rho}({s}_{ijn}|\hat{s}_{ijn}\frac{{w}_{nik}}{\hat{{w}}_{nik}}) \right] + \bar{\rho}(\hat{s}_{ijn}) \\ & + \left[ \hat{\rho}^\prime({s}_{ijn}|\hat{s}_{ijn})\sum_k({w}_{nik} - \hat{{w}}_{nik}){h}_{nkj} + \hat{\rho}({s}_{ijn}|\hat{s}_{ijn}) \right]
\end{split}
\end{equation}
is an auxiliary function to $Q(\mathbf{w}_{ni\cdot})$ at $\hat{{w}}_{nik}$ \cite{fevotte2011algorithms}, where
$\check{\rho}$ is a convex function with respect to $\hat{s}_{ijn}\frac{{w}_{nik}}{\hat{{w}}_{nik}}$, $\hat{\rho}$ is a concave function with respect to $\hat{s}_{ijn}$, and $\bar{\rho}$ is {a constant function with respect to ${s}_{ijn}$}.
$\hat{\rho}^\prime$ is the differential of $\hat{\rho}({s}_{ijn}|\hat{s}_{ijn})$ at $\hat{s}_{ijn}$.
Due to the {Itakura-Saito (IS)} divergence, we have $\check\rho(x|y) = xy^{-1}$, $\hat{\rho}(x|y) = \log y$, $\bar{\rho}(x) = x(\log x -1)$, $\hat{\rho}^\prime(x|y) = y^{-1}$.
\end{lemma}

Similar to m-MNMF, we use \eqref{TaylorExp} to construct a low-bound of the likelihood function with respect to $\mathbf{W}_n$:
\begin{equation}\label{Likelihood_W}
\begin{split}
\mathcal{L}_{w_{nik}} = & \sum_{n=1}^N \sum_{i=1}^I \Big[ Q(\mathbf{w}_{ni}^T|\hat{\mathbf{w}}_{ni}^T) + \gamma \big[ \log|\det(\mathbf{V}_n^{-1})| \\
& + \mathrm{tr}(\mathbf{V}_n\mathbf{W}_n^T\mathbf{W}_n) - K\big] \Big] \\
\end{split}
\end{equation}
where $\mathbf{V}_n = (\mathbf{Z}^T\mathbf{Z} + \delta\mathbf{I})^{-1}$ with $\delta \geq 0$, $\mathbf{Z} \in \mathbb{R}^{I\times K}$ is an arbitrary  positive definite matrix. We can set $\mathbf{Z} = \mathbf{W}_n$ in the experiments, since $\mathbf{W}_n$ is a positive definite matrix. Finally, the right side of \eqref{TaylorExp} is an auxiliary function for $ \log|\mathbf{W}_n^T\mathbf{W}_n|$. Because it is quadratic and inseparable, we use an approximation to represent the right side of \eqref{TaylorExp}. Specifically, let $\mathbf{V}_n = \mathbf{V}_n^+ - \mathbf{V}_n^-$ with $\mathbf{V}_n^+ = \max(\mathbf{V}_n,0)$ and $\mathbf{V}_n^- = \max(-\mathbf{V}_n, 0)$, Then, the right side of \eqref{Likelihood_W} can be written as:
\begin{equation}\label{Likelihood_W1}
\begin{split}
\mathcal{L}_{w_{nik}} = & - \sum_{n=1}^N \sum_{i=1}^I \Bigg[ Q(\mathbf{w}_{ni}^T|\hat{\mathbf{w}}_{ni}^T) + \gamma \Big[ \hat{\mathbf{w}}_{ni} \mathbf{V}_n \hat{\mathbf{w}}_{ni}^T  \\
& + 2\Delta{\hat{\mathbf{w}}_{ni}} \mathbf{V}_n \hat{\mathbf{w}}_{ni}^T + \Delta{\hat{\mathbf{w}}_{ni}} \boldsymbol{\Omega}{(\hat{\mathbf{w}}_{ni}^T)} \Delta{\hat{\mathbf{w}}_{ni}^T} \Big] \Bigg] \\
\end{split}
\end{equation}
with $\boldsymbol{\Omega}(\mathbf{w}_{ni}^T) = \mathrm{Diag} \Big( 2\frac{[\mathbf{V}^+\mathbf{w}_{ni}^T + \mathbf{V}^-\mathbf{w}_{ni}^T]}{[\mathbf{w}_{ni}^T]} \Big)$, where the operator ``$\frac{[\mathbf{A}]}{[\mathbf{B}]}$'' is the component-wise division between $\mathbf{A}$ and $\mathbf{B}$.

We let the partial derivative of $\mathcal{L}_{w_{nik}}$ equal to zero and derive the MU update rule of the factor $w_{nik}$ as follows:
\begin{equation}\label{DerivativeLikelihoodILRMA_W}
\begin{split}
\frac{\partial \mathcal{L}_{w_{nik}}}{\partial w_{nik}} = &\Big( \sum_j \frac{h_{nkj}}{\hat{s}_{ijn}} - \sum_j h_{nkj}\frac{\hat{{w}}^2_{nik}{s}_{ijn}}{{w}^2_{nik} \hat{{s}}^2_{ijn}}  \\
&  + 2\gamma [\mathbf{V}_n\hat{\mathbf{w}}_{ni}^T]_k  \\
& + 2 \gamma \bigg[ \mathrm{Diag}\Big( \frac{\mathbf{V}^+ \hat{\mathbf{w}}_{ni}^T + \mathbf{V}^- \hat{\mathbf{w}}_{ni}^T}{\hat{\mathbf{w}}_{ni}^T} \Big) \bigg]_k w_{nik} \\
& - 2 \gamma \bigg[ \mathrm{Diag} \Big( \frac{\mathbf{V}^+ \hat{\mathbf{w}}_{ni}^T + \mathbf{V}^- \hat{\mathbf{w}}_{ni}^T}{\hat{\mathbf{w}}_{ni}^T} \Big) \bigg]_k \hat{w}_{nik}
\end{split}
\end{equation}
To make the above objective function easier, we let
\begin{equation}\label{ILRMACoeparameterW_a}
\begin{split}
a = 2 \gamma \bigg[ \mathrm{Diag}\Big( \frac{\mathbf{V}^+ \hat{\mathbf{w}}_{ni}^T + \mathbf{V}^- \hat{\mathbf{w}}_{ni}^T}{\hat{\mathbf{w}}_{ni}^T} \Big) \bigg]_k
\end{split}
\end{equation}
\begin{equation}\label{ILRMACoeparameterW_b}
\begin{split}
b = &  - \sum_j \frac{h_{nkj}}{\hat{s}_{ijn}} + 2\gamma [\mathbf{V}\hat{\mathbf{w}}_{ni}^T]_k\\
& - 2 \gamma \bigg[ \mathrm{Diag} \Big( \frac{\mathbf{V}^+ \hat{\mathbf{w}}_{ni}^T + \mathbf{V}^- \hat{\mathbf{w}}_{ni}^T}{\hat{\mathbf{w}}_{ni}^T} \Big) \bigg]_k \hat{w}_{nik}
\end{split}
\end{equation}
\begin{equation}\label{ILRMACoeparameterW_d}
\begin{split}
d = \sum_j h_{nkj}\frac{\hat{{w}}^2_{nik}{s}_{ijn}}{ \hat{{s}}^2_{ijn}}
\end{split}
\end{equation}
Setting the derivative to zero equals to the problem of computing the roots of the following degree-three polynomial:
 \begin{equation}\label{eq:xxxx}
\begin{split}
aw_{nik}^3 + bw_{nik}^2 + d = 0
\end{split}
\end{equation}
Similar to \eqref{eq:xx4}, we use the {{cubic roots}} procedure \cite{rechtschaffen200892} to solve the above polynomial problem.

Similar to m-MNMF, given $\mathbf{W}_n$ and $\mathbf{G}_{ni}$ fixed, the closed-form MU rules for $\mathbf{H}_n$ is:
\begin{equation}\label{ILRMAnumeratorH}
\begin{split}
a^h = \sum_{i=1}^I w_{nik} |s_{ijn}|^2 \lambda_{ijn}^{-2}
\end{split}
\end{equation}
\begin{equation}\label{ILRMAdenominatorH}
\begin{split}
b^h = \sum_{i=1}^I w_{nik} \lambda_{ijn}^{-1}
\end{split}
\end{equation}
\begin{equation}\label{UpdateILRMA_H}
\begin{split}
h_{nkj} \leftarrow h_{nkj}^\star \sqrt{\frac{a_h}{b_h}}
\end{split}
\end{equation}

Given $\mathbf{W}_n$ and $\mathbf{H}_n$ fixed, an IVA-based auxiliary function \cite{ono2011stable} is used to optimize the spatial model $\mathbf{G}_{ni}$, which results in the following solution:
\begin{equation}\label{SpatialUpdates}
\begin{split}
 \mathbf{G}_{ni} & = \frac{1}{J}\sum_j\frac{1}{\lambda_{ijn}}\mathbf{x}_{ij}\mathbf{x}_{ij}^h \\
 \mathbf{d}_{in} & \leftarrow (\mathbf{D}_i\mathbf{G}_{ni})^{-1}\mathbf{e}_m \\
 \mathbf{d}_{in} & \leftarrow \mathbf{d}_{in}(\mathbf{d}_{in}^h\mathbf{G}_{ni}\mathbf{d}_{in})^{-\frac{1}{2}}
\end{split}
\end{equation}
where $\mathbf{e}_m$ denotes the $n$th column vector of an $M\times M$-dimensional identity matrix.

The above algorithm is summarized in Algorithm \ref{alg:sDNMF}.

\begin{algorithm}[t]
\label{alg:sDNMF}
\small
\setlength{\tabcolsep}{0.1mm}{
  \SetKwInOut{Input}{\textbf{Input}}   
  \SetKwInOut{Output}{\textbf{Output}}
  \caption{m-ILRMA. }
  \Input{
         Mixture $\mathbf{x}_{ij}$,
         number of sources $N$,
         \textit{MaxIteration},
         hyperparameter $\eta \geq  0$.\\
        }
    \Output{Separated signal $\mathbf{s}_{ij}$.}
    \BlankLine
    Initialize: $\mathbf{W}_n$,
          $\mathbf{H}_n$,
          $\mathbf{G}_{ni}$;\\
    \For{\text{iteration} = 1 \text{to MaxIteration}}
        {
        \For{n=1 \text{to} N}
        {
            \For{i=1 \text{to} I}
            {
                \For{k=1 \text{to} K}
                {
                    Update $w_{nik}$ by solving \eqref{eq:xxxx}; \\
                    Update $h_{nkj}$ by \eqref{UpdateILRMA_H};\\
                }
            }
            \For{i=1 \text{to} I}
            {
                Update $\mathbf{d}_{in}$ by \eqref{SpatialUpdates}; \\
            }
        }

          }
          $s_{ij,n} \leftarrow \mathbf{d}_{in}^h \mathbf{x}_{ij}$
    }
\end{algorithm}

\subsection{On the hyper-parameter selection and estimation}

The objectives \eqref{Likelihood_MNMF} and \eqref{Likelihood_ILRMA} have two hyper-parameters $\eta$ and $\gamma$.

The hyper-parameter $\eta$ in the objectives \eqref{Likelihood_MNMF} and \eqref{Likelihood_ILRMA} is a small positive constant that prevents the term $\log\left|\mathbf{W}_n^T\mathbf{W}_n + \eta\mathbf{I}\right|$ from $-\infty$.
It should not be chosen too small, otherwise $\mathbf{W}_n^T\mathbf{W}_n + \eta \mathbf{I}$ might be badly conditioned which results in the optimization problems hard to solve.

The regularization coefficient $\gamma$ strongly affects the model performance. Here we update $\gamma$ automatically. First, the variables $\hat{\mathbf{X}}_{ij}$ and $\mathbf{W}_n$ are initialized with the successive nonnegative projection algorithm \cite{leplat2019minimum}, then
$\gamma$ is updated by:
\begin{equation}\label{parameter_lambda}
\begin{split}
 \gamma \leftarrow {\gamma}^{\star} \frac{\sum_{i,j} \left[ \mathrm{tr}\left( \mathbf{X}_{ij} \hat{\mathbf{X}}_{ij}^{-1} \right)+ \log|\hat{\mathbf{X}}_{ij}|\right]}{\log|{\mathbf{W}_n^T\mathbf{W}_n}+\delta \mathbf{I}|}
\end{split}
\end{equation}
where ${\gamma}^{\star}$ is the value of $\gamma$ at the previous iteration. We recommend to select ${\gamma}^{\star}$ from a range of $[10^{-3},1]$ at the first iteration.

\section{Experiments}
\label{Experiment}

In this section, we compare the proposed m-MNMF and m-ILRMA methods with 5 representative multichannel BSS methods in both simulated environments and real-world environments.


\subsection{Experimental settings}\label{subsec:expset}

\begin{figure*}[t]
\centering
\vspace{-0.7cm} 
\subfloat[Speech separation, $T_{60} = 130$ms]{\includegraphics[width=2.35in]{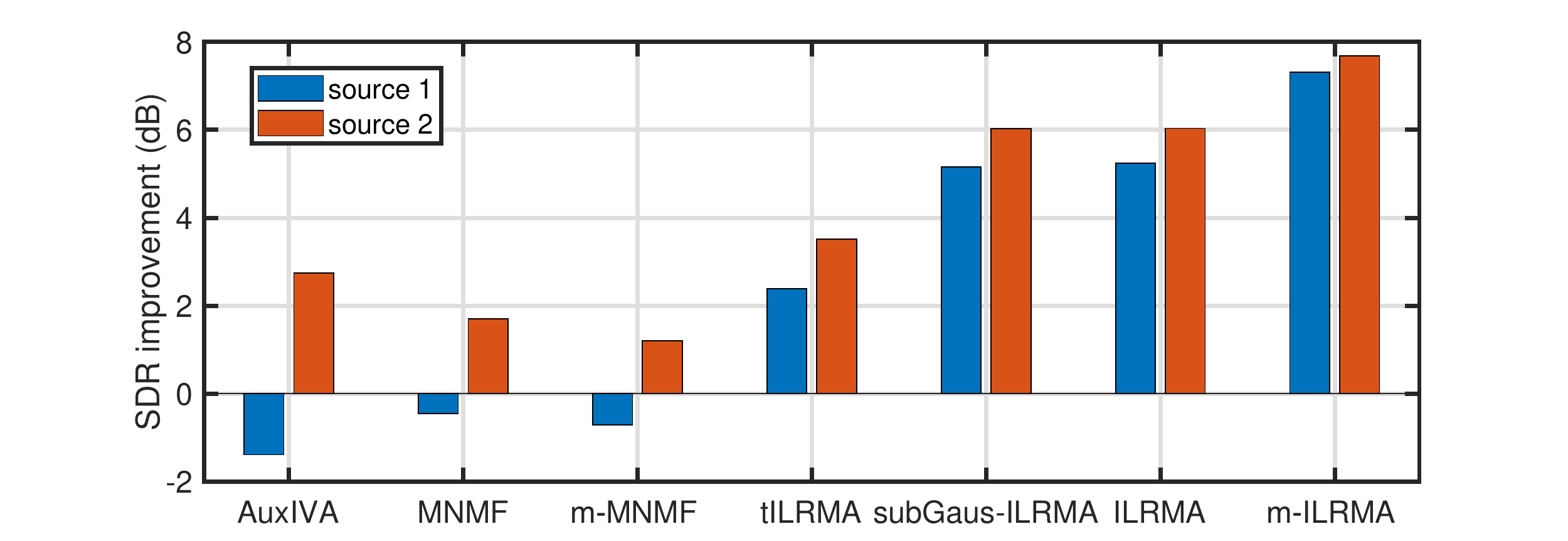}%
\label{fig_SISEC2011_1a}}
\hfil
\subfloat[Speech separation, $T_{60} = 250$ms]{\includegraphics[width=2.35in]{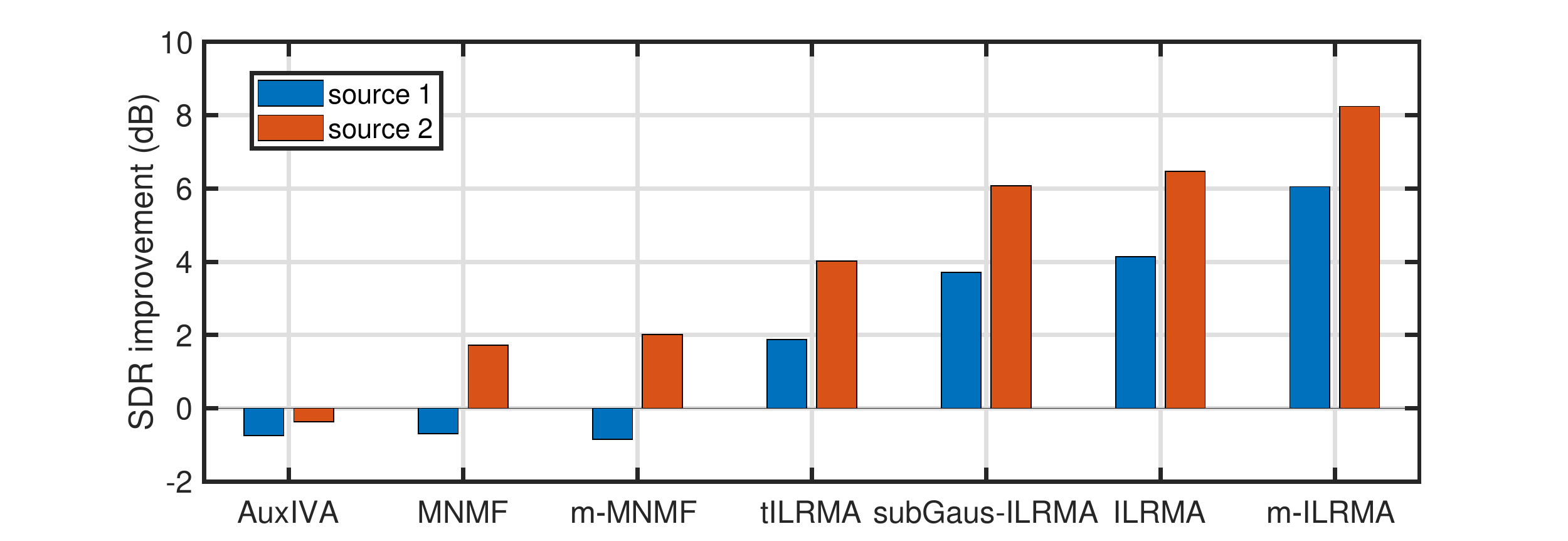}%
\label{fig_SISEC2011_1b}}
\hfil
\subfloat[Music separation, $T_{60} = 250$ms]{\includegraphics[width=2.35in]{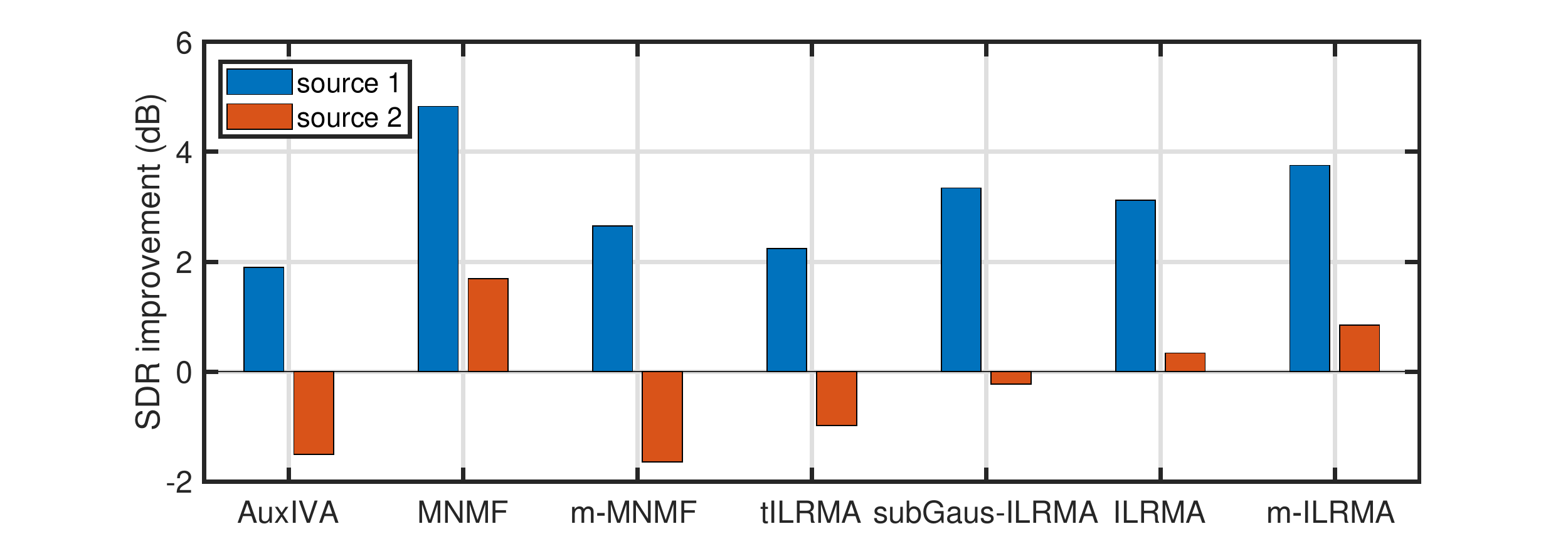}%
\label{fig_SISEC2011_1c}}
\hfil

\subfloat[Speech separation, $T_{60} = 130$ms]{\includegraphics[width=2.35in]{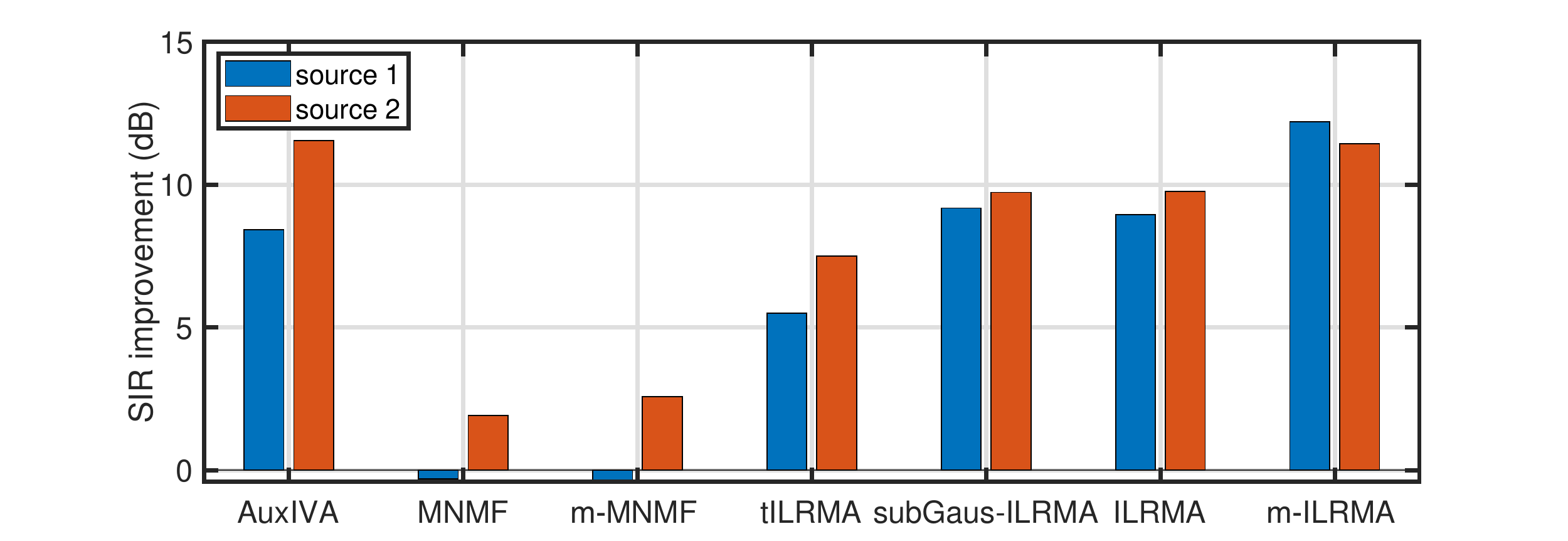}%
\label{fig_SISEC2011_1d}}
\hfil
\subfloat[Speech separation, $T_{60} = 250$ms]{\includegraphics[width=2.35in]{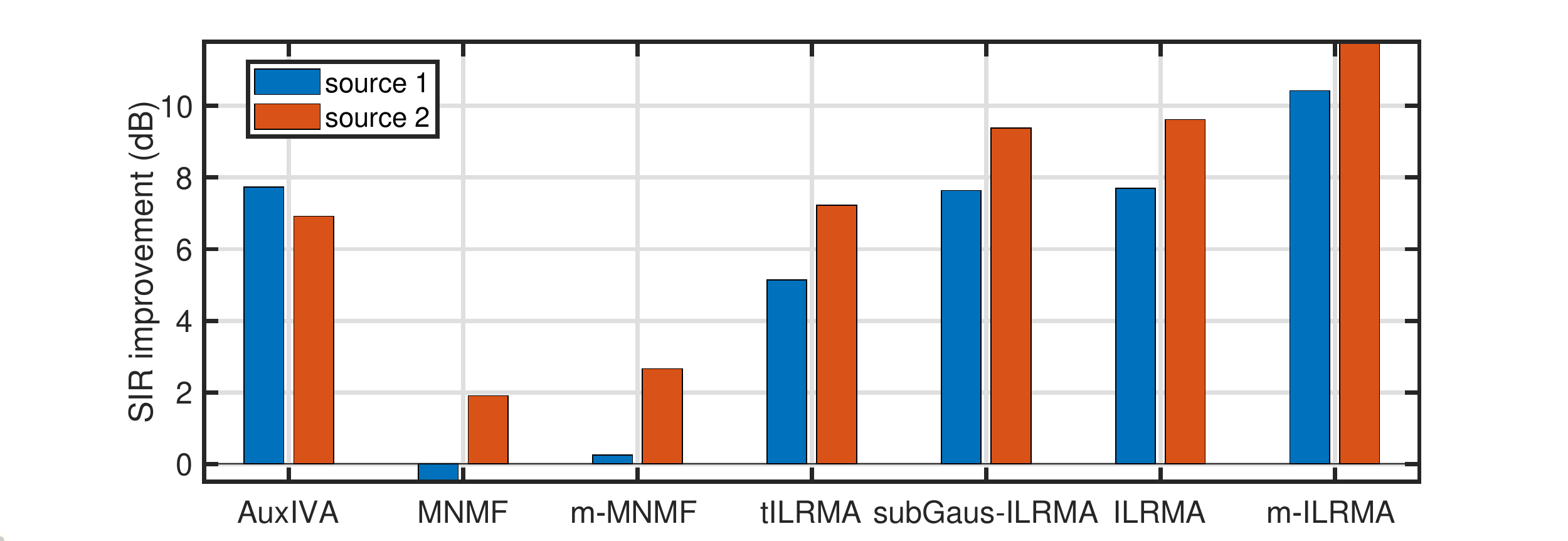}%
\label{fig_SISEC2011_1e}}
\hfil
\subfloat[Music separation, $T_{60} = 250$ms]{\includegraphics[width=2.35in]{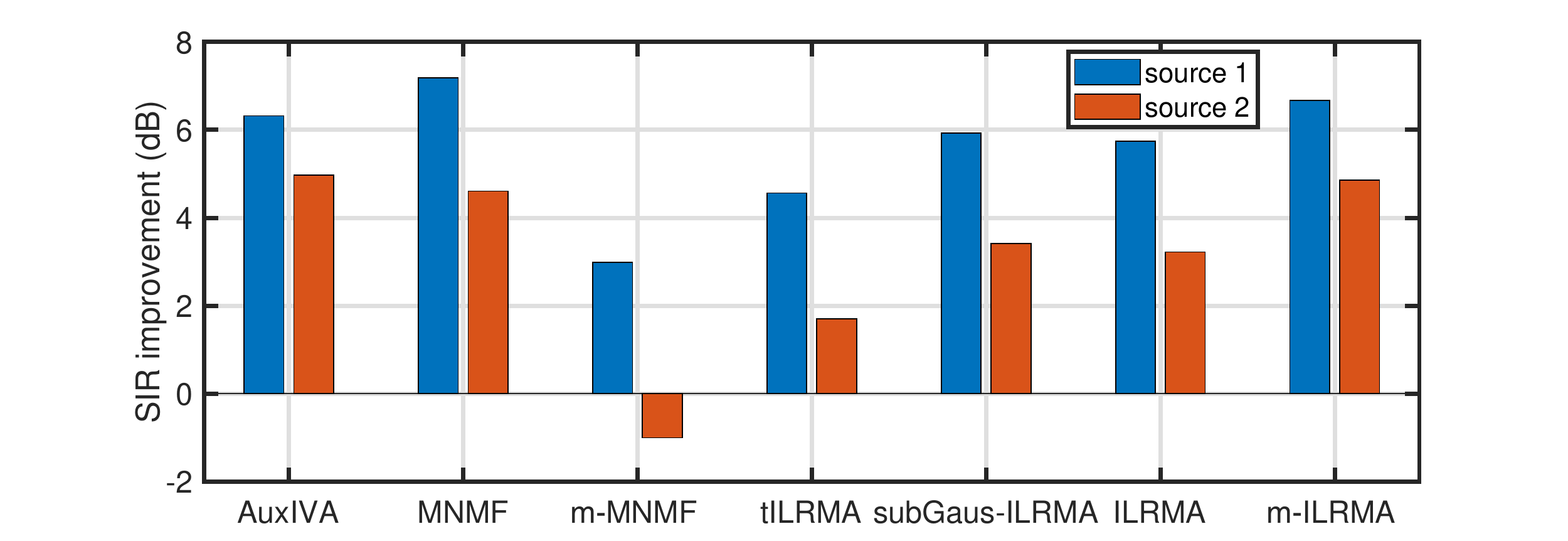}%
\label{fig_SISEC2011_1f}}
\hfil
\caption{Performance of the comparison methods on SISEC2011 when the distance between the sources and the microphones is 1m.}
\label{fig_SISEC2011_1}
\end{figure*}

\begin{figure*}[t]
\centering
\vspace{-0.7cm} 
\subfloat[Speech separation, $T_{60} = 130$ms]{\includegraphics[width=2.35in]{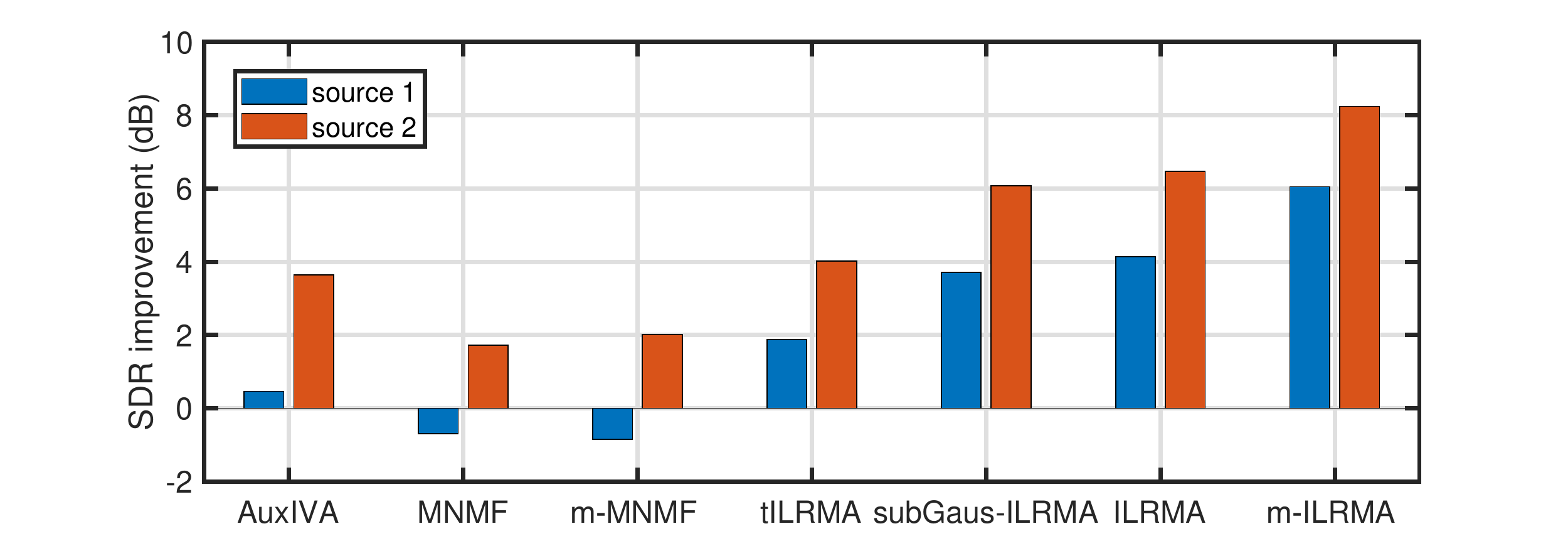}%
\label{fig_SISEC2011_5a}}
\hfil
\subfloat[Speech separation, $T_{60} = 250$ms ]{\includegraphics[width=2.35in]{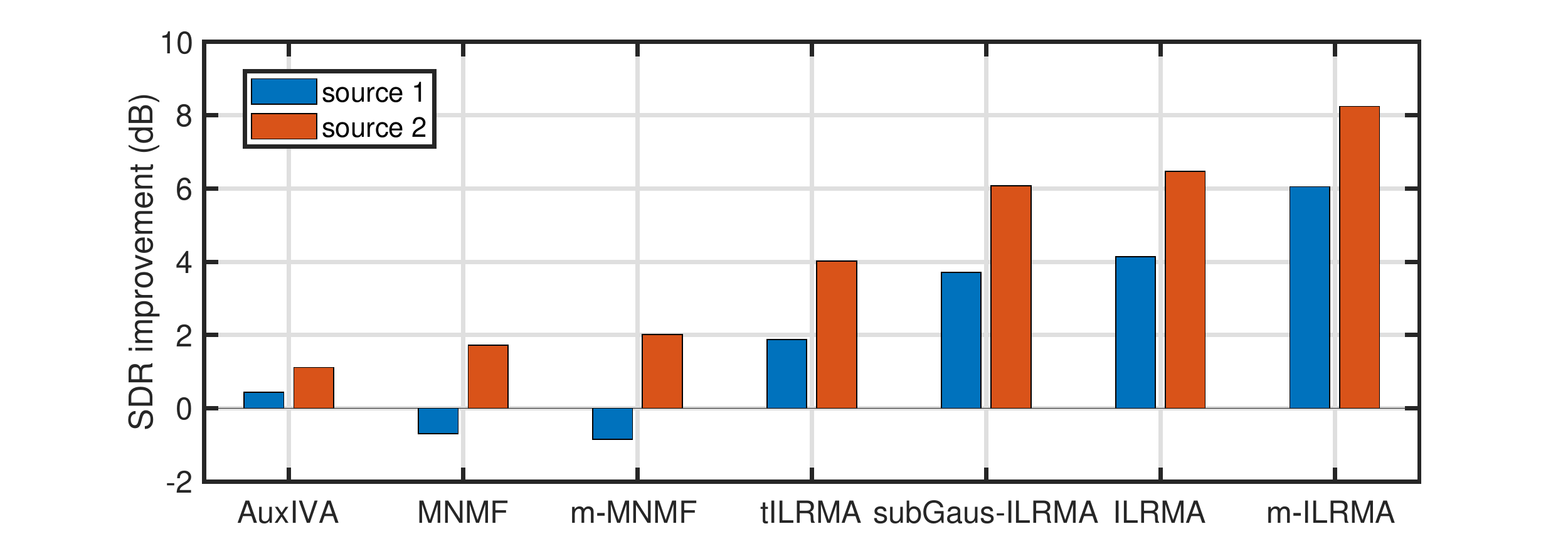}%
\label{fig_SISEC2011_5b}}
\hfil
\subfloat[Music separation, $T_{60} = 250$ms ]{\includegraphics[width=2.35in]{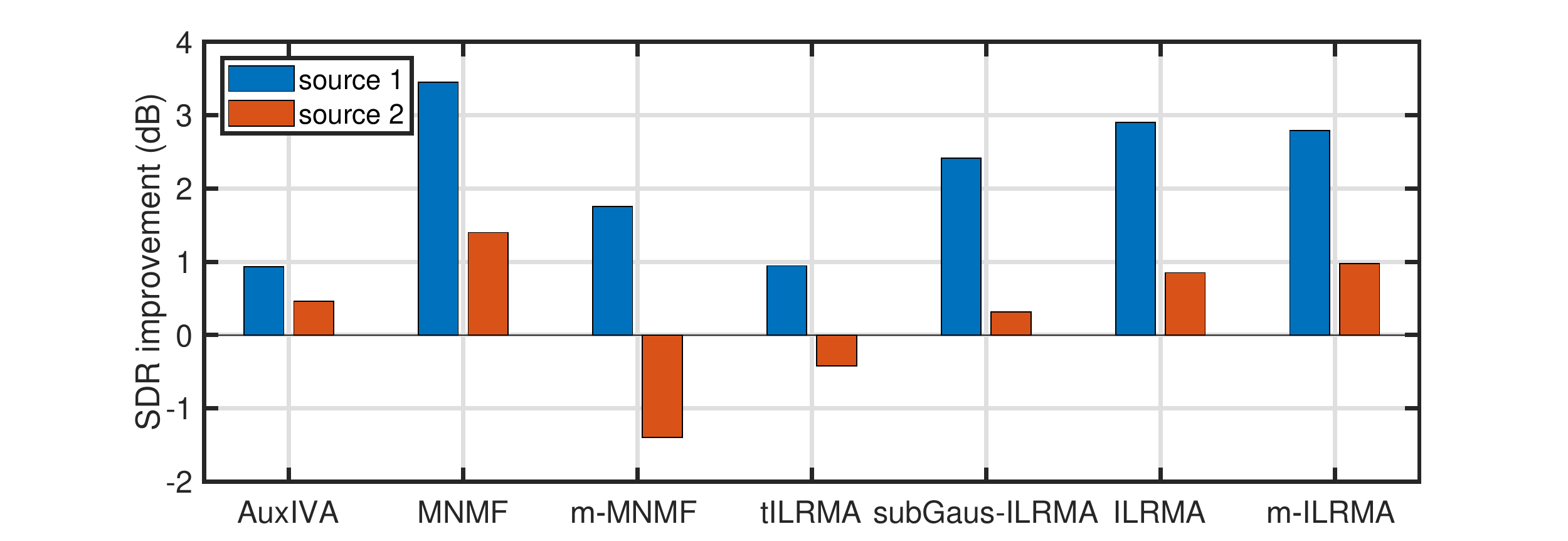}%
\label{fig_SISEC2011_5c}}
\hfil
\subfloat[Speech separation, $T_{60} = 130$ms ]{\includegraphics[width=2.35in]{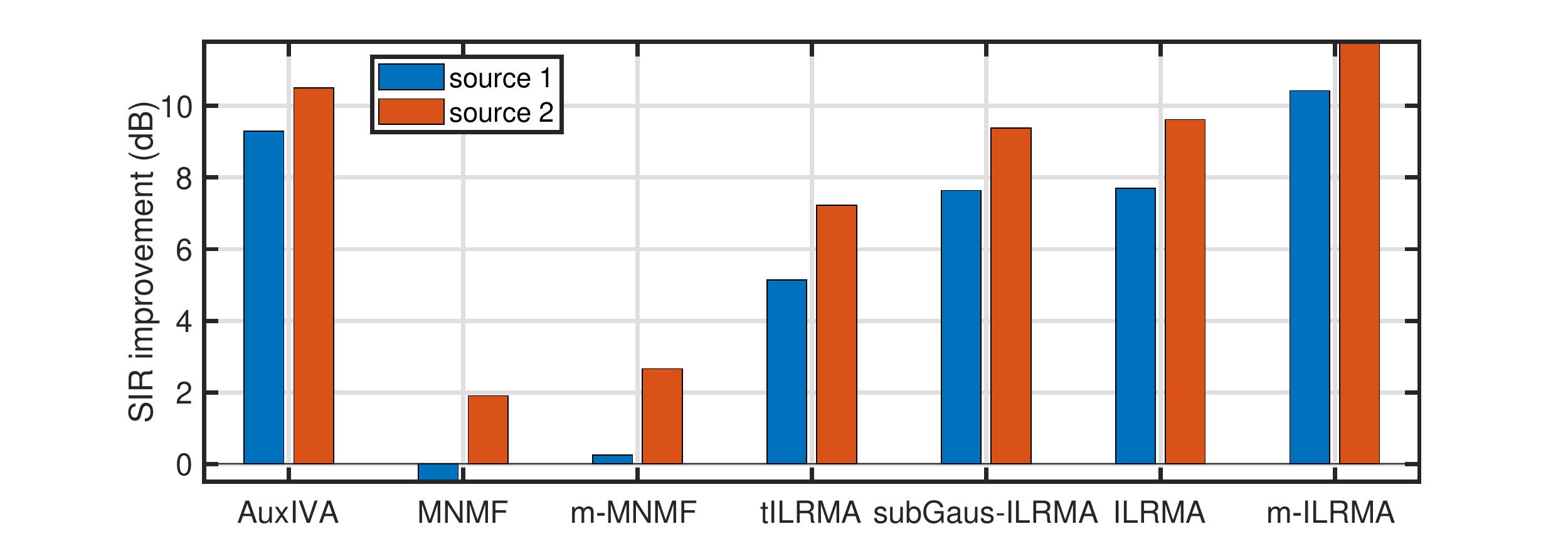}%
\label{fig_SISEC2011_5d}}
\hfil
\subfloat[Speech separation, $T_{60} = 250$ms ]{\includegraphics[width=2.35in]{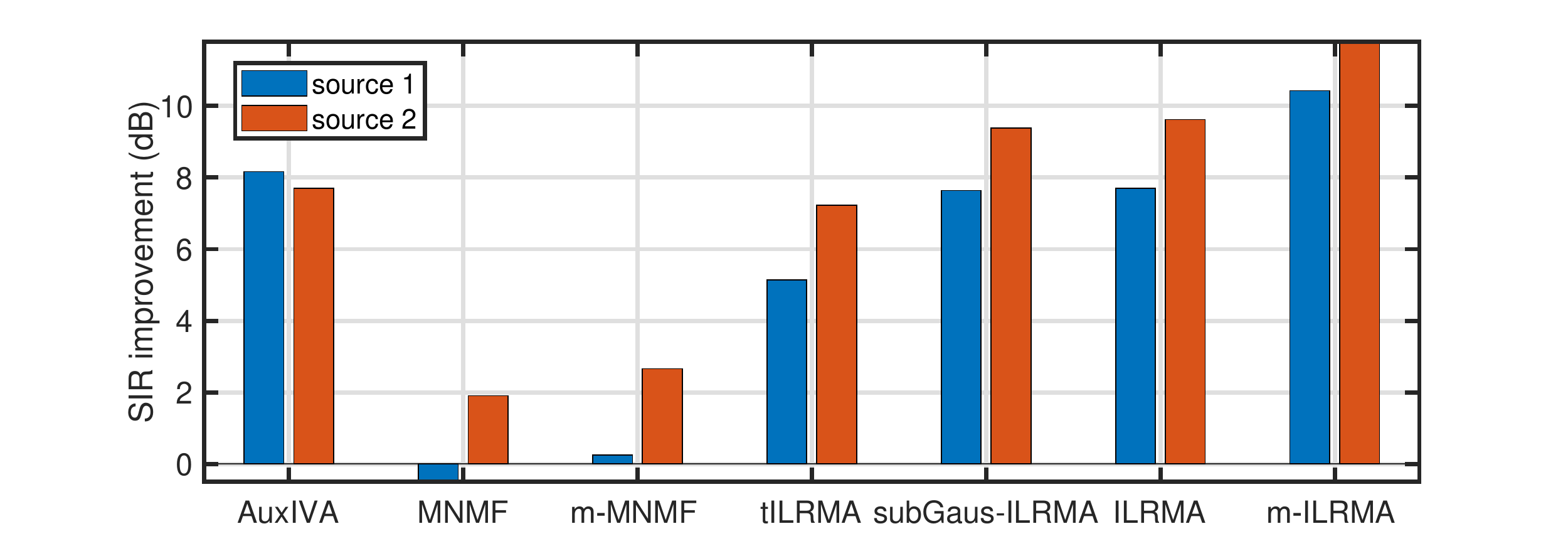}%
\label{fig_SISEC2011_5e}}
\hfil
\subfloat[Music separation, $T_{60} = 250$ms ]{\includegraphics[width=2.35in]{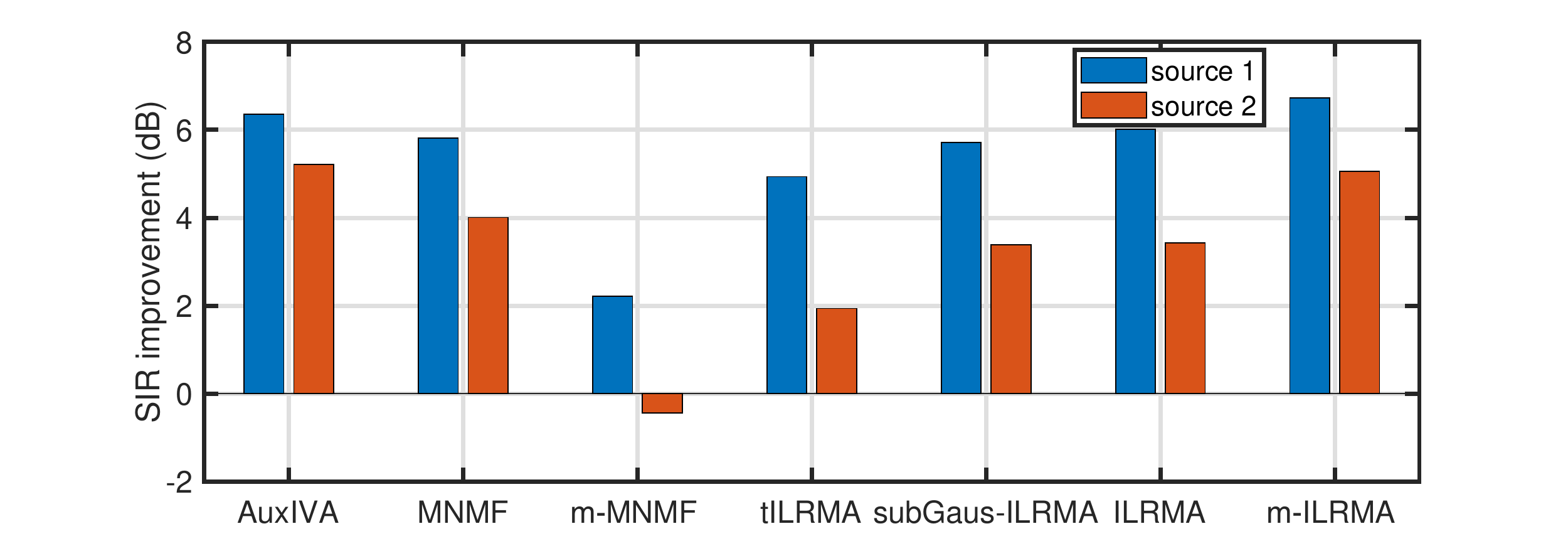}%
\label{fig_SISEC2011_5f}}
\caption{Performance of the comparison methods on SISEC2011, where the distance between source and microphone is 5cm.}
\label{fig_SISEC2011_5}
\end{figure*}

\begin{figure}[!t]
\begin{minipage}[b]{1.0\linewidth}
  \centering
  \centerline{\includegraphics[width=6.5cm]{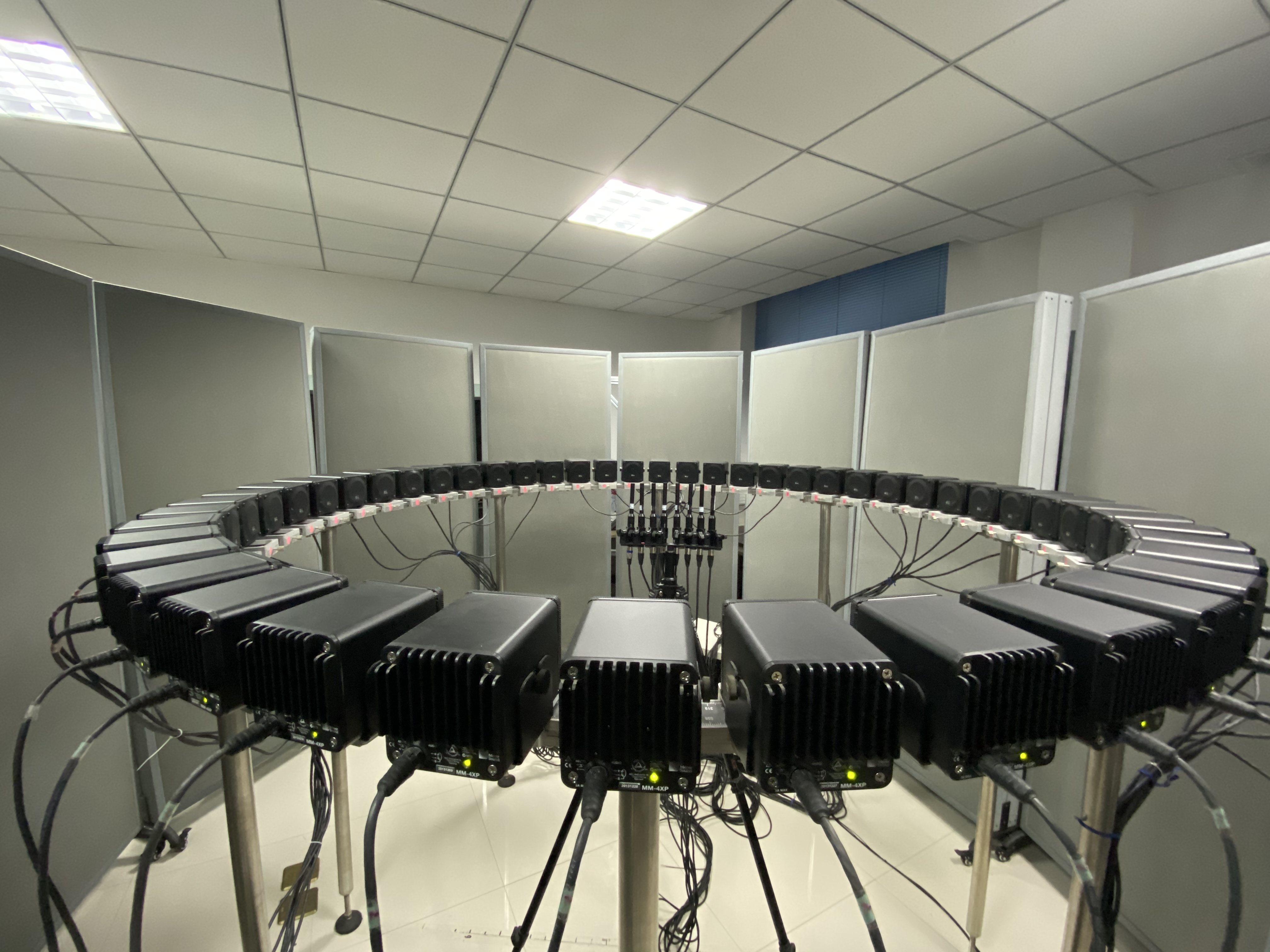}}
\end{minipage}

\begin{minipage}[b]{1.0\linewidth}
  \centering
  \centerline{\includegraphics[width=6.5cm]{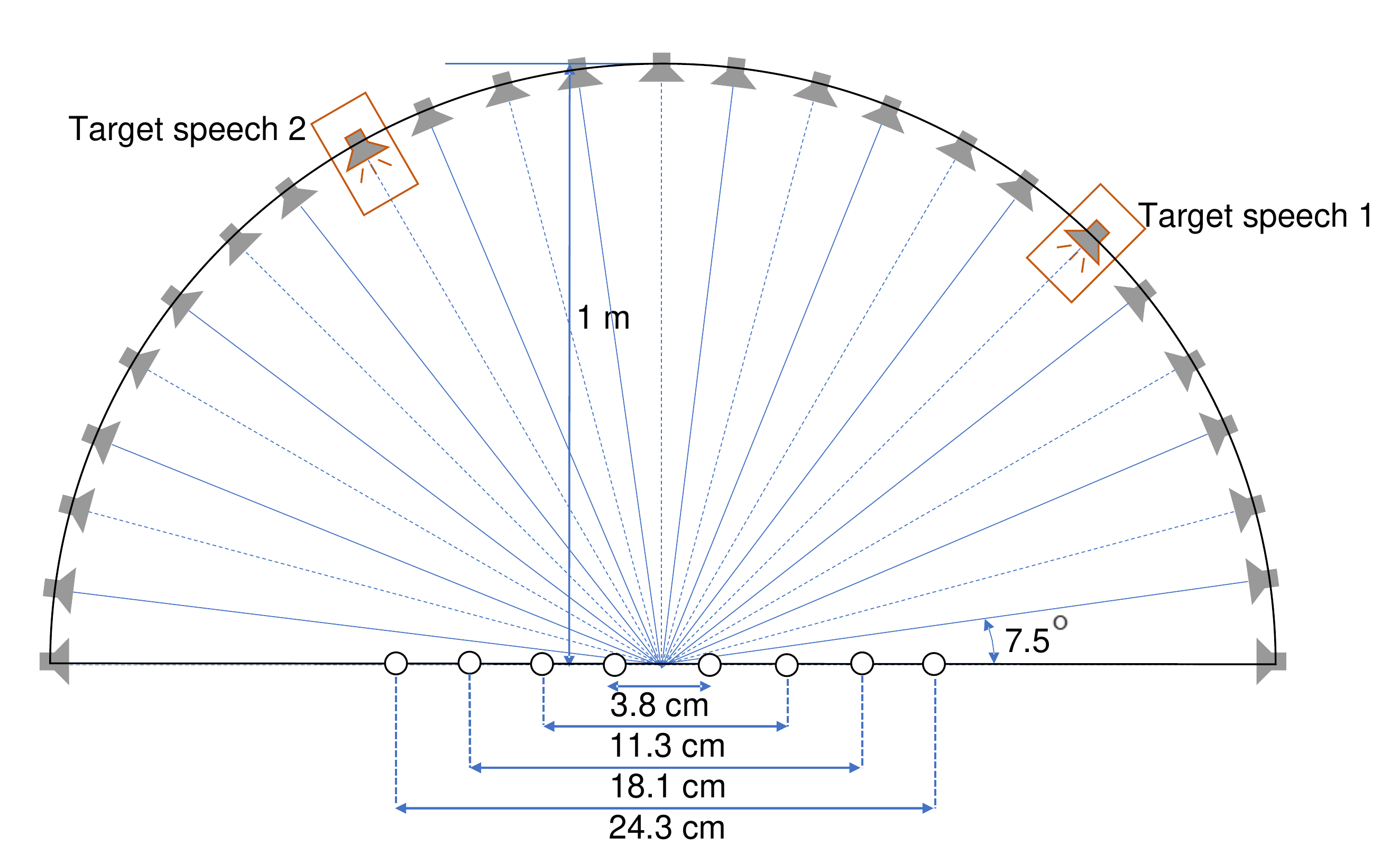}}
\end{minipage}

\caption{Recording conditions of impulse responses.}
\label{fig:model1}
\end{figure}
\subsubsection{Simulated databases}

{We used SISEC2011 \cite{araki20122011} as the first experimental dataset. SISEC2011 consists of three subsets named \textit{dev1}, \textit{dev2} and \textit{dev3}. For the speech separation problem, the clean speech was used to construct an underdetermined BSS task. After mixing up, 192 stereo mixture signals with female and male speech were generated, where the microphone spacing is 1m or 5cm and the reverberation time is 130ms or 250ms. For the music separation problem, we used non-percussive music sources and the music sources including drums in the \textit{dev1} and \textit{dev2} datasets, which has 12 music mixtures in total, where the experiments setting is consistent with that of the speech separation problem.
}

{Then, we used SISEC2018 \cite{stoter20182018} as the second experimental dataset. Specifically, we used the clean speech in the \textit{asynchronous recordings of speech mixtures} of SISEC2018 \cite{stoter20182018} as the speech source.
After mixing up, SISEC2018 includes 72 mixture signals (\textit{dev} and \textit{dev2} datasets) with female and male speech, where the microphone spacing is 2.15cm or 7.65cm and the reverberation time is 150ms or 300ms.
}

{In our experiments, we followed the environment of the SISEC challenge \cite{araki20122011} to construct a determined multichannel speech separation task with the number of channels $M=2$ and number of speakers per mixture $N=2$.}

\begin{figure*}[t]
\centering
\vspace{-0.7cm} 
\subfloat[Speech separation, $T_{60} = 150$ms]{\includegraphics[width=3.5in]{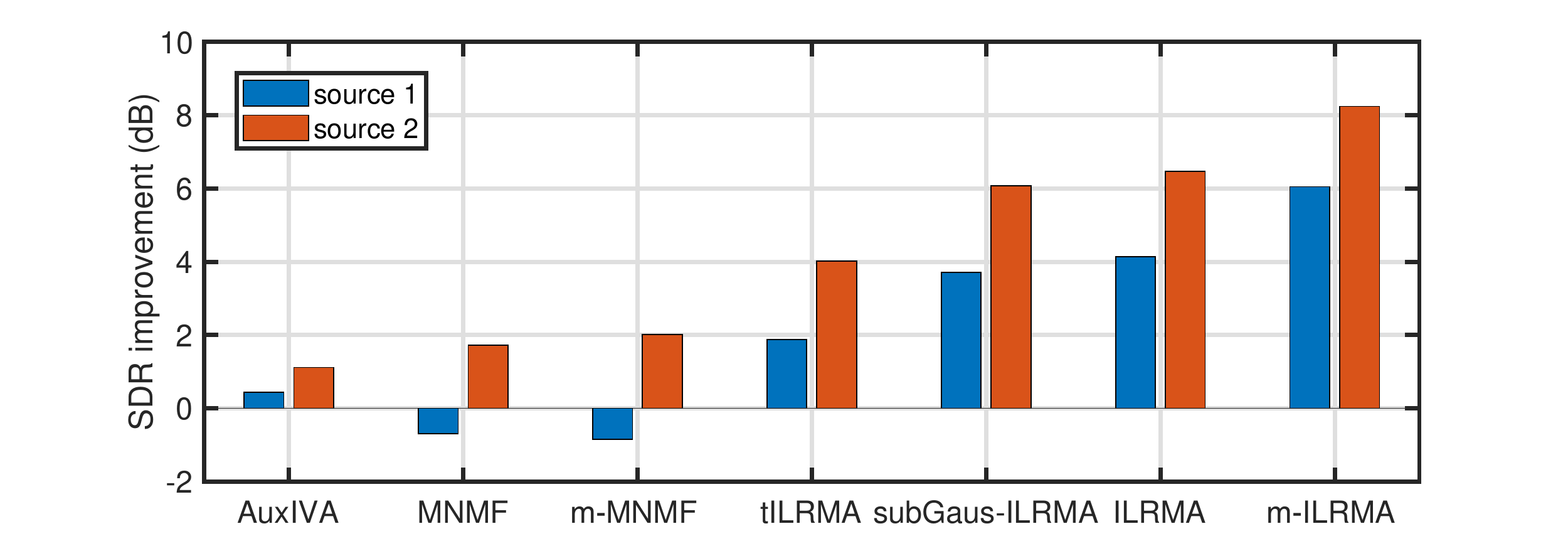}%
\label{fig_SISEC2018a}}
\hfil
\subfloat[Speech separation, $T_{60} = 300$ms]{\includegraphics[width=3.5in]{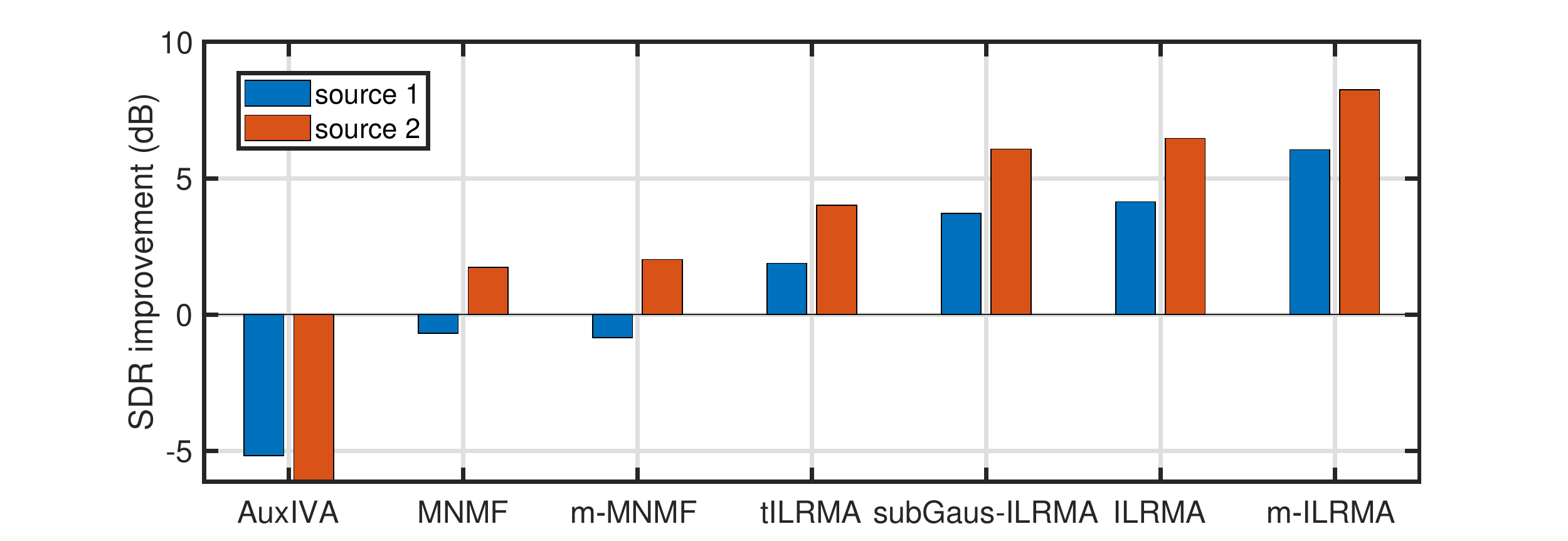}%
\label{fig_SISEC2018b}}
\hfil
\subfloat[Speech separation, $T_{60} = 150$ms]{\includegraphics[width=3.5in]{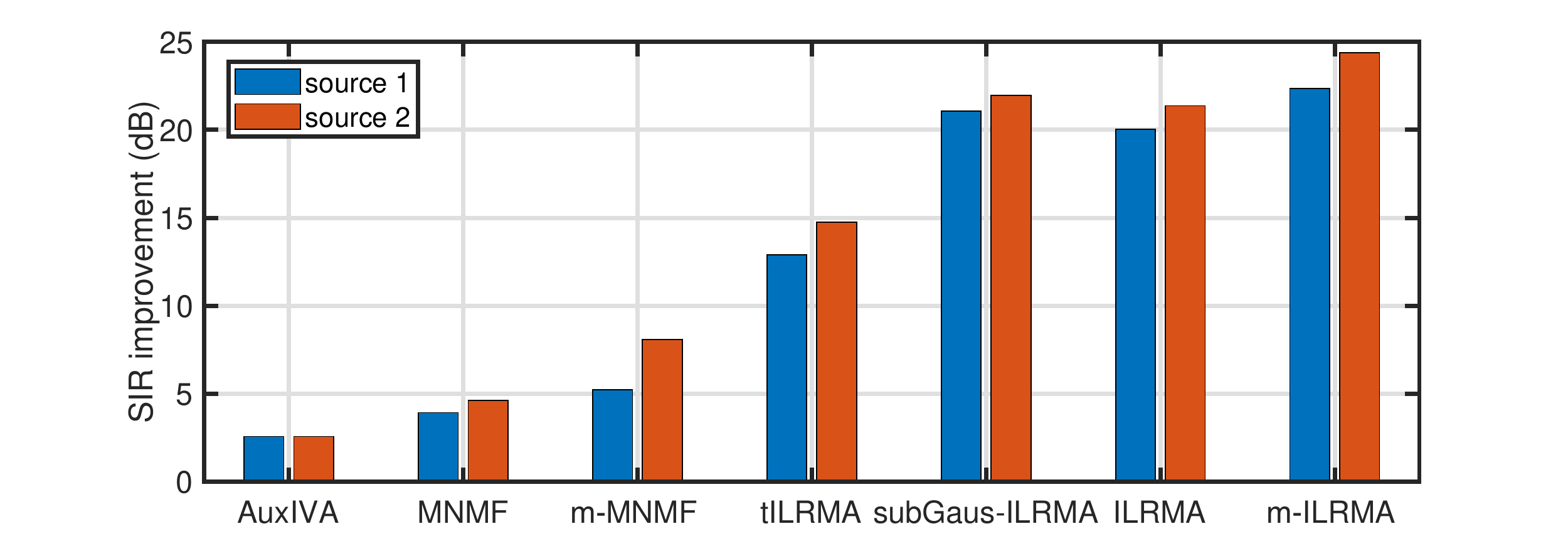}%
\label{fig_SISEC2018c}}
\hfil
\subfloat[Speech separation, $T_{60} = 300$ms]{\includegraphics[width=3.5in]{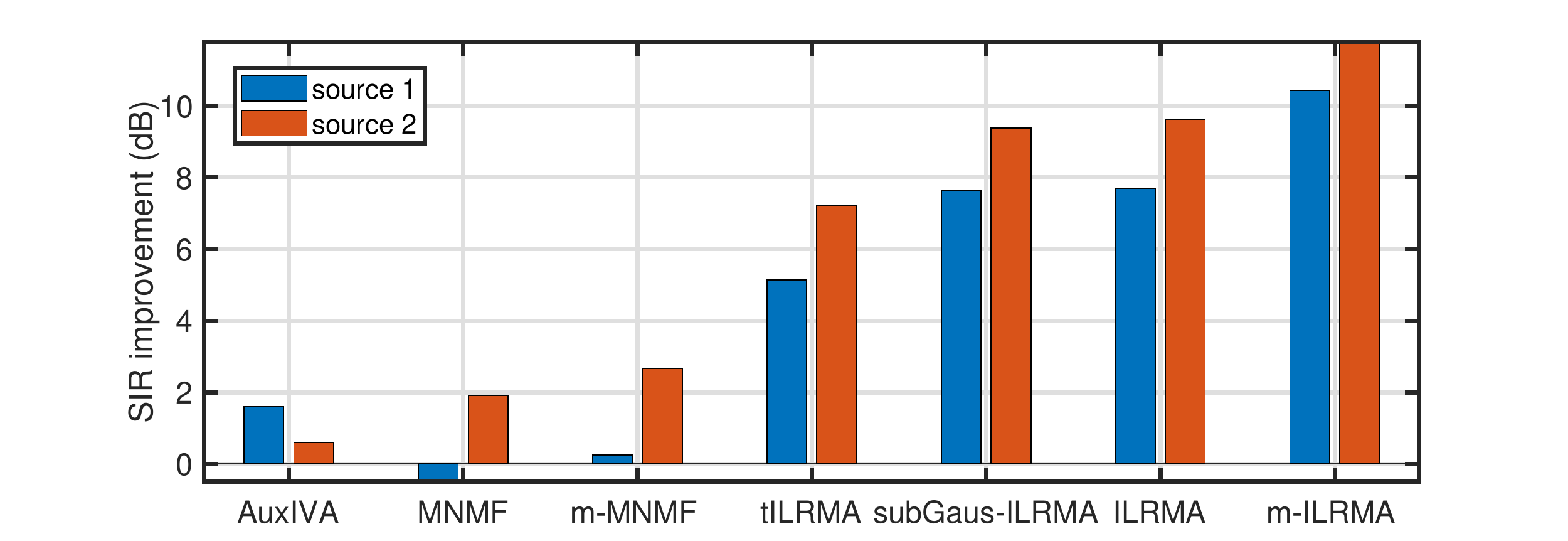}%
\label{fig_SISEC2018d}}
\caption{Performance of the comparison methods on SISEC2018.}
\label{fig_SISEC2018}
\end{figure*}

Besides the above two test corpora, we also followed the environment of the SISEC challenge \cite{araki20122011} to construct a determined multichannel speech separation task with $M=N=2$, where we used the Wall Street Journal (WSJ0) corpus \cite{garofolo1993csr} as the clean speech source. We evaluated the comparison methods on all gender combinations. We generated two test conditions for this test corpus, denoted as condition 1 and condition 2. In both conditions, the room size was set to $6\times 6\times 3$ m; the two speakers were positioned 2 m from the center of the two microphones. The differences between the two conditions are that (i) the distance between the two microphones are 5.66 cm and 2.83 cm respectively, and (ii) the incident angles of the two speakers follow \cite[Figs. 9a and 9b]{mogami2019independent}.
The image source model \cite{allen1979image} was used to generate the room impulse response with the reverberation time $T_{60}$ selected from $[130, 150, 200, 250, 300, 350, 400, 450, 500]$ ms. For each condition, we generated 200 mixtures for each gender combination at each $T_{60}$, which amounts to 7200 mixtures. The sampling rate was set to 16 kHz. We named the simulated data without reverberation as \textit{WSJ0-anechoic}, and the simulated data with reverberation as \textit{WSJ0-reverb}.

\subsubsection{Semi-real database}

As shown in Fig. \ref{fig:model1}, we also conducted an experiment in a real world environment. Specifically,
we used a circular array of 48 equiangular-placed loudspeakers with a radius of 1m and a height of 1m to produce a desired sound field.
Then, we transcribed SISEC2011 as shown in Fig. \ref{fig:model1}.
A linear array of 8 microphones, indexed as mic1 to mic8 from left to right, was placed at the center of the circular loudspeaker array.
The target speech was located at the $45^\circ$ and $120^\circ$ of the linear array respectively.
We named this semi-real database as \textit{semi-real-SISEC2011}.

\subsubsection{Comparison algorithms}\label{subsec:comparison}

{
The hyperparameters of m-MNMF and m-ILRMA in all experiments were set as follows:
$K=10$, $\eta=0.5$, and the number of iterations was set to 100.
The frame-length and frame-shift were set to 1024 and 512, respectively.
We compared m-MNMF and m-ILRMA with AuxIVA and four NMF-based multichannel BSS methods which are described as follows:
}
\begin{itemize}
 \itemsep=0.0pt
  \item \textbf{Auxiliary-function-based Independent Vector Analysis (AuxIVA)\cite{ono2011stable}}: It introduces an auxiliary function for IVA, which is solved by a stable and fast update rule.
  \item \textbf{Multichannel Nonnegative Matrix Factorization (MNMF)\cite{sawada2013multichannel}}: It is modeled by the spatial covariance of a zero-mean multivariate Gaussian distribution. It can be considered as a natural extension of NMF, since the Hermitian positive semi-definite is utilized as a multichannel counterpart of nonnegativity. The number of basis vectors $K$ was set to 10 as default.
  \item \textbf{Independent Low-Rank Matrix Analysis (ILRMA)\cite{kitamura2016determined}}: It is a unification of IVA and NMF, which assumes both the statistical independence between sources and a low-rank time-frequency structure for each source. The demixing systems of ILRMA are estimated without encountering the permutation problem. The iteration was set to 100. The number of basis vectors $K$ was set to 10.
  \item \textbf{T-distribution for Independent Low-Rank Matrix Analysis (tILRMA)\cite{mogami2017independent}}: It generalizes the source generative model of ILRMA from the complex Gaussian distribution to a complex Student's $t$-distribution, which is expected to further improve the performance as well as the stability of the parameter initialization. In our experiment, we set the hyperparameters $\nu = 1000$ and $\rho = 10$ respectively.
  \item \textbf{Sub-Gaussian Independent Low-Rank Matrix Analysis (subGaus-ILRMA) \cite{ikeshita2018independent}}: The generalization of subGaus-ILRMA is similar to $t$-ILRMA. SubGaus-ILRMA differs from $t$-ILRMA in that the distribution of its source generative model is a generalized Gaussian distribution. We set the hyperparameters $\beta = 1.99$ and $\rho = 0.5$ respectively.
\end{itemize}

\subsubsection{Evaluation metrics}

{
We used the source-to-distortion ratio (SDR) and source-to-interference ratio (SIR) \cite{vincent2006performance} to evaluate the \textit{quality} of the separated speech, which are defined as follow:
\begin{equation}\label{SDR}
\begin{split}
\mathrm{SDR} := 10 \log_{10} \frac{\| s_{\mathrm{target}} \|^2}{\| e_{\mathrm{interf}} + e_{\mathrm{noise}} + e_{\mathrm{artif}} \|^2}
\end{split}
\end{equation}
\begin{equation}\label{SIR}
\begin{split}
 \mathrm{SIR} :=  10 \log_{10} \frac{\| s_{\mathrm{target}} \|^2}{\| e_{\mathrm{interf}} \|^2}
\end{split}
\end{equation}
where $s_{\mathrm{target}}$ is a version of the wanted source modified by an allowed distortion, and $e_{\mathrm{interf}}$, $e_{\mathrm{noise}}$, and $e_{\mathrm{artif}}$ are respectively the interferences, noise, and artifacts error terms.
}

\begin{figure*}[!t]
\centering
\vspace{-0.7cm} 
\subfloat[female+female]{\includegraphics[width=2.35in]{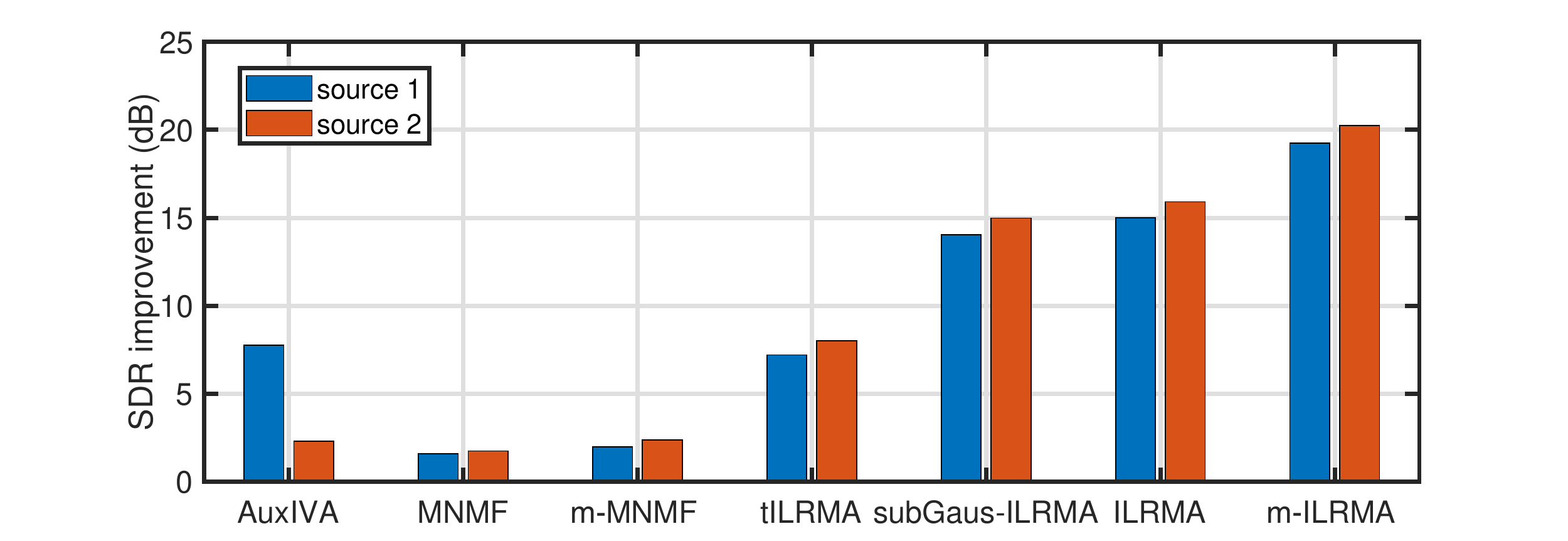}%
\label{fig_sim4a}}
\hfil
\subfloat[male+male]{\includegraphics[width=2.35in]{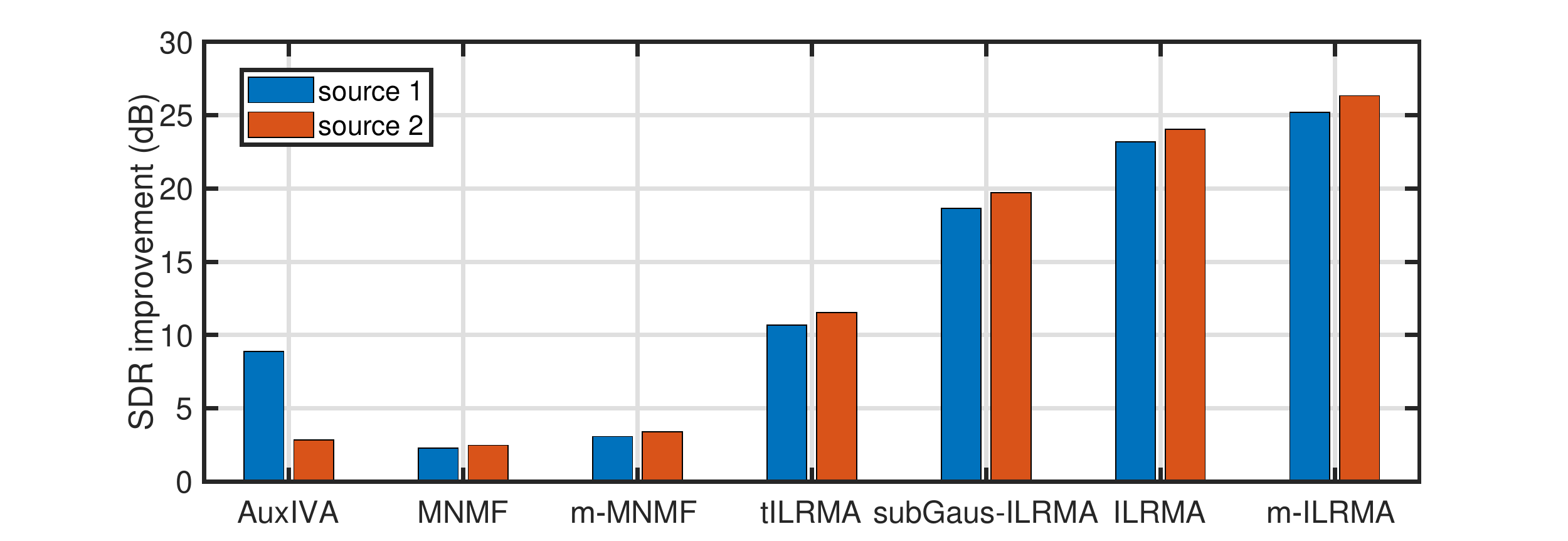}%
\label{fig_sim4b}}
\hfil
\subfloat[female+male]{\includegraphics[width=2.35in]{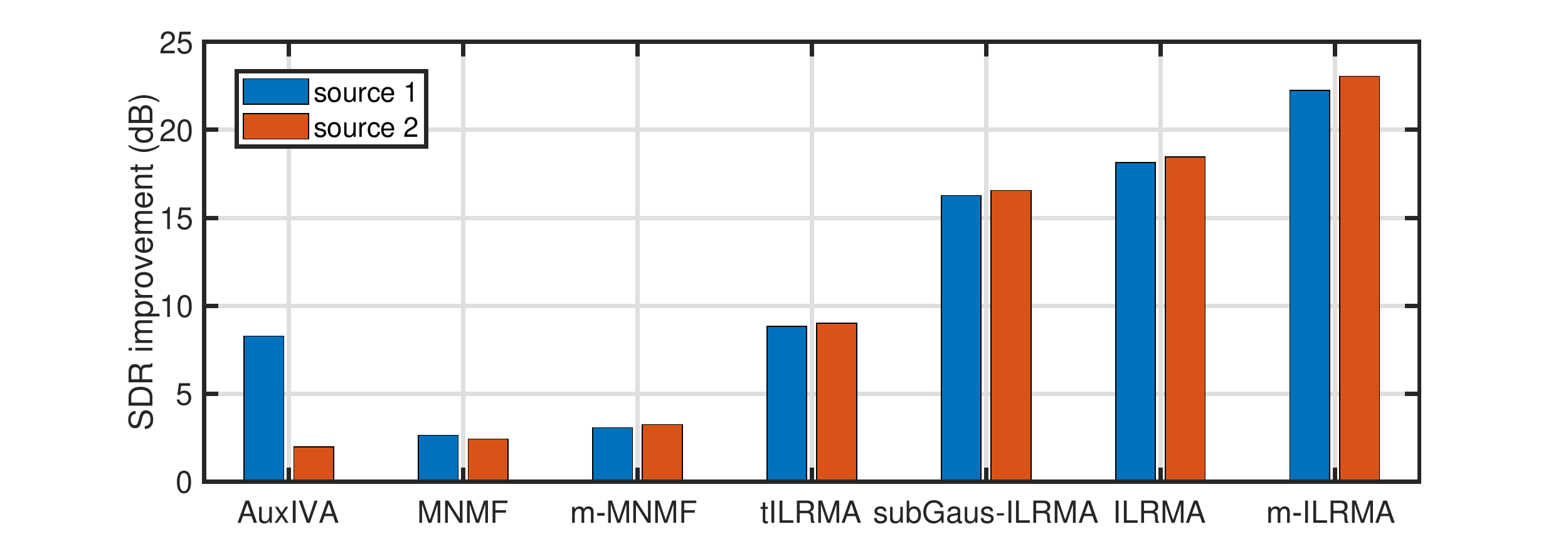}%
\label{fig_sim4c}}
\hfil
\subfloat[female+female]{\includegraphics[width=2.35in]{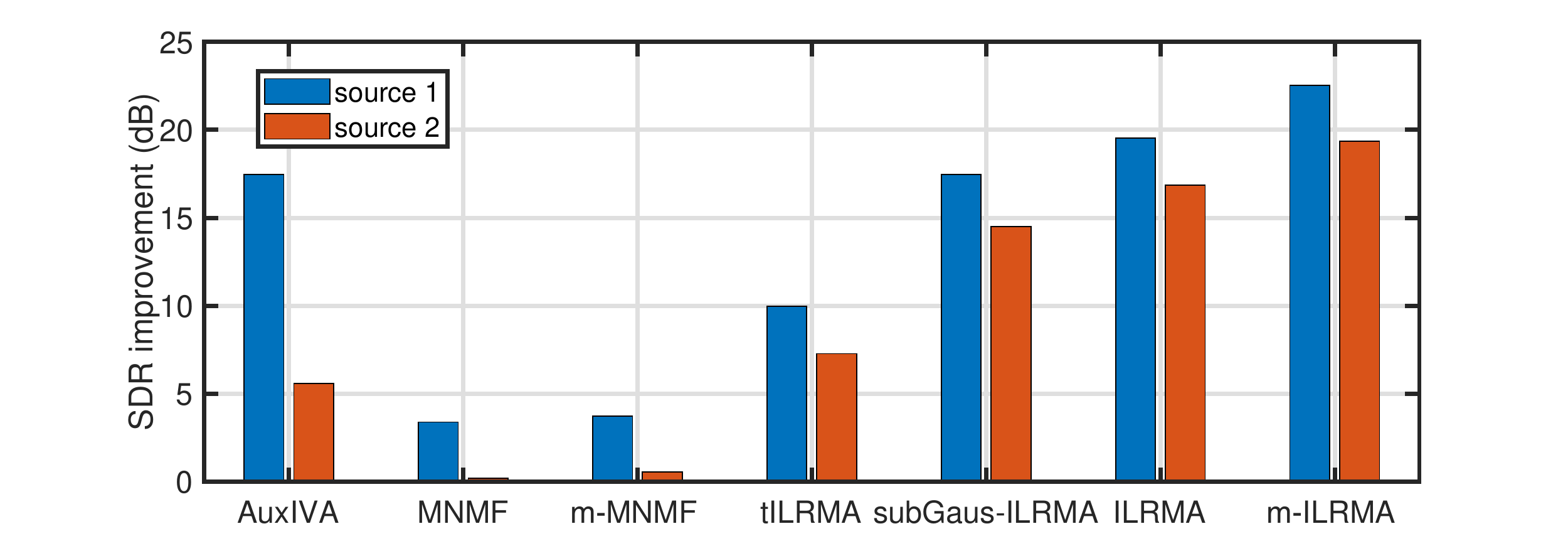}%
\label{fig_sim4d}}
\hfil
\subfloat[male+male]{\includegraphics[width=2.35in]{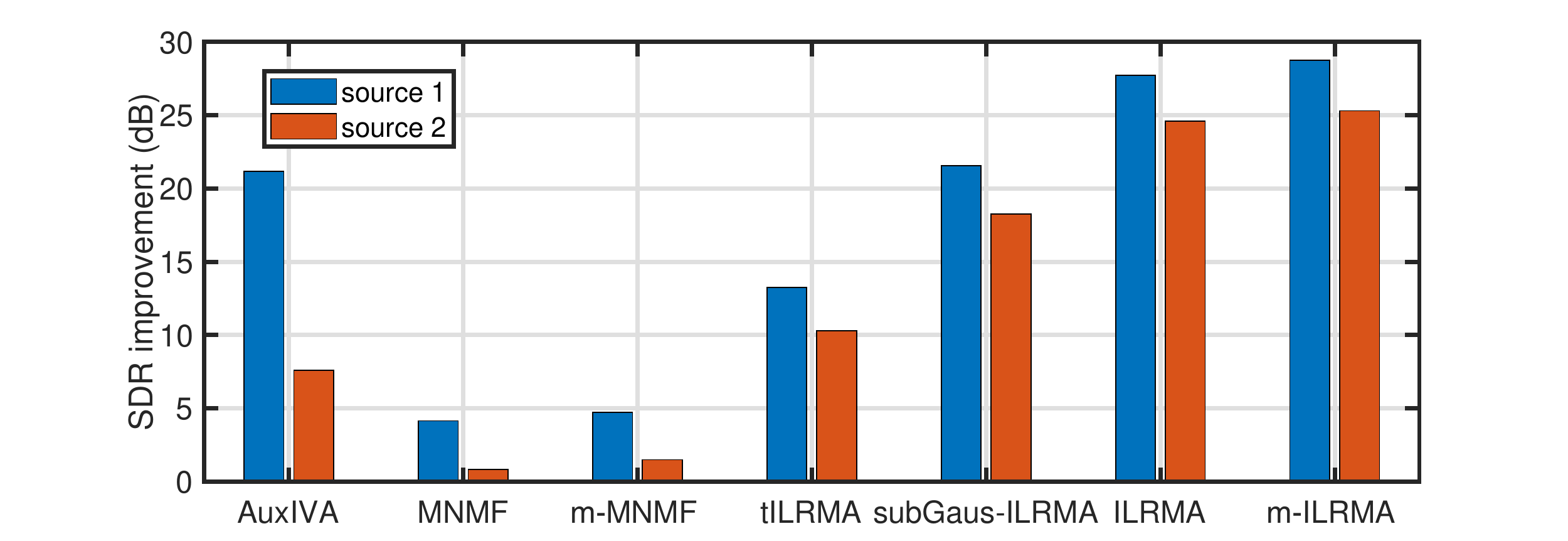}%
\label{fig_sim4e}}
\hfil
\subfloat[female+male]{\includegraphics[width=2.35in]{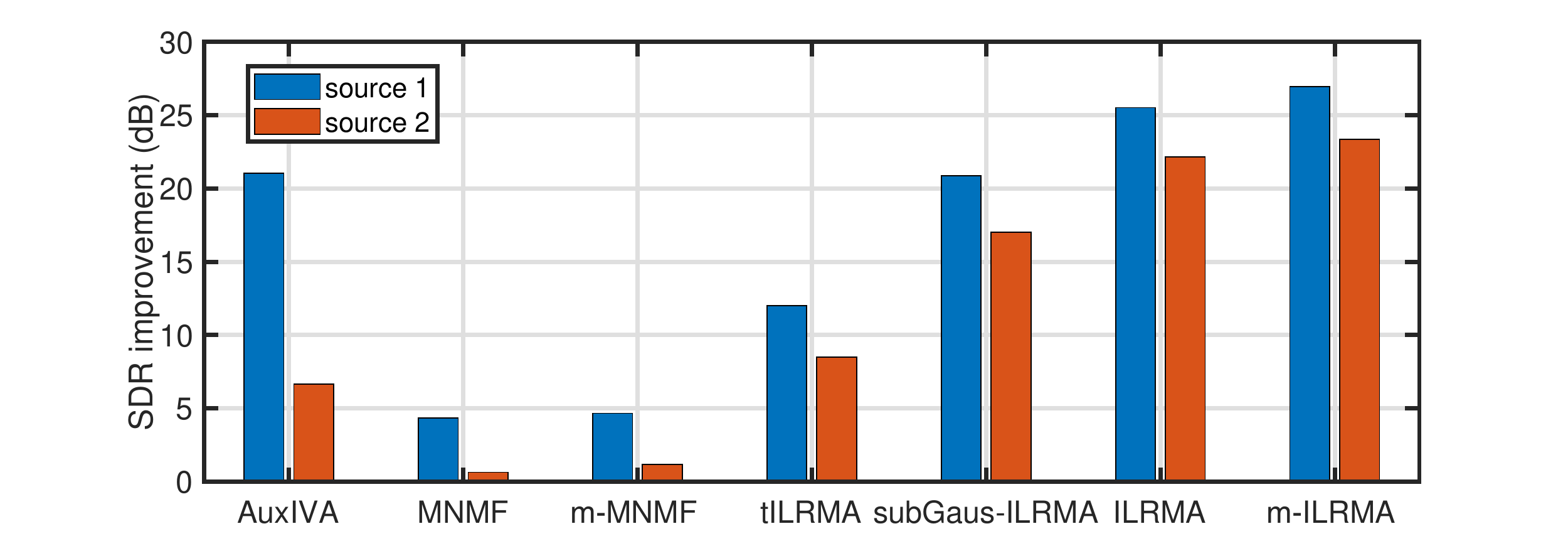}%
\label{fig_sim4f}}
\caption{SDR improvement of the comparison methods on the WSJ0-anechoic corpus. (a), (b), (c) are the results in condition 1. (d), (e), (f) are the results in condition 2.}
\label{fig_sim4}
\end{figure*}

\begin{figure*}[!t]
\centering
\vspace{-0.7cm} 
\subfloat[female+female]{\includegraphics[width=2.35in]{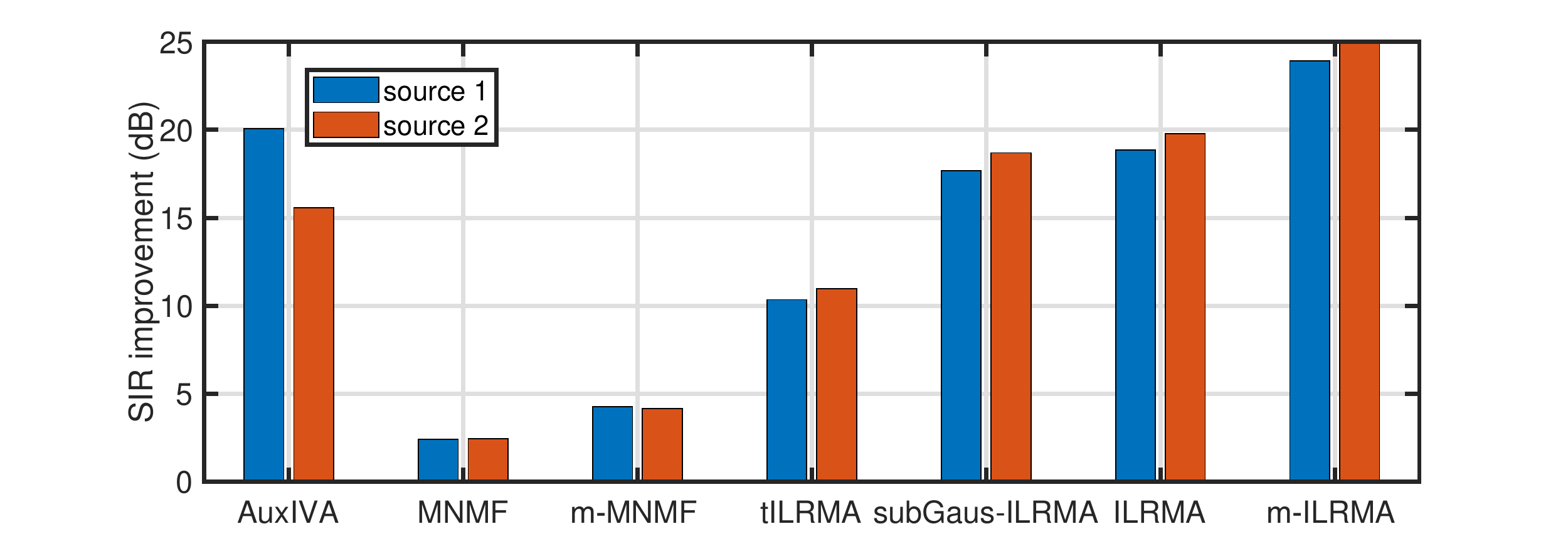}%
\label{fig_sim5a}}
\hfil
\subfloat[male+male]{\includegraphics[width=2.35in]{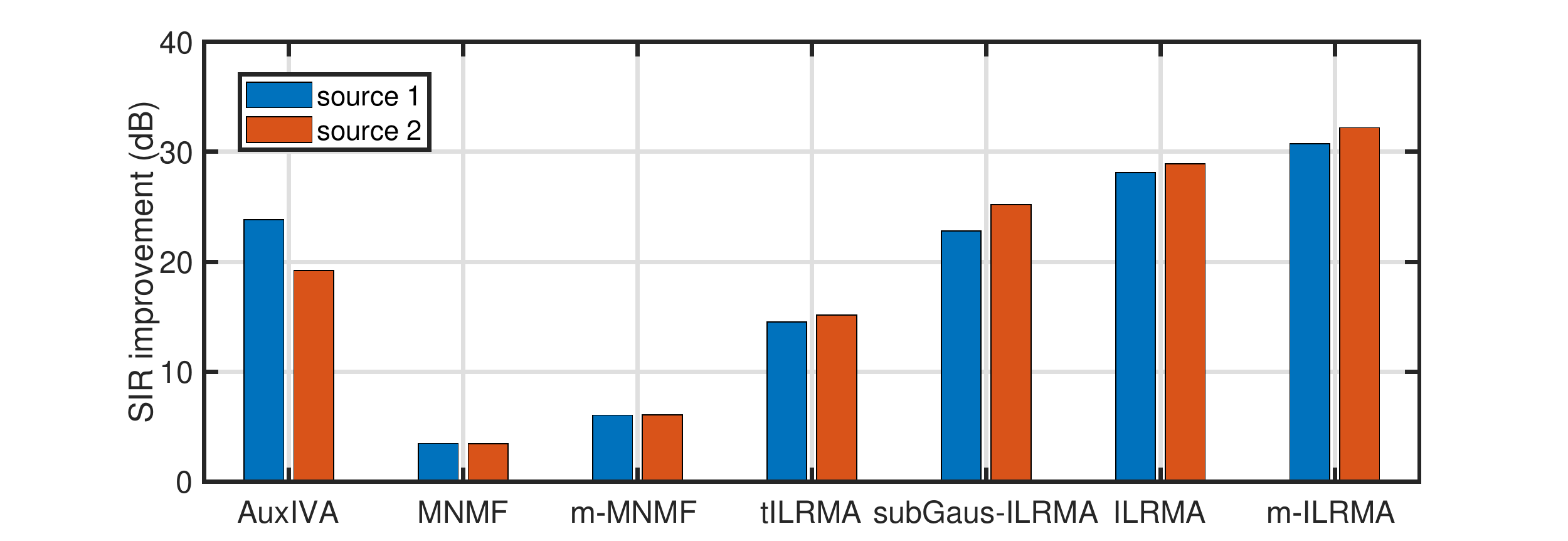}%
\label{fig_sim5b}}
\hfil
\subfloat[female+male]{\includegraphics[width=2.35in]{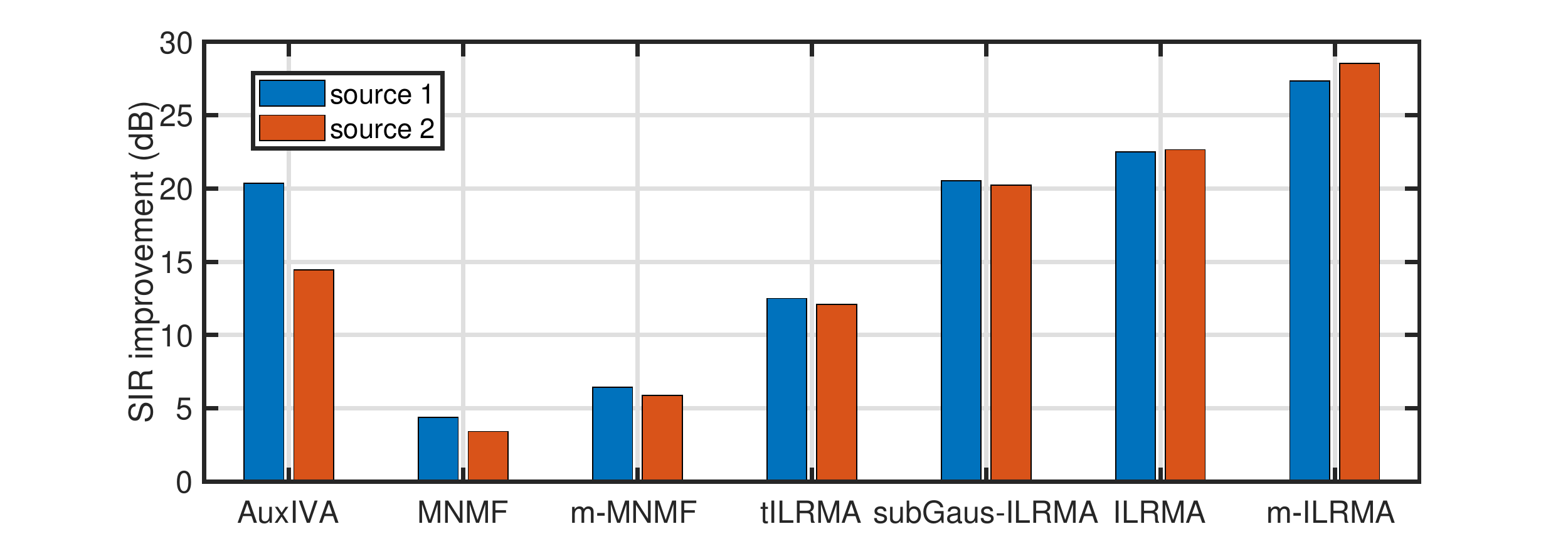}%
\label{fig_sim5c}}
\hfil
\subfloat[female+female]{\includegraphics[width=2.35in]{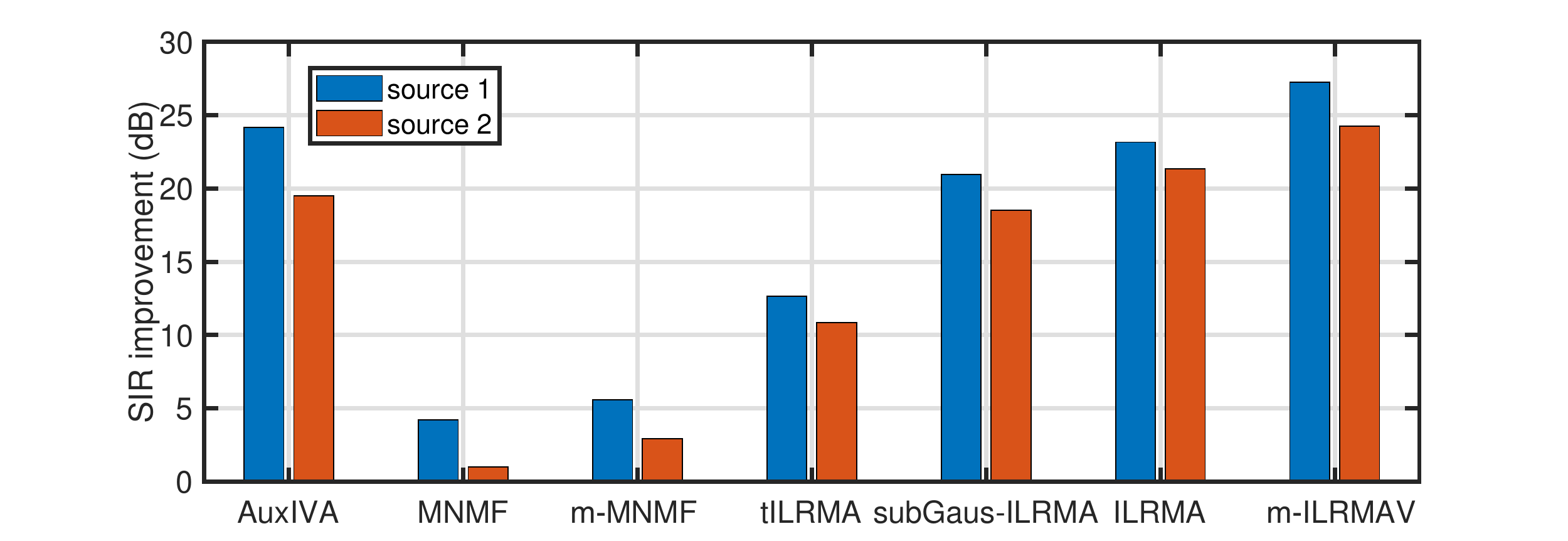}%
\label{fig_sim5d}}
\hfil
\subfloat[male+male]{\includegraphics[width=2.35in]{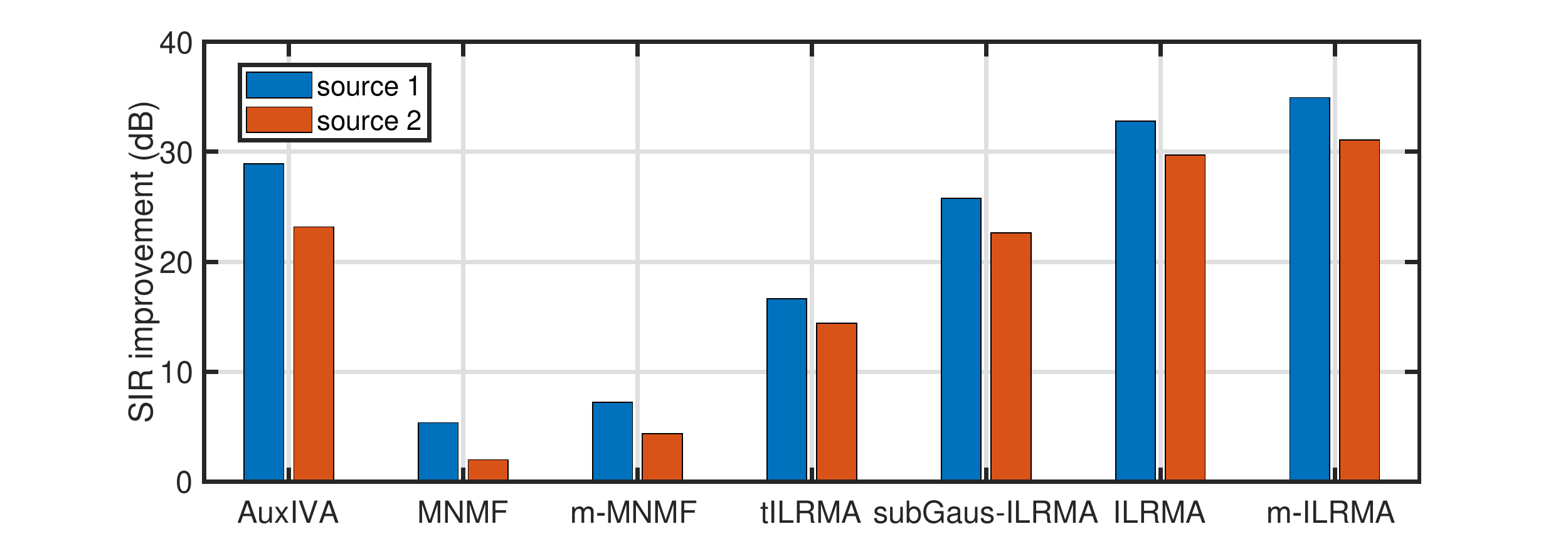}%
\label{fig_sim5e}}
\hfil
\subfloat[female+male]{\includegraphics[width=2.35in]{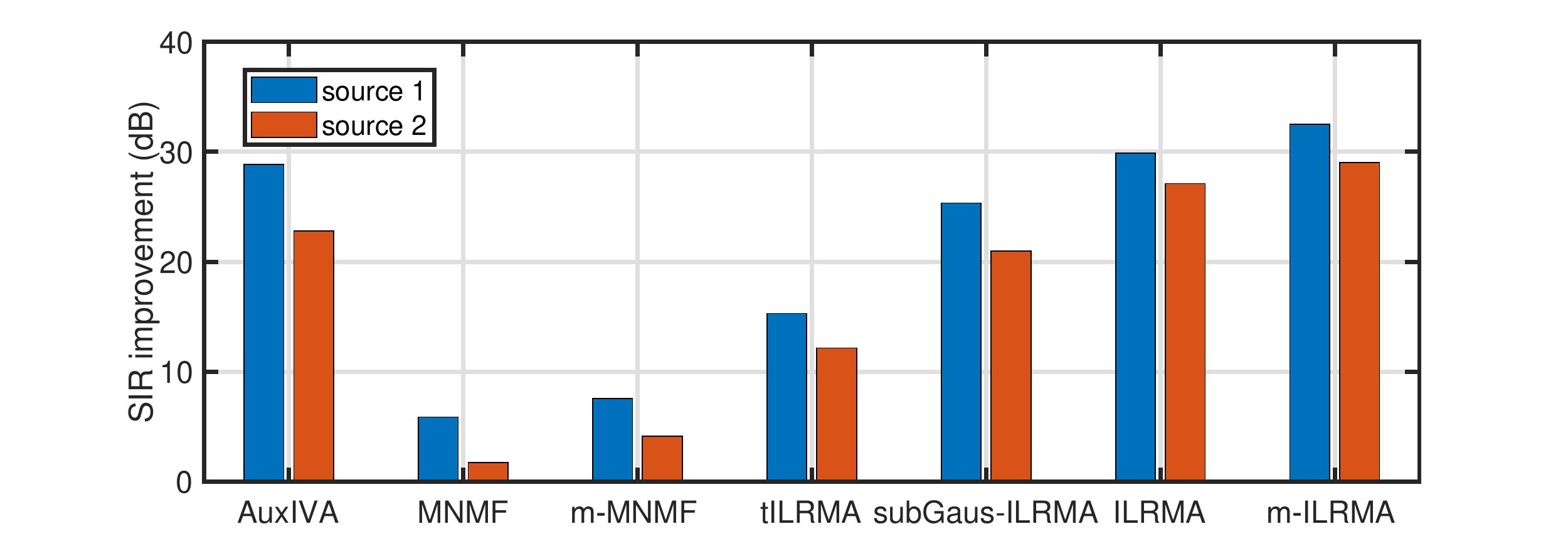}%
\label{fig_second_case}}
\caption{SIR improvement of the comparison methods on the WSJ0-anechoic corpus. (a), (b), (c) are the results in condition 1. (d), (e), (f) are the results in condition 2.}
\label{fig_sim5}
\end{figure*}

\begin{table}[t]
\vspace{-0cm}
\centering
\caption{Average SDR improvement (dB) of the comparison methods over different reverberation time on WSJ0-reverb.}\label{tab:avg_list1}
\scalebox{1}{
\setlength{\tabcolsep}{2mm}{
\begin{tabular}{lccc|ccc}
  \toprule[1pt]
  \multirow{2}{*}{Methods}
    & \multicolumn{3}{c}{Condition 1} & \multicolumn{3}{c}{Condition 2} \\
    \cmidrule[1pt]{2-7}
    & {f+f} & {m+m} & {f+m} & {f+f} & {m+m} & {f+m} \\
   \midrule[1pt]
    AuxIVA {\cite{ono2011stable}} & {{2.98}}  & {{3.40}}  & {{2.95}}  & {{5.92}}  & {{7.55}}  & {{7.60}}  \\
    MNMF  {\cite{sawada2013multichannel}} & 1.25  & 1.84  & 1.97  & 1.47  & 2.00  & 2.11  \\
    t-ILRMA {\cite{mogami2017independent}} & 3.30 & 5.10 & 3.95 & 3.29 & 4.95 & 3.92 \\
    subGaus-ILRMA {\cite{ikeshita2018independent}} & 5.13 & 7.08 & 5.81 & 5.27 & 7.40 & 6.14 \\
    ILRMA {\cite{kitamura2016determined}} & 5.03  & 6.89  & 5.72  & 5.17  & 7.31  & 6.00  \\
    {m-MNMF} & 1.05 & 1.55 & 1.69 & 1.37 & 1.87 & 1.90 \\
    {m-ILRMA} & 7.39  & 8.77  & 7.87  & 8.31  & 10.06  & 9.29  \\
  \bottomrule[1pt]
\end{tabular}}
}
\vspace{-0cm}
\end{table}

\begin{table}[!t]
\vspace{-0cm}
\centering
\caption{Average SIR improvement (dB) of the comparison methods over different reverberation time on WSJ0-reverb.}\label{tab:avg_list2}
\scalebox{1}{
\setlength{\tabcolsep}{1.5mm}{
\begin{tabular}{lccc|ccc}
  \toprule[1pt]
  \multirow{2}{*}{Methods}
    & \multicolumn{3}{c}{Condition 1} & \multicolumn{3}{c}{Condition 2} \\
    \cmidrule[1pt]{2-7}
    & {f+f} & {m+m} & {f+m} & {f+f} & {m+m} & {f+m} \\
   \midrule[1pt]
    AuxIVA {\cite{ono2011stable}} & {10.09}  & {11.86}  & {10.20}  & {12.19}  & {14.58}  & {13.69}  \\
    MNMF  {\cite{sawada2013multichannel}} & 1.58  & 2.34  & 2.57  & 1.87  & 2.59  & 2.76  \\
    t-ILRMA {\cite{mogami2017independent}} & 6.02 & 8.35 & 6.91 & 5.80 & 8.00 & 6.69 \\
    subGaus-ILRMA {\cite{ikeshita2018independent}} & 8.10 & 10.55 & 8.96 & 7.96 & 10.75 & 9.06 \\
    ILRMA {\cite{kitamura2016determined}} & 7.83  & 10.11  & 8.67  & 7.65  & 10.31  & 8.68  \\
    {m-MNMF} & 1.60 & 2.33 & 2.60 & 2.17 & 2.85 & 2.96 \\
    {m-ILRMA} & 10.80  & 12.63  & 11.47  & 11.69  & 14.06  & 12.98  \\
  \bottomrule[1pt]
\end{tabular}}
}
\vspace{-0cm}
\end{table}

\subsection{Main results}\label{subsec:results}

\subsubsection{Results on SISEC2011}
The comparison results on SISEC2011 are summarized in Figs. \ref{fig_SISEC2011_1} and \ref{fig_SISEC2011_5}. Specifically, Figs. \ref{fig_SISEC2011_1a}, \ref{fig_SISEC2011_5a} and Figs. \ref{fig_SISEC2011_1b}, \ref{fig_SISEC2011_5b} show the SDR scores of the comparison methods on the speech separation problem with the reverberation time of 130ms and 250ms respectively. Figs. \ref{fig_SISEC2011_1d}, \ref{fig_SISEC2011_1e} and Figs. \ref{fig_SISEC2011_5d}, \ref{fig_SISEC2011_5e} show the corresponding SIR scores of the comparison methods.
From the figures, we see that the performance of the proposed {m-ILRMA} is significantly better than the other methods. For example, it achieves an SDR improvement of about 2 dB higher than the best baselines, i.e. {ILRMA} and {subGaus-ILRMA}, in both of the test environments.

Figs. \ref{fig_SISEC2011_1c}, \ref{fig_SISEC2011_1f} and Figs. \ref{fig_SISEC2011_5c}, \ref{fig_SISEC2011_5f} show the comparison result on the music separation problem. From the figures, we see that {m-ILRMA} achieves better performance than the other methods except {MNMF}.

\subsubsection{Results on SISEC2018}

Fig. \ref{fig_SISEC2018} shows the comparison results on speech separation in terms of the average SDR and SIR improvement.
From the figure, we see that m-MNMF outperforms MNMF, and m-ILRMA outperforms ILRMA, which demonstrate the effectiveness of the proposed MinVol prior for the multichannel BSS.

\subsubsection{Results on WSJ0-anechoic}

 Figs. \ref{fig_sim4} and \ref{fig_sim5} show respectively the average SDR and SIR improvement of the comparison methods over the mixed speech in the simulated anechoic environment of  WSJ0-anechoic. From the figures, we see that the performance of the proposed {m-ILRMA} is significantly better than that of the other methods. For example, {m-ILRMA} achieves an SDR improvement of about 3 dB higher than the best reference method, i.e. ILRMA.

\subsubsection{Results on WSJ0-reverb}

\begin{figure}[t]

\begin{minipage}[b]{.46\linewidth}
  \centering
  \centerline{\includegraphics[width=4.8cm]{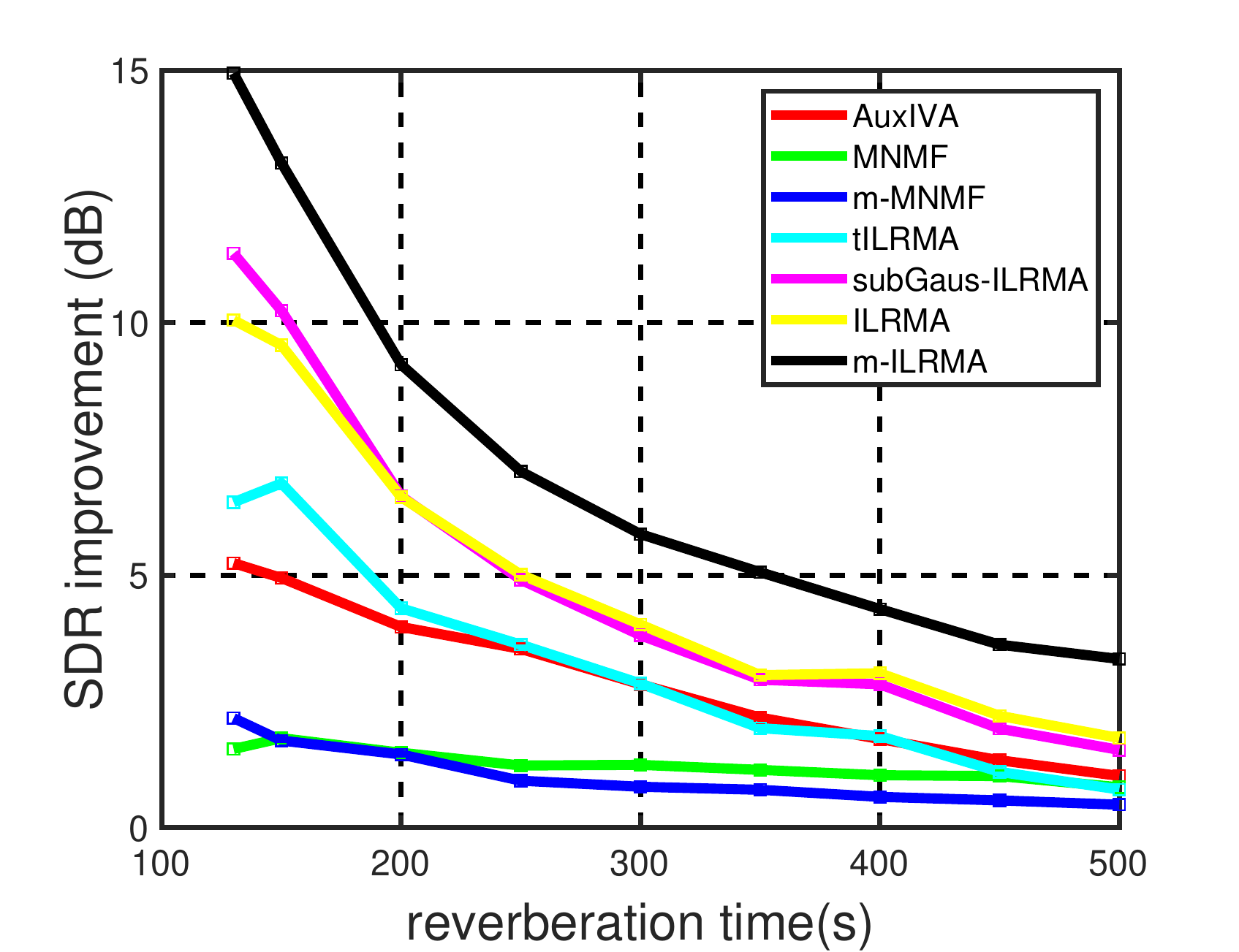}}
  \centerline{(a) female+female}\medskip
\end{minipage}
\begin{minipage}[b]{.55\linewidth}
  \centering
  \centerline{\includegraphics[width=4.8cm]{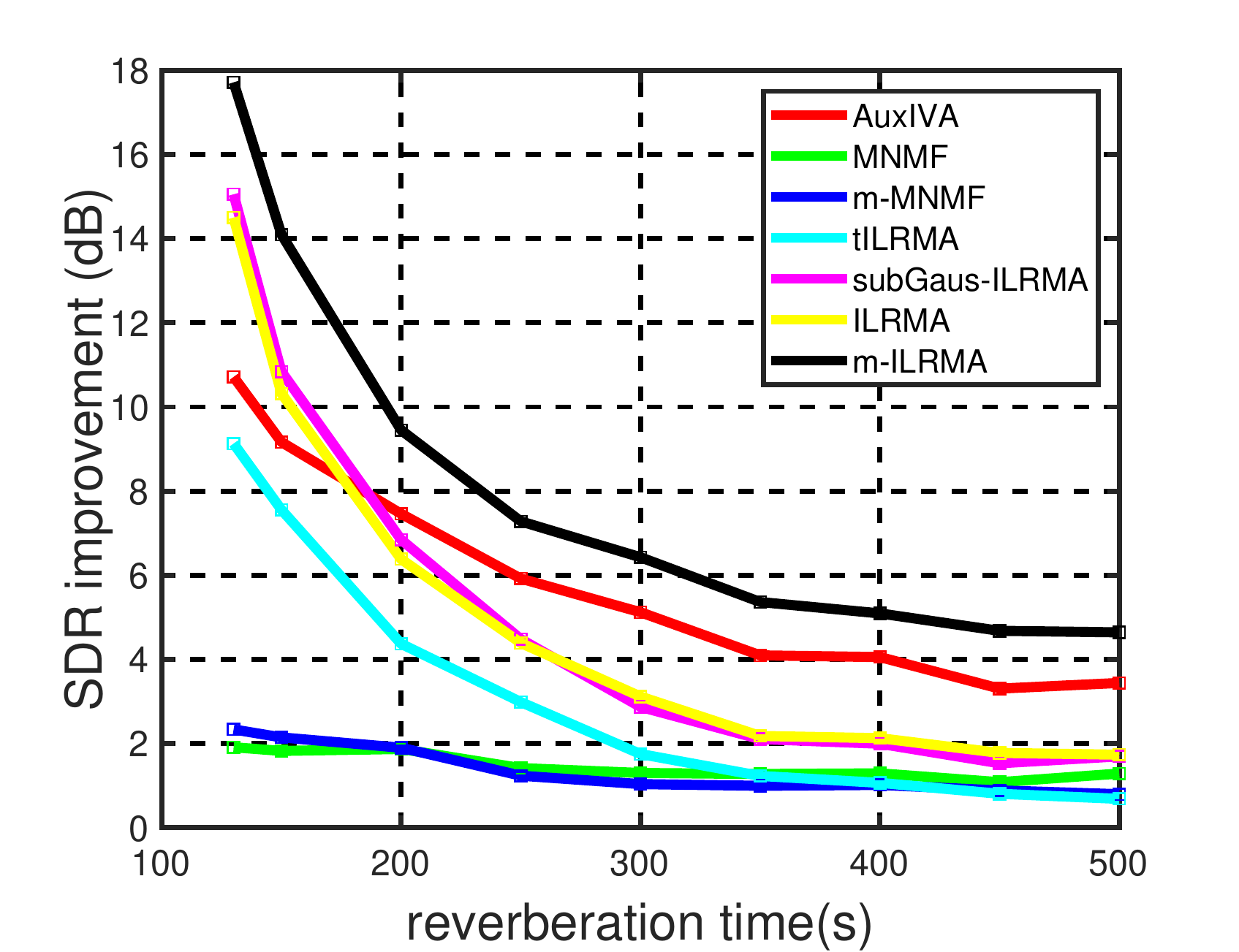}}
  \centerline{(b) female+female}\medskip
\end{minipage}
\vspace{-0.5cm}

\begin{minipage}[b]{0.46\linewidth}
  \centering
  \centerline{\includegraphics[width=4.8cm]{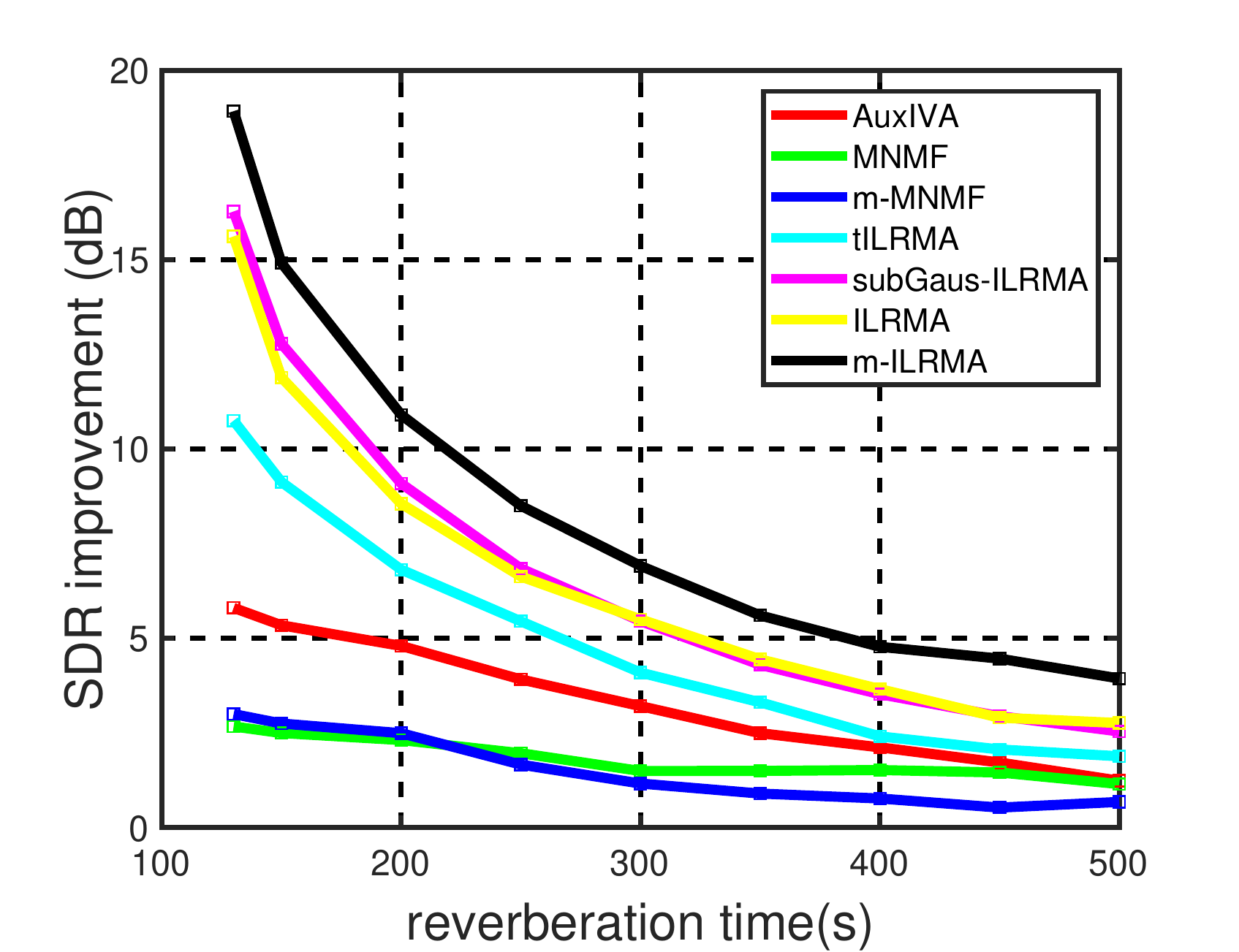}}
  \centerline{(c) male+male}\medskip
\end{minipage}
\begin{minipage}[b]{.55\linewidth}
  \centering
  \centerline{\includegraphics[width=4.8cm]{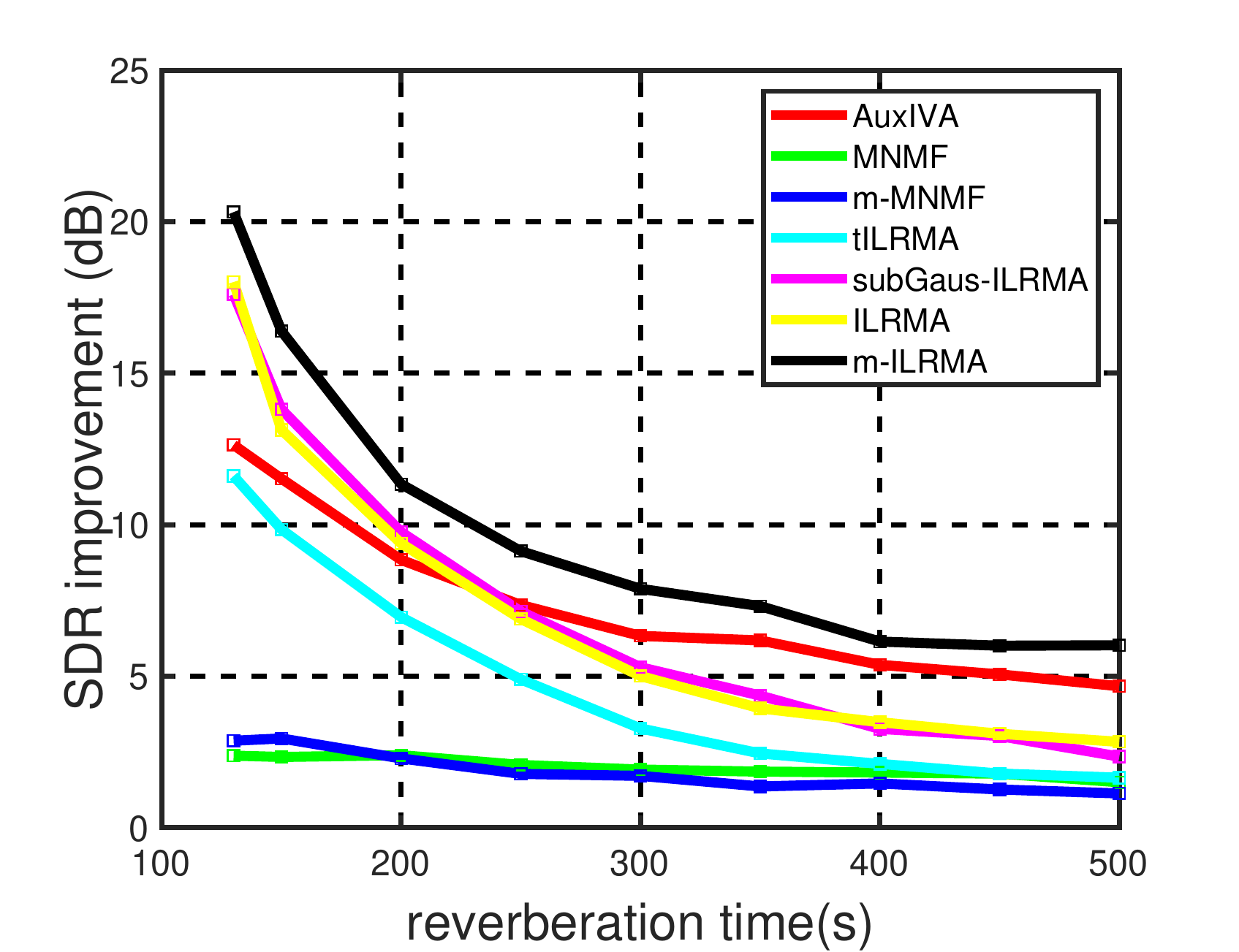}}
  \centerline{(d) male+male}\medskip
\end{minipage}
\vspace{-0.5cm}

\begin{minipage}[b]{.46\linewidth}
  \centering
  \centerline{\includegraphics[width=4.8cm]{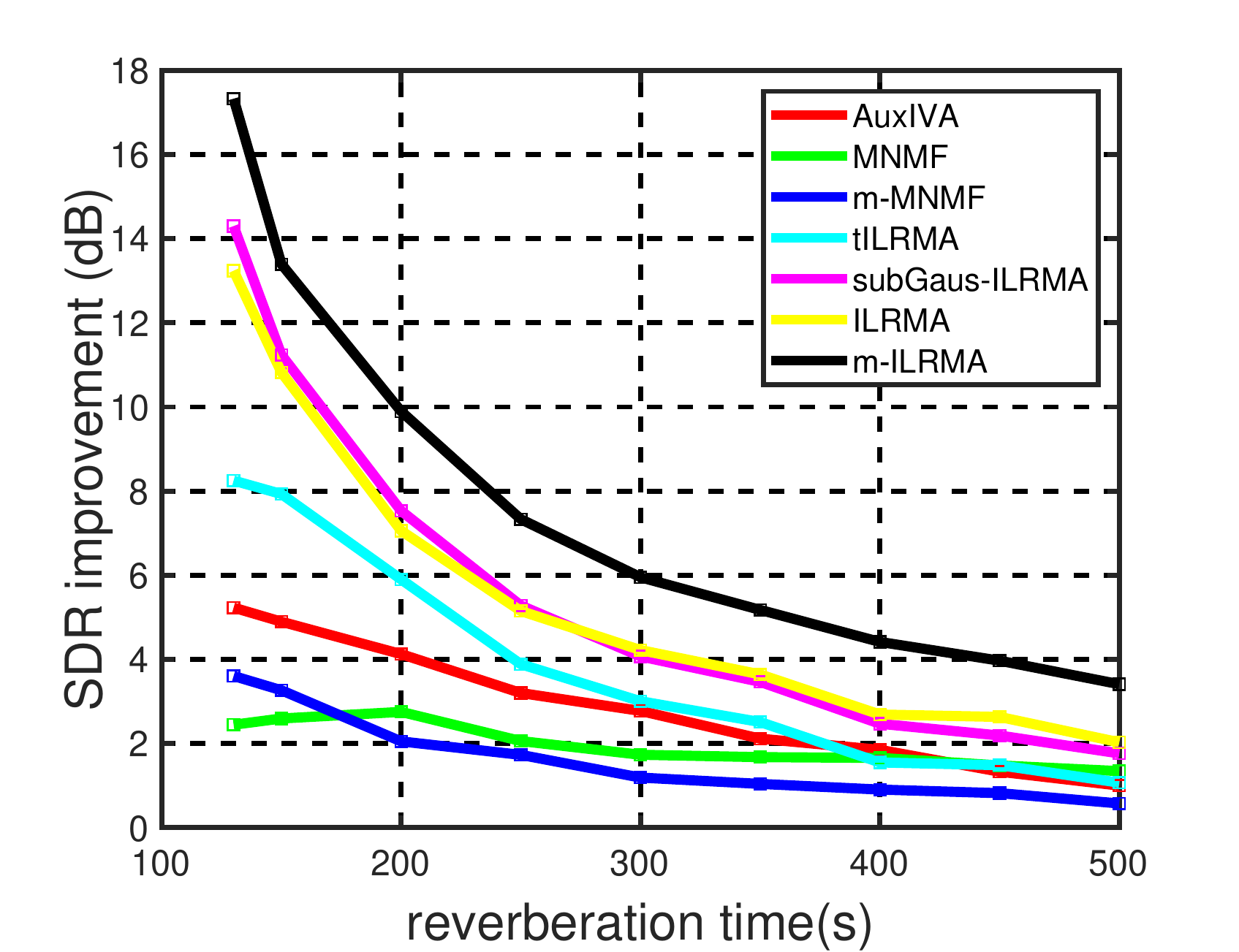}}
  \centerline{(e) female+male}\medskip
\end{minipage}
\begin{minipage}[b]{0.55\linewidth}
  \centering
  \centerline{\includegraphics[width=4.8cm]{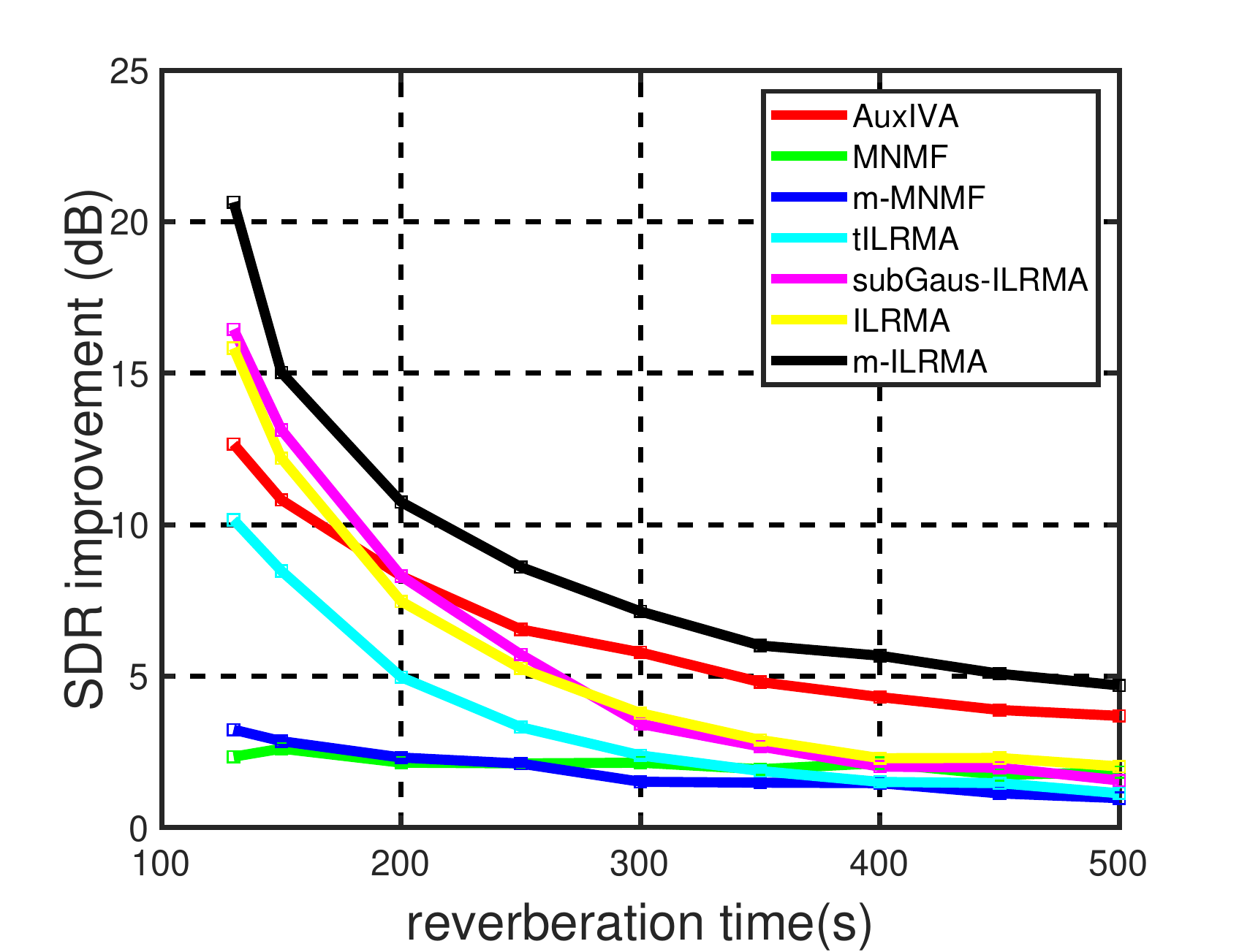}}
  \centerline{(f) female+male}\medskip
\end{minipage}
\vspace{-0.8cm}
\caption{SDR improvement of the comparison methods on WSJ0-reverb. (a), (c), (e) are the results in condition 1. (b), (d), (f) are the results in condition 2.}
\label{fig:7}
\vspace{-0cm}
\end{figure}

\begin{figure}[t]

\begin{minipage}[b]{.46\linewidth}
  \centering
  \centerline{\includegraphics[width=4.8cm]{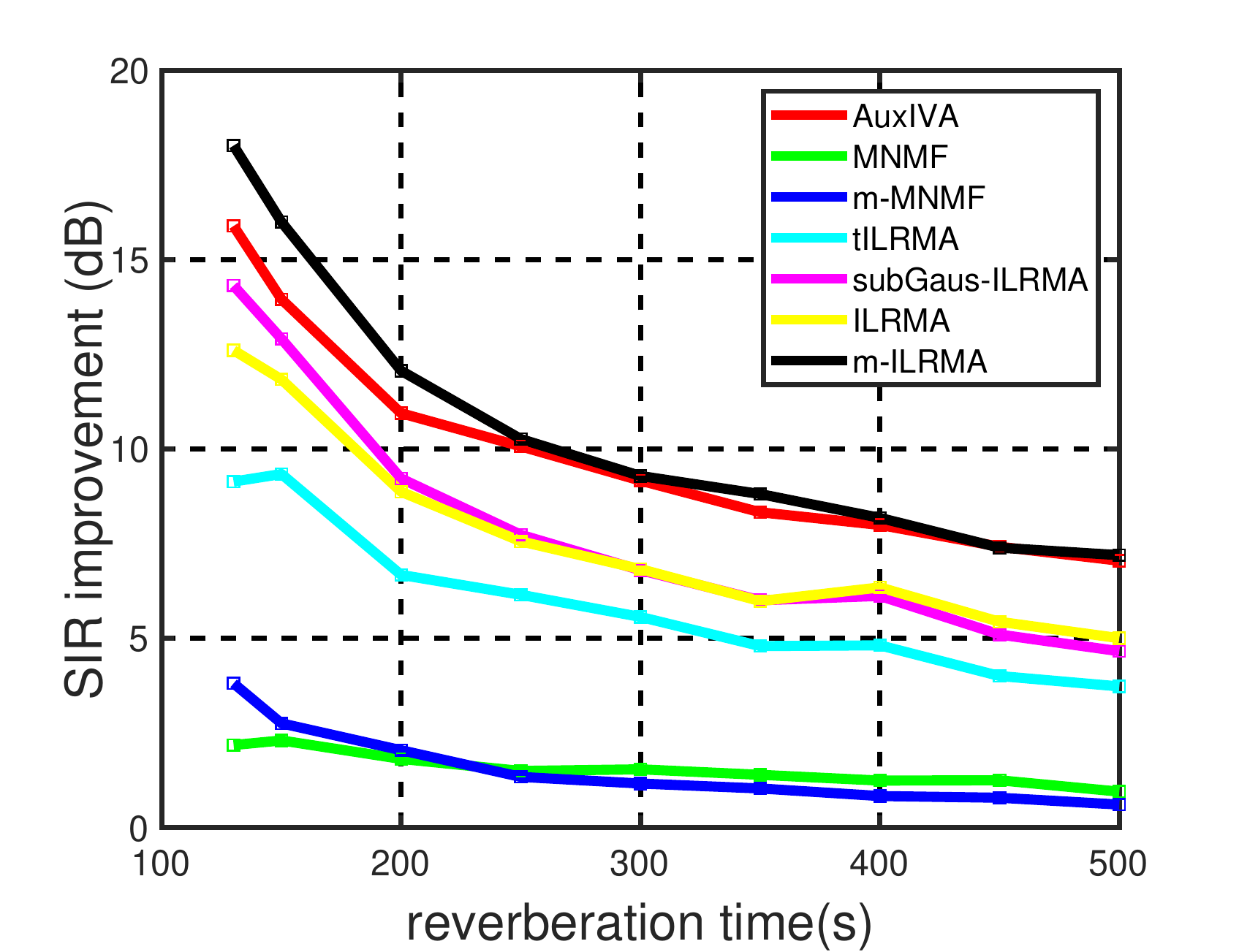}}
  \centerline{(a) female+female}\medskip
\end{minipage}
\begin{minipage}[b]{.55\linewidth}
  \centering
  \centerline{\includegraphics[width=4.8cm]{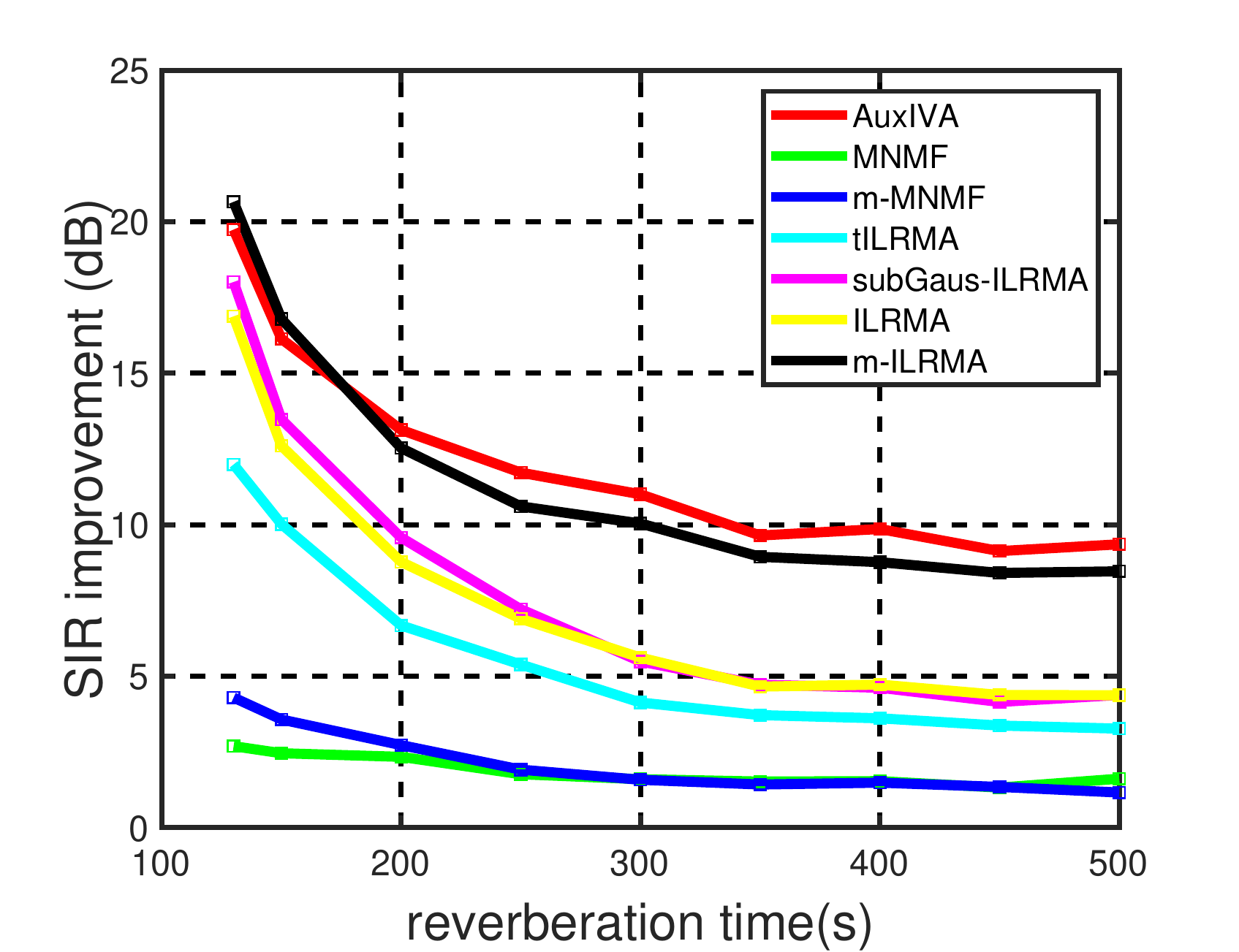}}
  \centerline{(b) female+female}\medskip
\end{minipage}
\vspace{-0.5cm}

\begin{minipage}[b]{0.46\linewidth}
  \centering
  \centerline{\includegraphics[width=4.8cm]{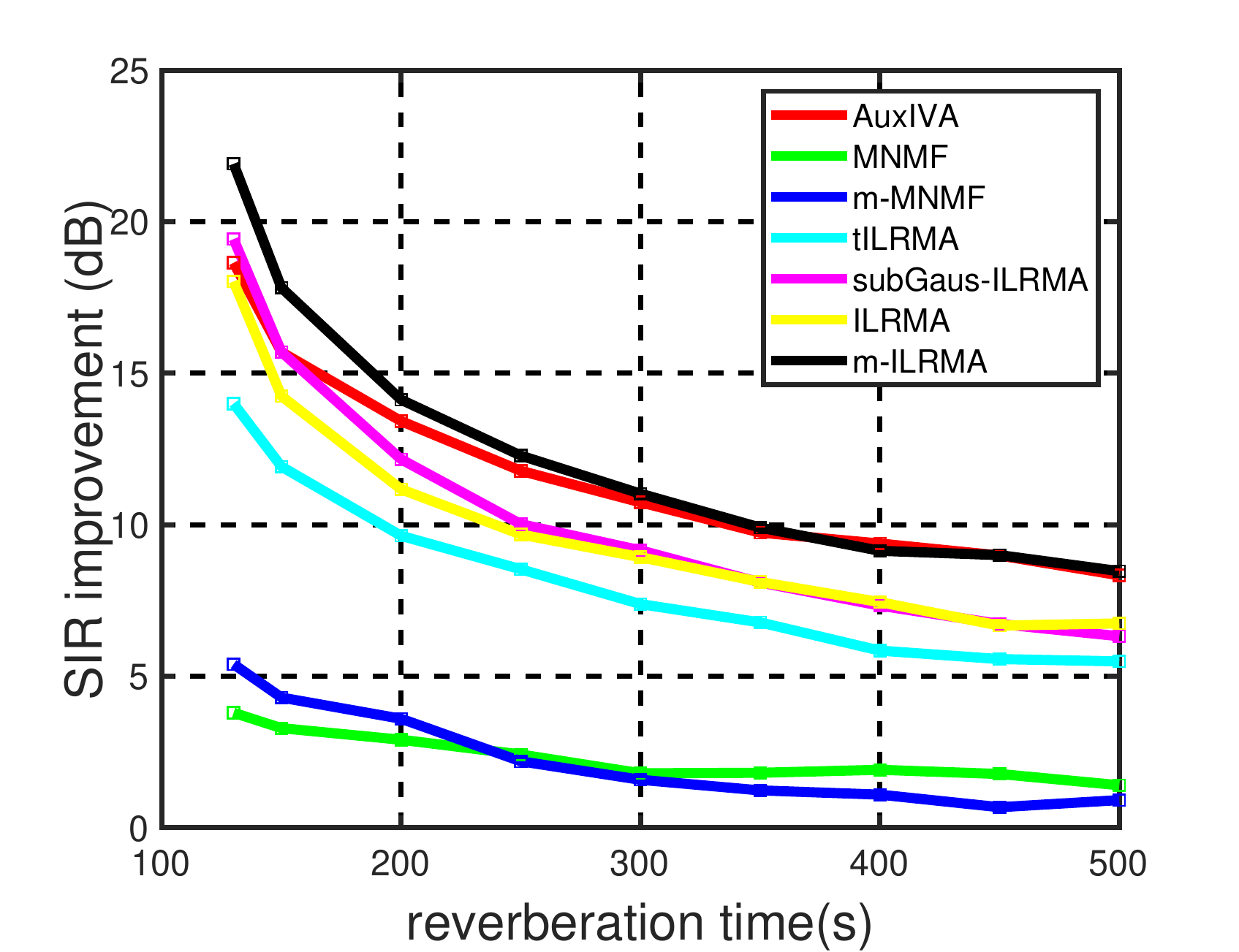}}
  \centerline{(c) male+male}\medskip
\end{minipage}
\begin{minipage}[b]{.55\linewidth}
  \centering
  \centerline{\includegraphics[width=4.8cm]{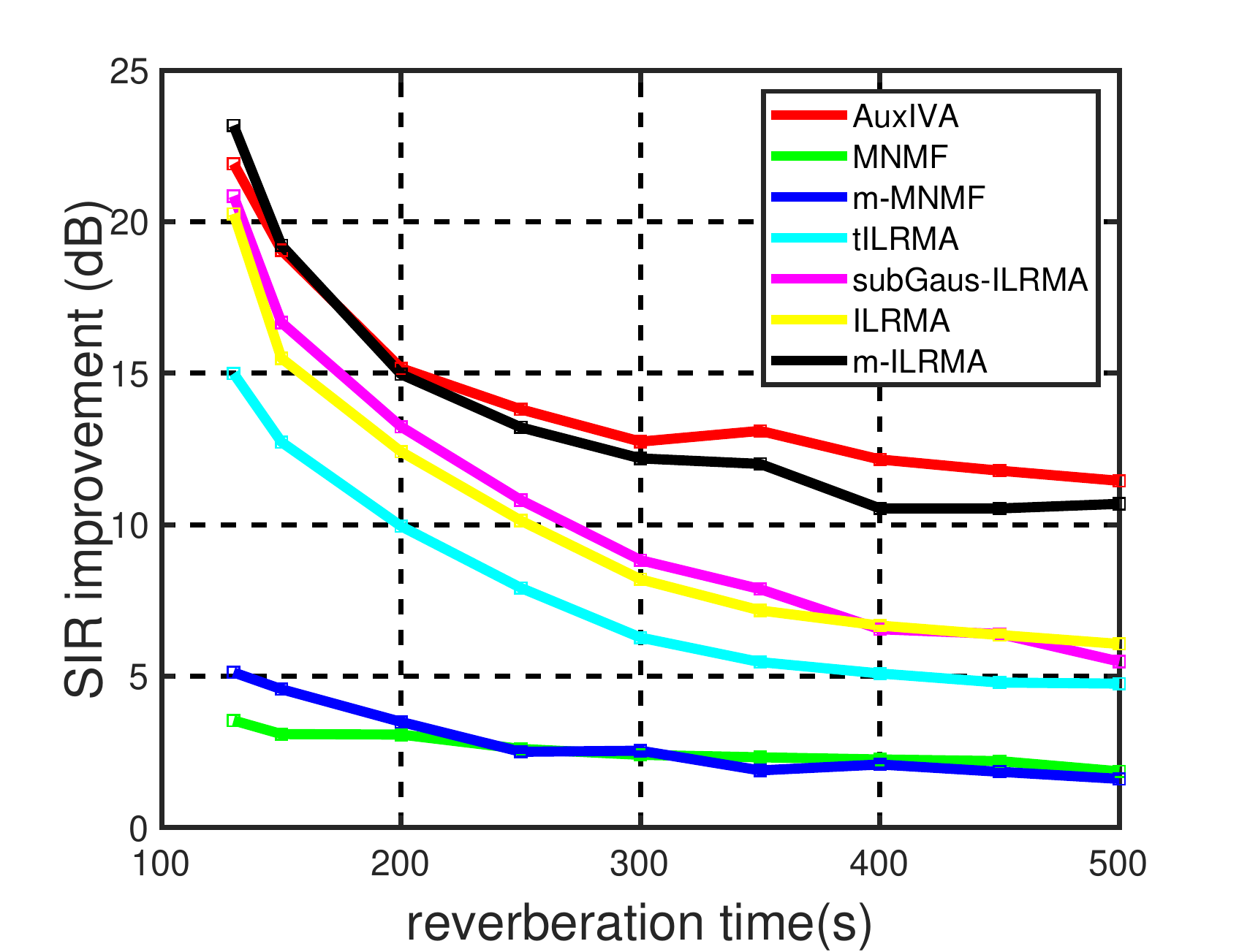}}
  \centerline{(d) male+male}\medskip
\end{minipage}
\vspace{-0.5cm}

\begin{minipage}[b]{.46\linewidth}
  \centering
  \centerline{\includegraphics[width=4.8cm]{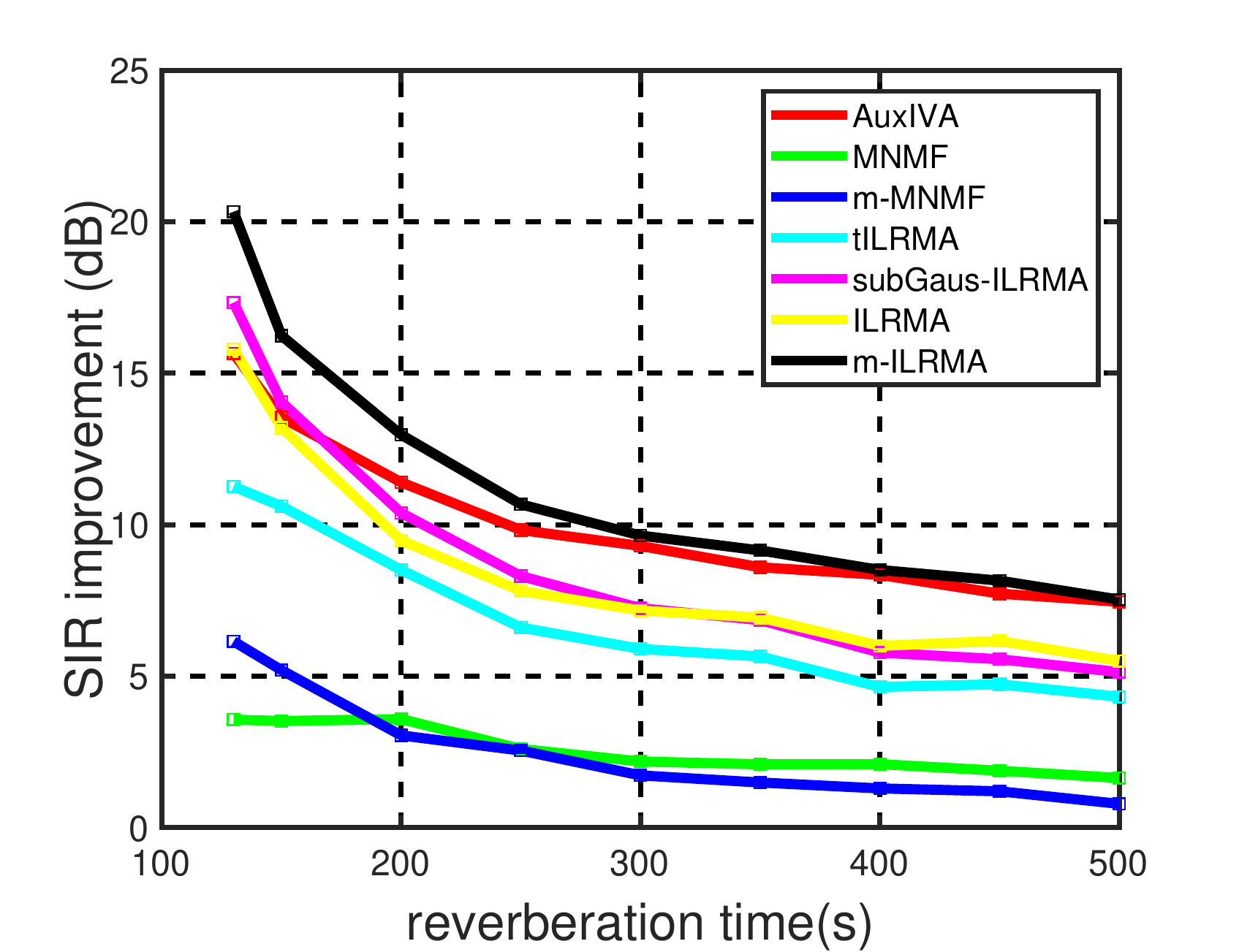}}
  \centerline{(e) female+male}\medskip
\end{minipage}
\begin{minipage}[b]{0.55\linewidth}
  \centering
  \centerline{\includegraphics[width=4.8cm]{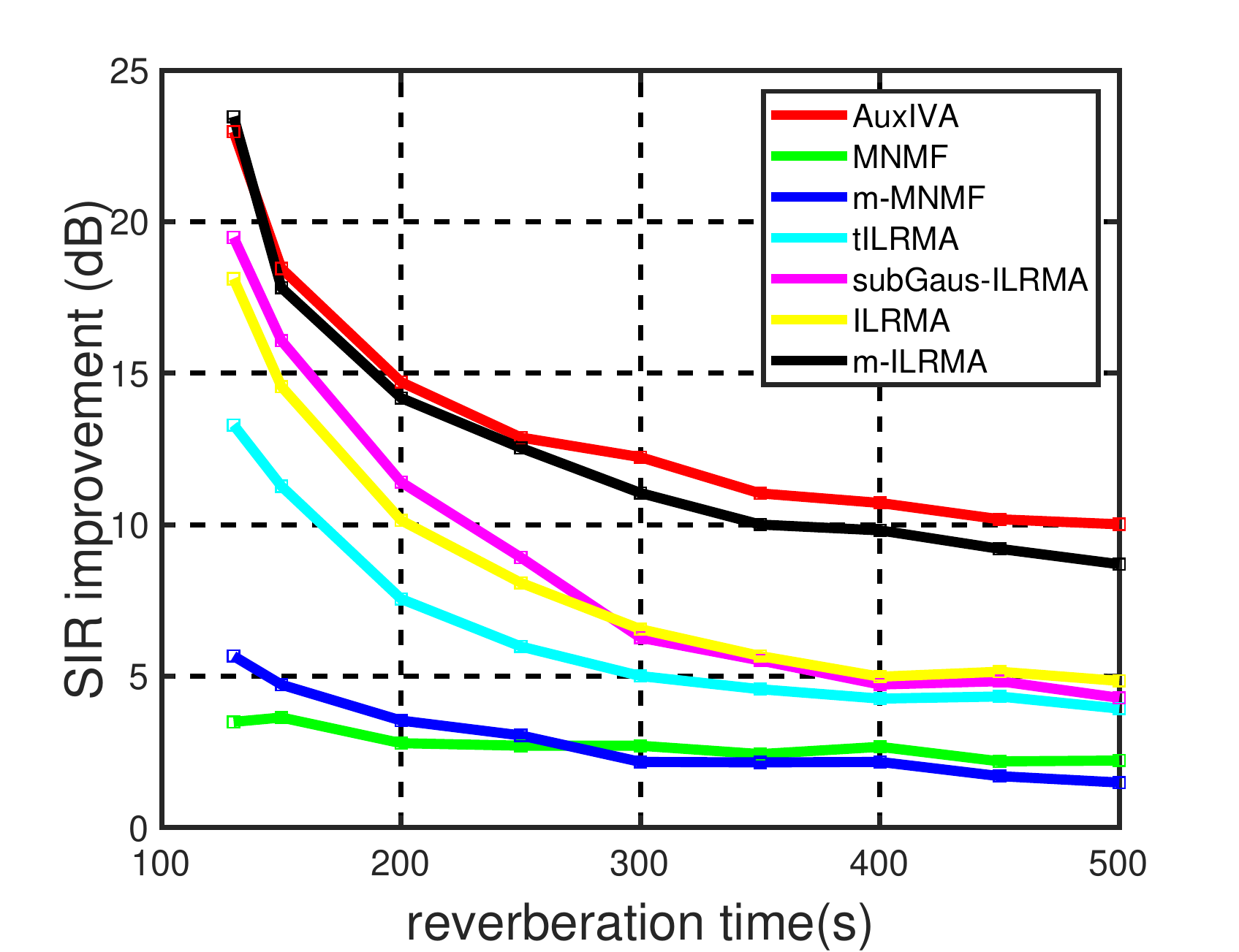}}
  \centerline{(f) female+male}\medskip
\end{minipage}
\vspace{-0.8cm}
\caption{SIR improvement of the comparison methods on WSJ0-reverb. (a), (c), (e) are the results in condition 1. (b), (d), (f) are the results in condition 2.}
\label{fig:8}
\vspace{-0cm}
\end{figure}

Figs. \ref{fig:7} and \ref{fig:8} show the SDR and SIR improvement respectively over the mixed speech in the simulated reverberant environment of  WSJ0-reverb. From the figures, we see that the curves of the SDR improvement produced by {m-ILRMA} are always higher than those produced from the comparison methods. The minimum improvement of m-ILRMA over the comparison methods is 2 dB.

To clearly show the general improvement of {m-ILRMA} over the referenced methods, we average the SDR improvement with respect to different gender combinations and $T_{60}$ for each condition. The average results are listed in Tables \ref{tab:avg_list1} and \ref{tab:avg_list2}, respectively. From the tables, we see that the average SDR improvement brought by the proposed {m-ILRMA} is 2 dB higher than ILRMA in condition 1, and 3 dB higher than the latter in condition 2. The average SIR improvement of {m-ILRMA} is comparable to AuxIVA, and outperforms the other methods.

\subsubsection{Results on semi-real-SISEC2011}

Tables \ref{tab:avg_list3} and \ref{tab:avg_list4} show the separation performance of the comparison methods on the real-world recording environment of semi-real-SISEC2011.
From Table \ref{tab:avg_list3}, we see that the SDR improvement of m-ILRMA is 2 dB higher than that of ILRMA on average in all four situations.
From Table \ref{tab:avg_list4}, we see that the SIR improvement of m-ILRMA is competitive with the best comparison method.

\begin{table}[t]
\vspace{-0cm}
\centering
\caption{SDR improvement (dB) of the comparison methods on semi-real SISEC2011.}\label{tab:avg_list3}
\scalebox{1}{
\setlength{\tabcolsep}{1mm}{
\begin{tabular}{l|cccc}
  \toprule[1pt]
  \multirow{1}{*}{Methods}
    & {mic4 / mic5} & {mic3 / mic6} & {mic2 / mic7} & {mic1 / mic8}  \\
   \midrule[1pt]
    AuxIVA {\cite{ono2011stable}}   & {{2.59} / {2.41}}  & {0.45 / 3.04} & {-0.01 / 2.79}  & {1.84 / 1.25}  \\
    MNMF  {\cite{sawada2013multichannel}}  & {-0.12 / 1.72}  & {-1.34 / 1.17}  & {-2.06 / 1.51}  & {-0.22 / -0.46}\\
    t-ILRMA {\cite{mogami2017independent}} & {2.81 / 3.46} & {0.38 / 2.97} & {0.10 / 3.69} & {1.76 / 1.60} \\
    subG-ILRMA {\cite{ikeshita2018independent}} & {3.43 / 4.63} & {0.75 / 3.42} & {0.73 / 4.58} & {1.90 / 1.86}\\
    ILRMA {\cite{kitamura2016determined}} & {4.20 / 5.16}  & {1.34 / 3.94}  & {0.87 / 4.62} & {2.94 / 2.62} \\
    {m-MNMF} & {-0.99 / 0.91} & {-2.16 / 0.33} & {-2.75 / 0.91} & {-0.81 / -1.04} \\
    {m-ILRMA} & {5.98 / 6.90}  & {1.50 / 4.53}  & {2.57 / 7.04} &  {3.87 / 4.35} \\
  \bottomrule[1pt]
\end{tabular}}
}
\vspace{-0cm}
\end{table}

\begin{table}[!t]
\vspace{-0cm}
\centering
\caption{SIR improvement (dB) of the comparison methods on semi-real SISEC2011.}\label{tab:avg_list4}
\scalebox{1}{
\setlength{\tabcolsep}{1mm}{
\begin{tabular}{l|cccc}
  \toprule[1pt]
  \multirow{1}{*}{Methods}
    & {mic4 / mic5} & {mic3 / mic6} & {mic2 / mic7} & {mic1 / mic8}  \\
   \midrule[1pt]
    AuxIVA {\cite{ono2011stable}}   & {{10.81} / {9.93}}  & {8.49 / 8.74}  & {8.65 / 9.41} & {11.09 / 7.94} \\
    MNMF  {\cite{sawada2013multichannel}}  & {0.60 / 2.65}  & {-0.15 / 3.43}  & {-0.74 / 3.86} & {1.17 / 1.90} \\
    t-ILRMA {\cite{mogami2017independent}} & {7.16 / 6.47} & {4.26 / 5.90} & {4.18 / 6.70} & {6.11 / 4.65} \\
    subG-ILRMA {\cite{ikeshita2018independent}} & {8.38 / 8.12} & {5.09 / 6.74} & {5.41 / 8.26} & {6.65 / 5.24} \\
    ILRMA {\cite{kitamura2016determined}} & {8.64 / 8.05}  & {5.40 / 6.86}  & {5.61 / 7.79} & {7.62 / 5.55} \\
    {m-MNMF} & {-0.23 / 1.42} & {-0.92 / 2.11} & {-1.38 / 2.76} & {0.64 / 0.84} \\
    {m-ILRMA} & {12.47 / 10.71}  & {5.78 / 8.06}  & {8.53 / 10.93} & {9.40 / 8.25} \\
  \bottomrule[1pt]
\end{tabular}}
}
\vspace{-0cm}
\end{table}

\subsection{Discussion}

In this section, we demonstrate the effectiveness of the MinVol prior on the sparsity, orthogonality, and uniqueness of the spectra matrix by comparing ILRMA with m-ILRMA. Before analysis, we first define the sparsity, orthogonality, and uniqueness of a matrix as follows:
\begin{Def}
  Sparseness measurement \cite{hoyer2004non}: The sparseness of a matrix is built on the relationship between the $L_1$ norm and the $L_2$ norm:
\begin{equation}\label{sparseness}
\begin{split}
 \zeta(\mathbf{w}_{k}) = \frac{\sqrt{n}-(\sum_i|w_{ik}|)/\sqrt{\sum_i{{w}_{ik}^2}}}{\sqrt{n}-1}
\end{split}
\end{equation}
\begin{equation}\label{sparseness}
\begin{split}
 \hat{\zeta}(\mathbf{W}) = \frac{1}{K}\sum_k \zeta(\mathbf{w}_{k})
\end{split}
\end{equation}
where $\mathbf{w}_{k} = [w_{1k},\dots,w_{ik},\dots,w_{Ik}]^T$ is the $k$th column of the matrix $\mathbf{W}$, $\zeta(\mathbf{w}_{k})$ calculates the sparseness of the vector $\mathbf{w}_{k}$, and $\hat{\zeta}(\mathbf{W}_{n})$ defines the sparseness of the matrix $\mathbf{W}_{n}$.

The higher the sparsity score is, the stronger the part-based representation ability of the matrix $\mathbf{W}$ is.
\end{Def}

\begin{Def}
Orthogonality measurement \cite{choi2008algorithms,yuan2009projective}: Two nonnegative vectors are orthogonal if and only if they do not have the same non-zero elements, and we measure the orthogonality of a matrix by:
\begin{equation}\label{orthogonality}
\begin{split}
 \operatorname{Orthogonality}(\mathbf{W}) = \| \mathbf{W}^T\mathbf{W} - \mathbf{I} \|
\end{split}
\end{equation}
where $\mathbf{I}$ is an identity matrix.

The lower the orthogonality score is, the stronger the orthogonality between the basis vectors of the matrix $\mathbf{W}$ is.
\end{Def}

\begin{Def}
Uniqueness measurement \cite{theis2005first}:
Assume that $\mathbf{T}$ can be approximated by $\mathbf{T} \approx \mathbf{W}\mathbf{H}$. In the ideal case, we have $\mathbf{T} = \mathbf{W}'\mathbf{H}$. When the equality holds, we have $\mathbf{W}' = \mathbf{T}\mathbf{H}^{-1}$. However, the equality relation can hardly be achieved in practice.  Therefore, the closer the two different solutions $\mathbf{W}$ and $\mathbf{W}^\prime$ are to degeneracy, the better the unique solution of the source model $\mathbf{T}$ is.
Here, we use the squared Frobenius norm to measure their difference:
\begin{equation}\label{uniqueness}
\begin{split}
  \operatorname{Dif}(\mathbf{W},{\mathbf{W}}^\prime) = \| \mathbf{W} - {\mathbf{W}}^\prime \|^2_F
\end{split}
\end{equation}
The lower the uniqueness score is, the stronger the identifiability of the matrix $\mathbf{W}$ is.
\end{Def}

To analyze the sparsity, orthogonality and uniqueness of the spectra matrix generated by ILRMA and m-ILRMA, we averaged the results of 50 spectra matrices in terms of the three measurements. The results are that (i) the sparsity scores of ILRMA and m-ILRMA are 0.65 and 0.68 respectively, (ii) the orthogonality scores are 0.99 and 0.74 respectively, and (iii) the uniqueness scores are 68.53 and 2.73, respectively.
The results show that the spectral matrix of m-ILRMA has stronger representation ability than that of ILRMA, which proves the effectiveness of the MinVol prior for the multichannel BSS.

\section{Conclusion}
\label{Conclusion}
In this paper, we have proposed a MinVol prior for the source model of multichannel BSS methods. To our knowledge, this is the first MinVol prior regularized multichannel BSS model.
The novelty of the MinVol prior lies in the following aspect.
First, we propose a novel MinVol prior distribution for the source model which improves the identifiability, sparseness, and orthogonality of the separated spectrograms produced from the source model.
It performs as a regularization of the source model in the objective functions of the multichannel BSS.
To evaluate its effectiveness, we implement two multichannel MinVol-based BSS algorithms, denoted as m-MNMF and m-ILRMA.
The optimization of the two proposed methods is intractable since that the objective functions contain logarithmic determinant terms.
To overcome this problem, we relax the logarithmic determinant terms with their tightened lower bounds.
Finally, we apply multiplicative update rules to solve the optimization problems.
We have conducted an extensive experimental comparison with five representative comparison methods on four simulated datasets and a real dataset, which are SISEC2011, SISEC2018, WSJ0-anechoic, WSJ0-reverb, and semi-real-SISEC2011, respectively.
Experimental results show that the proposed m-ILRMA outperforms the comparison methods significantly in terms of SDR and SIR. Although m-MNMF does not reach the top performance, it performs better than its counterpart MNMF.
Moreover, we analyzed the identifiability, sparseness, and orthogonality of the spectral matrix produced by ILRMA and m-ILRMA. The results show that the spectral matrix of m-ILRMA has stronger representation ability than that of ILRMA, which proves the effectiveness of the MinVol prior for the multichannel BSS.



%

\appendices

\ifCLASSOPTIONcaptionsoff
  \newpage
\fi

\bibliographystyle{IEEEtran}
\bibliography{refs}

%








\end{document}